\documentclass[a4paper]{article}

\usepackage[margin=1in]{geometry} 

\usepackage{amsmath}
\usepackage{amsthm}
\usepackage{amssymb}
\usepackage{arydshln}
\usepackage{mathtools}
\usepackage{multirow}
\usepackage{bm}
\usepackage{pgfplots}
\usepackage{subcaption}
\usepackage{tikz}

\usepackage[linesnumbered,ruled,vlined]{algorithm2e}
\usepackage{algpseudocode}

\SetKwInput{KwInput}{Input} 
\SetKwInput{KwOutput}{Output}

\usepackage[utf8]{inputenc}
\usepackage{hyperref}
\hypersetup{
	unicode,
	pdfauthor={Tri P. Nguyen, Ilon Joseph, Mayya Tokman},
	pdftitle={Exploring Exponential Time Integration for Strongly Magnetized Charged Particle Motion},
	pdfsubject={A simple article template},
	pdfkeywords={article, template, simple},
	pdfproducer={LaTeX},
	pdfcreator={pdflatex}
}

\usepackage[sort&compress,numbers,square]{natbib}
\theoremstyle{plain}
\newtheorem{theorem}{Theorem}

\theoremstyle{definition}

\usepackage{graphicx, color}
\graphicspath{{fig/}}

\usepackage{mathrsfs} 

\usepackage{lipsum}

\title{Exploring Exponential Time Integration for Strongly Magnetized Charged Particle Motion}
\author{Tri P. Nguyen$^1$ \and Ilon Joseph$^2$ \and Mayya Tokman$^1$}

\date{
	$^1$School of Natural Sciences, University of California, Merced \\ \texttt{\{tnguyen478, mtokman\}@ucmerced.edu}\\%
	$^2$Lawrence Livermore National Laboratory \\ \texttt{joseph5@llnl.gov}\\[2ex]%
}

\begin{document}
	\maketitle
	
\begin{abstract}
A fundamental task in particle-in-cell (PIC) simulations of plasma physics is solving for charged particle motion in electromagnetic fields. This problem is especially challenging when the plasma is strongly magnetized due to numerical stiffness arising from the wide separation in time scales between highly oscillatory gyromotion and overall macroscopic behavior of the system. In contrast to conventional finite difference schemes, we investigated exponential integration techniques to numerically simulate strongly magnetized charged particle motion. Numerical experiments with a uniform magnetic field show that exponential integrators yield superior performance for linear problems (i.e. configurations with an electric field given by a quadratic electric scalar potential) and are competitive with conventional methods for nonlinear problems with cubic and quartic electric scalar potentials.\\

\end{abstract}
\vfill
\paragraph{Acknowledgements} This work was supported in part by the Department of Energy [Contract DE-AC52 07NA27344] and the National Science Foundation [Award Numbers 1840265, 2012875].

\newpage
\section{Introduction}
\label{Introduction}
Solving for charged particle dynamics is a key problem in particle-in-cell (PIC) simulations of plasma physics, a task known as the particle pushing problem. Since realistic simulations call for the modeling of a vast number of particles, the problem is computationally intensive. This task is especially challenging when the plasma is strongly magnetized in which case charged particles gyrate about magnetic field lines in highly oscillatory gyromotion. In contrast, the macroscopic evolution of the system occurs on a time scale orders of magnitude slower. The presence of such a wide range of time scales in the system results in the numerical stiffness of the equations modeling the dynamics. Moreover, accuracy requirements of simulations typically demand resolution at the scale of the gyromotion, which necessitate small time steps for conventional time integration schemes. These difficulties, therefore, call for computationally efficient numerical particle pushing methods.

The standard approach to solving the particle pushing problem numerically is to discretize the equations of motion with a finite-difference model from which the dynamical state of the particle is advanced by a time stepping algorithm \cite{Birdsall, Hockney}. Two well-known examples of this conventional approach are the Boris \cite{Boris} and Buneman \cite{Buneman} particle pushers. Both methods stagger particle position and velocity by one-half time step resulting in a leapfrog-like centered-difference scheme that gives second-order accuracy in time. The Boris algorithm, in particular, currently enjoys status as the de facto particle pusher \cite{Qin}. A fundamental requirement in this framework is that electromagnetic fields be approximately constant over each time step size. Consequently, problems with large field gradients demand small step sizes to maintain accurate solutions, which results in excessive computational expense.

The investigation of computationally efficient numerical particle pushers continues to be an active research field. For example, investigation into numerical methods that address this time step size restriction include \cite{Brackbill, ChenG, Cohen, Filbet1, Filbet2, Genoni, Vu}.

Notable among the more recent developments are the energy-conserving, asymptotic preserving scheme \cite{Ricketson} and the filtered Boris algorithm \cite{Hairer}. The first method is a modified implicit Crank-Nicolson scheme that conserves energy and incorporates an effective force in the velocity update that captures the leading order drift motion (called grad-$B$ drift) in non-uniform magnetic fields. This effective force is carefully chosen such that it approximates the grad-$B$ force acting on the guiding center in a gyro-averaged sense. The filtered Boris algorithm, on the other hand, modifies the standard Boris pusher by introducing so-called filter functions to more accurately resolve the fast oscillations in particle velocity due to strong magnetic fields. Different choices of filter functions and choices of the positions where the magnetic field is evaluated yield different variants of the filtered Boris algorithm. It is interesting to note that for problems with constant magnetic and electric fields, the filtered Boris algorithm reduces to a type of an exponential integrator that solves the problem exactly. However, we shall demonstrate that our exponential integrators presented in this paper are exact solvers for problems with a constant magnetic field and electric fields that are linear functions of the particle position in addition to constant electric fields. Both the modified Crank-Nicolson scheme and the filtered Boris algorithm (for the general case of arbitrary magnetic fields) are implicit methods and are more complex to implement than the standard Boris algorithm. Hence, they have the advantage of allowing for larger time step sizes for problems with non-uniform electromagnetic fields, but at the cost of being more computationally expensive than the standard Boris pusher.

This paper explores an alternative approach to numerical particle pushing using a technique called exponential integration. Exponential integrators approximate the solution of a nonlinear dynamical system in terms of exponential-like functions of matrices which are either Jacobians of the system or their approximations. Exponential methods offer several desirable features. By construction, exponential time integrators solve the linear portion of the particle pushing problem exactly thus accounting for the electric field gradient component of the solution.  While traditional particle pushers such as the Boris and Buneman algorithms assume that the electric field gradient is nearly zero over the course of the time step, exponential integration methods allow for a non-zero gradient and enable larger time steps to be taken.  In addition, since computing individual particle trajectories is a low dimensional problem, it is possible to evaluate the exponential-like functions of the Jacobians required by exponential methods with relatively low computational cost. In this paper we exploit the good stability properties of the exponential integration methods and the low-dimensionality of the particle pushing problem to propose exponential integrators for calculating the dynamics of particles under the influence of a strong constant magnetic field and spatially varying electric fields. Similar to the modified Crank-Nicolson scheme and the filtered Boris algorithm, the exponential integrators presented here can compute accurate solutions using larger step sizes but are more complex and computationally expensive than conventional particle pushers. However, these exponential integrator particle pushers are \emph{explicit} methods in contrast to the Crank-Nicolson and filtered Boris pushers. We emphasize that since this is an initial exploration into the relatively novel approach of numerical particle pushing by exponential integration, this study focuses on problems with a uniform magnetic field as a first step and defer investigation into problems with non-uniform magnetic fields for future work.

The organization of this paper is as follows: Section 2 describes the equations of motion of the particle pushing problem. Section 3 presents exponential integrators used for solving these equations. A computational technique to evaluate the exponential-like matrix functions which constitute the main computational expense of an exponential integrator is discussed in section 4. Numerical experiments for several test problems comparing exponential integrators with the Boris and Buneman algorithms are presented in section 5. Finally, section 6 summarizes and concludes this paper. Two appendices are provided that describe in detail the Boris and Buneman particle pushing algorithms and prove a theorem justifying our method to compute matrix functions.

\section{The Particle Pushing Problem}
\label{ParticlePushingProblem}
For typical applications of plasma physics, the dominant forces of the system are due to electromagnetic fields. In the presence of electric field $\bm{E}$ and magnetic field $\bm{B}$ the force acting on a particle of mass $m$ and electric charge $q$ is given by the Lorentz force equation
\[
m\frac{d\bm{v}}{dt} = q(\bm{E} + \bm{v} \times \bm{B}),
\]
where $\bm{v}$ is the particle's velocity. Denoting the particle's position by $\bm{x}$, the particle pushing problem is expressed by the Newtonian equations of motion: 
\begin{subequations}\label{Newtonform}
\begin{align}
\frac{d\bm{x}}{dt} & = \bm{v}, \label{v}\\[0.5em]
\frac{d\bm{v}}{dt} & = \frac{q}{m}(\bm{E} + \bm{v}\times\bm{B}). \label{LorentzF}
\end{align}
\end{subequations}

An equivalent formulation can be derived in terms the particle's position $\bm{x}$ and conjugate momentum $\bm{p}$ by considering the Hamiltonian of the system
\begin{equation}\label{Hamiltonian}
H(\bm{x},\bm{p}) = \frac{1}{2m}\|\bm{p} - q\bm{\mathbf{A}}(\bm{x})\|^2 + qV(\bm{x}).
\end{equation}
Here, $\mathbf{A}(\bm{x})$ is the magnetic vector potential and $V(\bm{x})$ is the electric scalar potential such that the magnetic and electrics fields are given by
\[
\bm{B} = \nabla_{\bm{x}}\times\bm{\mathbf{A}}
\]
and
\[
\bm{E} = -\nabla_{\bm{x}} V(\bm{x})
\]
respectively. Hamilton's equations thus give the equations of motion:
\begin{subequations}\label{Heq}
\begin{align}
\frac{d\bm{x}}{dt} & = \phantom{-}\nabla_{\bm{p}}H(\bm{x},\bm{p}), \label{Heq1}\\[0.5em]
\frac{d\bm{p}}{dt} & = -\nabla_{\bm{x}}H(\bm{x},\bm{p}). \label{Heq2}
\end{align}
\end{subequations}

\section{Exponential Integrator Particle Pusher}
Observe that if the particle state is known at time $t = t_n$, then the particle pushing problem is of the form
\begin{equation}\label{IVP}
\frac{d\bm{u}}{dt} = \bm{f}(\bm{u}), \quad \bm{u}_n = \bm{u}(t_n),
\end{equation}
where $\bm{f}(\bm{u})$ is the right-hand side function of the equations of motion. Taking a first-order Taylor expansion of the right-hand side function of \eqref{IVP} about the known state $\bm{u}_n$, we get
\begin{equation}\label{1storderTaylor}
\frac{d\bm{u}}{dt} = \bm{f}(\bm{u}_n) + A_n(\bm{u} - \bm{u}_n) + \bm{r}(\bm{u}),
\end{equation}
where
\[
A_n = \frac{\partial\bm{f}}{\partial\bm{u}}\bigg|_{\bm{u} = \bm{u}_n}
\]
is the Jacobian matrix and
\begin{equation}\label{nonlinRterm}
\bm{r}(\bm{u}) = \bm{f}(\bm{u}) - \bm{f}(\bm{u}_n) - A_n(\bm{u} - \bm{u}_n)
\end{equation}
is the nonlinear remainder term. Multiplying equation \eqref{1storderTaylor} by the integrating factor $\exp(-tA_n)$ and then integrating over the time interval $[t_n, t_n + h]$, we obtain the integral equation
\begin{equation}\label{intform1}
\bm{u}(t_n + h) = \bm{u}_n + h\,\varphi_1(hA_n)\bm{f}(\bm{u}_n) + \int\limits_{t_n}^{t_n + h} e^{A_n(t_n + h - t)}\bm{r}(\bm{u}(t))\,dt,
\end{equation}
where $\varphi_1(hA_n)$ is a matrix function defined by the MacLaurin series expansion of the scalar analytic function 
\[
\varphi_1(z) = \frac{e^z - 1}{z} = \int\limits_0^1 e^{z(1 - \tau)}\,d\tau
\]
applied to the matrix argument $hA_n$. Letting $t = t_n + \tau h$, equation \eqref{intform1} is equivalently expressed by
\begin{equation}\label{intform2}
\bm{u}(t_n + h) = \bm{u}_n + h\,\varphi_1(hA_n)\,\bm{f}(\bm{u}_n) + h\int\limits_0^1 e^{hA_n(1 - \tau)}\,\bm{r}(\bm{u}(t_n + \tau h))\,d\tau.
\end{equation}
Equation \eqref{intform2} is a starting point from which an exponential integrator can be derived as follows. Let $\bm{u}_n$ be a numerical solution obtained at a previous integration step and let $h$ be a specified time step size. Then formula \eqref{intform2} gives the exact solution at the next time step $\bm{u}(t_n + h)$. To approximate $\bm{u}(t_n + h)$ we can construct an exponential integrator by accomplishing the following two tasks:
\begin{enumerate}
\item[(i)] Develop a quadrature rule to approximate the nonlinear integral term 
\[
h\int\limits_0^1 e^{hA_n(1 - \tau)}\bm{r}(\bm{u}(t_n + \tau h))\,d\tau;
\]
\item[(ii)] Construct a technique to compute exponential-like matrix functions, called $\varphi$ functions.
\end{enumerate}

The Exponential Propagation Iterative Methods of Runge-Kutta-type (EPIRK) framework has been shown to allow construction of efficient exponential methods that reduce computational cost per time step compared to other exponential integrators \cite{Tokman2006, Tokman2010}. These methods have been shown to be computationally efficient for several applications including MHD modeling \cite{Einkemmer}. Thus, this is the first class of exponential methods we will explore for solving the particle pushing problem.

The formulation for a general EPIRK method is given by the following ansatz:
\begin{subequations}
\begin{align}
\bm{U}_i & = \bm{u}_0 + a_{i1}\psi_{i1}(g_{i1}hA_0)\,h\bm{f}(\bm{u}_0) + \sum_{j=2}^i a_{ij}\psi_{ij}(g_{ij}hA_0)\,\Delta^{(j-1)}\bm{r}(\bm{u}_0), \\
& \qquad i = 1,2,\ldots,s\rm{-}1, \notag \\[0.25em]
\bm{u}_1 & = \bm{u}_0 + b_1\psi_{s1}(g_{s1}hA_0)\,h\bm{f}(\bm{u}_0) + \sum_{j=2}^s b_j\psi_{sj}(g_{sj}hA_0)\,h\Delta^{(j-1)}\bm{r}(\bm{u}_0),
\end{align}
\end{subequations}
where the matrix $\psi_{ij}$ functions are defined by the scalar functions
\[
\psi_{ij}(z) = \sum_{k=1}^s p_{ijk}\,\varphi_k(z)
\]
with
\[
\varphi_k(z) = \int\limits_0^1 e^{z(1 - \tau)} \frac{\tau^{k-1}}{(k - 1)!}\,d\tau, \qquad k = 1,2,\ldots,
\]
and the vectors $\Delta^{(j-1)}\bm{r}(\bm{u}_0)$ are the $j\rm{-}1$th forward differences of the nonlinear remainder function \eqref{nonlinRterm} computed on the nodes $\bm{u}_0,\bm{U}_1,\bm{U}_2,\ldots,\bm{U}_{s\rm{-}1}$. Here, the first through the $j\rm{-}1$th forward differences of the nonlinear remainder function are defined by:
\[
\begin{array}{lcl}
\Delta\bm{r}(\bm{u}_0) & = & \bm{r}(\bm{U}_1) - \bm{r}(\bm{u}_0), \\[0.25em]
\Delta^2\bm{r}(\bm{u}_0) & = & \Delta\bm{r}(\bm{U}_1) - \Delta\bm{r}(\bm{u}_0) \\
& = & \bm{r}(\bm{U}_2) - 2\bm{r}(\bm{U}_1) + \bm{r}(\bm{u}_0), \\[0.25em]
\multicolumn{1}{c}{\vdots} & \vdots & \,\vdots \\[0.5em]
\Delta^{j-1}\bm{r}(\bm{u}_0) & = & \Delta^{j-2}\bm{r}(\bm{U}_{j-1}) - \Delta^{j-2}\bm{r}(\bm{u}_0) \\
& = & \displaystyle\sum_{i=0}^{j-1}(-1)^i{j\rm{-}1\choose i}\bm{r}(\bm{U}_{j-1-i}), \quad\text{where } \bm{U}_0 = \bm{u}_0.
\end{array}
\]
The coefficients $a_{ij},g_{ij},b_j,p_{ijk}$ are determined by satisfying the desired order conditions. This procedure has been used to derive the following two methods:
\begin{itemize}
\item Second-order exponential propagation method \cite{Tokman2006}
\begin{equation}\label{EP2}\tag{EP2}
\bm{u}_{n+1} = \bm{u}_n + h\,\varphi_1(hA_n)\bm{f}(\bm{u}_n),
\end{equation}
\item Third-order exponential propagation, Runge-Kutta type method \cite{Stewart}
\begin{equation}\label{EPRK3}\tag{EPRK3}
\begin{array}{ccl}
\bm{U_1} & = & \bm{u}_n + h\,\varphi_1\left(\frac{3}{4}\,hA_n\right)\,\bm{f}(\bm{u}_n), \\[0.5em]
\bm{R}_1 & = & \bm{f}(\bm{U_1}) - \bm{f}(\bm{u}_n) - A_n(\bm{U_1} - \bm{u}_n), \\[0.5em]
\bm{u}_{n+1} & = & \bm{u}_n + h\,\varphi_1(hA_n)\,\bm{f}(\bm{u}_n) + 2\,h\,\varphi_3(hA_n)\,\bm{R}_1.
\end{array}
\end{equation}
\end{itemize}

Note that in principle exponential integrators solve linear differential equations exactly. As a consequence, the region of stability for exponential integrators is the left-half of the complex plane; i.e. exponential integrators are $A$-stable.

\section{Computing the Matrix $\varphi$ Functions}\label{MatrixFunctions}
In any exponential integration scheme the most computationally expensive step is the evaluation of each action of an exponential-like matrix function $\varphi$. For small matrices, approximation techniques such as a finite Taylor polynomial, Pad\'{e} approximation, or scaling and squaring have been common approaches to compute the matrix function $\varphi(A)$. However, these methods are quite computationally expensive and are usually only used when the computational cost of evaluating matrix functions is not important. (For a detailed discussion on the computational issues of various methods to evaluate the exponential of a matrix, see \cite{Moler1, Moler2}.) Other methods include the Leja method \cite{Caliari} and Krylov subspace projection methods. Krylov subspace projection methods, in particular, have been shown to be computationally efficient techniques to approximate the action the matrix $\varphi$ function on a vector when the matrix is large \cite{Niesen, Gaudreault}.

The particle pushing problem, however, is a low dimensional problem. Even in three dimensions only six equations of motion have to be integrated simultaneously to advance a particle's trajectory.  Exploiting this low dimensionality, we propose an alternative approach to compute the matrix $\varphi$ functions by means of evaluating a finite degree matrix polynomial that yields an analytic result. We show that this direct analytic method is computationally efficient for such small problems.

The following theorem \cite{Sylvester, Buchheim, Gantmacher} asserts that any analytic matrix function has an exact expression in terms of a finite degree matrix polynomial.
\begin{theorem}[Lagrange-Sylvester Interpolation Formula]\label{matfuncthm}
Let $A$ be an $N \times N$ matrix and let $f$ be a scalar function analytic in a domain containing the spectrum of $A$. Then there exists a unique polynomial $p$ of (at most) degree $N \rm{-} 1$ such that:
\begin{enumerate}
\item If the eigenvalues of $A$ are all distinct, then $p$ is the polynomial that interpolates $f$ on the spectrum of $A$;
\item If $A$ has repeated eigenvalues, then $p$ is the polynomial that interpolates $f$ on the spectrum of $A$. In addition, for each eigenvalue $\lambda_j$ with multiplicity $r_j$, the polynomial $p$ also satisfies $r_j\rm{-}1$ osculating conditions in the sense that all derivatives up to order $r_j\rm{-}1$ of both $p$ and $f$ agree with each other at the interpolation node $\lambda_j$. In other words:
\[
\begin{array}{cccl}
p(\lambda_j) & = & f(\lambda_j) & \text{interpolation condition,} \\[0.25em]
p'(\lambda_j) & = & f'(\lambda_j) & \text{1st osculating condition,} \\[0.25em]
p''(\lambda_j) & = & f''(\lambda_j) & \text{2nd osculating condition,} \\[0.25em]
\vdots & \vdots & \vdots & \qquad\vdots \\[0.5em]
p^{(r_j\rm{-}1)}(\lambda_j) & = & f^{(r_j\rm{-}1)}(\lambda_j) & r_j\rm{-}1\textsuperscript{th}\text{ osculating condition,}
\end{array}
\]
where the superscript denotes the order of the derivative with respect to $\lambda$.
\end{enumerate}
In either case, the polynomial $p$ applied to the matrix argument $A$ is equivalent to the matrix function $f(A)$. That is,
\[
p(A) = f(A).
\]
\end{theorem}

\begin{proof}
See appendix B.
\end{proof}
Our method applies this theorem to calculate the matrix exponential-like $\varphi_k$ functions, which is presented in Algorithm \ref{Lagrange-Sylvester}. Note that the numerical computation of the scalar $\varphi_k(z)$ functions for $k \geq 1$ is subject to catastrophic cancellation for small argument values $z = h\lambda_j$. To overcome this issue, we employ the Cauchy integral formula suggested by Kassam and Trefethen \cite{Kassam}. Furthermore, our particular implementation of the Lagrange-Sylvester formula employs Newton divided differences \cite{Burden, Kincaid} to calculate the interpolation polynomial. That is, we seek the polynomial of the form
\begin{align*}
p(\lambda) & = b_0 + b_1(\lambda - \lambda_1) + b_2(\lambda - \lambda_1)(\lambda - \lambda_2) + \ldots \\
& \phantom{=} \quad + b_{N-1}(\lambda - \lambda_1)\cdots(\lambda - \lambda_{N-1})
\end{align*}
that agrees with the $\varphi_k$ function on the eigenvalues $\lambda_1, \lambda_2, \ldots, \lambda_N$, where the polynomial coefficients are given by the Newton divided differences:
\[
\begin{array}{lcl}
b_0 & = & \varphi_k[\lambda_1], \\
b_1 & = & \varphi_k[\lambda_1, \lambda_2], \vspace*{0.5em}\\
b_2 & = & \varphi_k[\lambda_1, \lambda_2, \lambda_3], \vspace*{0.25em}\\
\,\,\vdots & \vdots & \qquad \vdots \vspace*{0.25em}\\
b_{N-1} & = & \varphi_k[\lambda_1, \ldots, \lambda_N],
\end{array}.
\]
Here, the Newton divided differences on the right-hand side are defined as follows. The zeroth divided difference is
\[
\varphi_k[\lambda_i] = \varphi_k(\lambda_i).
\]
The first divided difference is
\[
\varphi_k[\lambda_i, \lambda_{i+1}] = \left\{\begin{array}{ll}
\varphi'(\lambda_{i+1}) & \text{if } \lambda_i = \lambda_{i+1}, \vspace*{1em}\\
\dfrac{\varphi_k[\lambda_{i+1}] - \varphi_k[\lambda_i]}{\lambda_{i+1} - \lambda_i} & \text{otherwise.}
\end{array}\right.
\]
The second divided difference is
\[
\varphi_k[\lambda_i, \lambda_{i+1}, \lambda_{i+2}] = \left\{\begin{array}{ll}
\dfrac{1}{2!}\varphi_k''(\lambda_i) & \text{if }\lambda_i = \lambda_{i+1} = \lambda_{i+2}, \vspace*{1em}\\
\dfrac{\varphi_k[\lambda_{i+1, i+2}] - \varphi_k[\lambda_i, \lambda_{i+1}]}{\lambda_{i+2} - \lambda_i} & \text{otherwise.}\end{array}\right.
\]
By recursive definition, the $j$\textsuperscript{th} divided difference is
\begin{align*}
& \varphi_k[\lambda_i, \ldots, \lambda_{i+j}] \\[0.5em]
& \, = \left\{\begin{array}{ll}
\dfrac{1}{j!}\varphi_k^{(j)}(\lambda_{i+j}) & \text{if } \lambda_i, \ldots, \lambda_{i+j} \text{ are all equal,} \vspace*{1em}\\
\dfrac{\varphi_k[\lambda_{i+1}, \ldots, \lambda_{i+j}] - \varphi_k[\lambda_i, \ldots, \lambda_{i+j-1}]}{\lambda_{i+j} - \lambda_i} & \text{otherwise,}
\end{array}\right.
\end{align*}
where the superscript denotes the order of the derivative of the $\varphi_k$ function with respect to $\lambda$. Thus, our algorithm is a generalization of Method 10 in \cite{Moler1, Moler2}, which computes the matrix exponential, using an interpolation polynomial calculated with Newton divided differences.

\begin{algorithm}
\caption{Lagrange-Sylvester Formula to compute the matrix function $\varphi_k(hA)$}\label{Lagrange-Sylvester}
\begin{algorithmic}[1]
\State Solve for the eigenvalues of $A$.
\State Solve for the interpolation polynomial $p$ such that for each eigenvalue $\lambda_j$:
\[
\begin{array}{ccc}
p(\lambda_j) & = & \varphi_k(h\lambda_j), \\[0.25em]
p'(\lambda_j) & = &\varphi_k'(h\lambda_j), \\[0.25em]
p''(\lambda_j) & = & \varphi_k''(h\lambda_j), \\[0.25em]
\vdots & \vdots & \vdots \\[0.5em]
p^{(r_j-1)}(\lambda_j) & = & \varphi_k^{(r_j-1)}(h\lambda_j),
\end{array}
\]
where $r_j \geq 1$ is the multiplicity of $\lambda_j$ and the superscript denotes the order of the derivative with respect to $\lambda$.
\State Evaluate the matrix polynomial $p(A)$, which is equal to $\varphi_k(hA)$ by theorem \eqref{matfuncthm}.
\end{algorithmic}
\end{algorithm}

To illustrate the computational efficiency of Algorithm \ref{alg:lsf} we compared two implementations of the second-order EP2 and third-order EPRK3 exponential integrators using (i) a Krylov subspace projection method called KIOPS \cite{Gaudreault}, and (ii) the Lagrange-Sylvester formula in MATLAB. Both implementations of the exponential integrators are used to solve the Hamiltonian equations of motion over the time interval $[0, 100]$ for a particle of unit mass and unit charge in a uniform magnetic field $\bm{B} = 100 \, \hat{\bm{z}}$ with electric fields
\[
\bm{E} = -\frac{100}{3}\begin{bmatrix}
x^3 \\
y^3
\end{bmatrix}\quad\text{for the two-dimensional model}
\]
and
\[
\bm{E} = -\frac{1}{3}\begin{bmatrix}
100x^3 \\
100y^3 \\
10z^3
\end{bmatrix}\quad\text{for the three-dimensional model.}
\]
The computed solutions were compared against a reference solution obtained by the MATLAB \texttt{ode113} solver with error tolerances set to $10^{-12}$ for \texttt{RelTol} (relative error tolerance) and $10^{-12}$ for \texttt{AbsTol} (absolute error tolerance). Relative error of the exponential integrator solution is defined as
\[
\text{error} = \frac{\|\bm{x}^* - \bm{x}\|}{\|\bm{x}^*\|},
\]
where $\bm{x}^*$ is the particle position of the reference solution and $\bm{x}$ is the particle position of the exponential integrator solution, both evaluated at the final time $t = 100$, and $\|\cdot\|$ denotes the Euclidean norm.

\begin{figure}[h]
\centering
\begin{tabular}{cc}
2D Model & 3D Model \\
\includegraphics[scale=0.6]{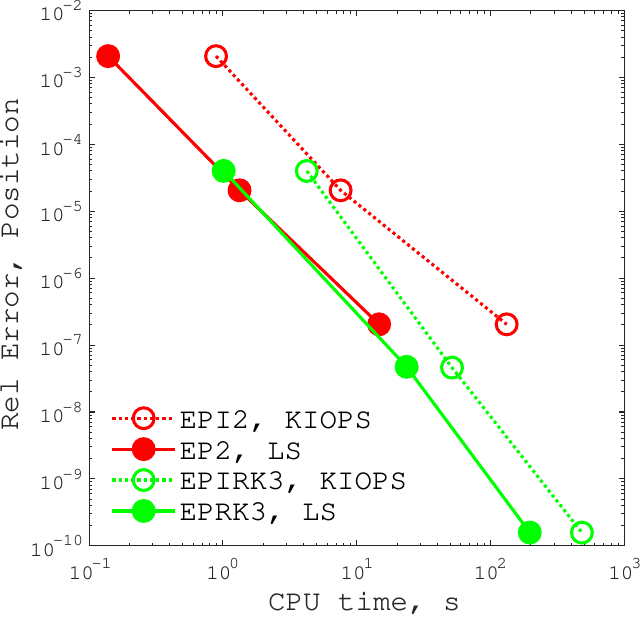} & \includegraphics[scale=0.6]{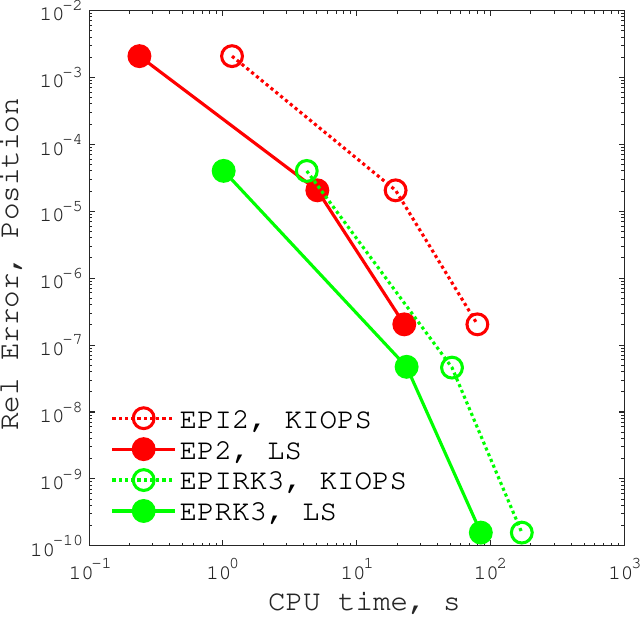}
\end{tabular}
\caption{Precision diagram showing performances of 2nd and 3rd order exponential integrators subspace projection (dotted lines) and Lagrange-Sylvester formula (solid lines) for step sizes $h = 0.01, 0.001, 0.0001$ over the time interval [0, 100].}\label{LSvsKIOPS}
\end{figure}

Figure \ref{LSvsKIOPS} shows precision diagrams (CPU time vs error) comparing two implementations of the EP2 and EPRK3 integrators. As we can see from the figure, the Lagrange-Sylvester formula enables significant computational savings for exponential integration compared to the KIOPS methods (with iteration convergence tolerance set to $1\rm{e-}9$). For the two-dimensional test problem, the EPI2 integrator using KIOPS takes on average seven times longer than the EP2 integrator using the Lagrange-Sylvester formula to compute the final solution. Similarly, the EPIRK3 integrator using KIOPS takes on average 2.7 times longer than the EPRK3 integrator using the Lagrange-Sylvester formula to compute the final solution for the two-dimensional test problem. For the three-dimensional test problem, the EPI2 integrator using KIOPS takes on average four times longer than the EP2 integrator using the Lagrange-Sylvester formula to compute the final solution. Likewise, the EPIRK3 integrator using KIOPS takes on average 2.8 times longer than the EPRK3 integrator using the Lagrange-Sylvester formula to compute the final solution for the three-dimensional test problem. This is expected since the KIOPS technique is designed for large scale problems and we expect the Lagrange-Sylvester formula to be more efficient for these low-dimensional systems.

We note that the Jacobian matrices of the problem possess some structure. For example, the Newtonian formulation of the problem yields a zero block matrix and an identity block matrix inside the Jacobian. Symmetries also exist in the Hamiltonian form of the Jacobian as well. It is possible that additional computational savings can be derived for both the Lagrange-Sylvester and KIOPS algorithms. We will investigate this direction in the future.

\section{Numerical Experiments}
\label{NumericalResults}
To assess the performance of exponential integrators for the particle pushing problems we use a series of test configurations and compare these integrators to the widely used Boris and Buneman algorithms.  We selected the second and third order exponential methods EP2 and EPRK3 to integrate the Hamiltonian form of the equations of motion \eqref{Heq}. Both exponential integrators are implemented with Lagrange-Sylvester interpolation formula to compute the matrix $\varphi$ functions as described in Algorithm \ref{alg:lsf}. 

The test problems under examination model a particle of unit mass and unit charge in a uniform in time and space magnetic field aligned in the $z$ direction, $\bm{B} = 100 \, \hat{\bm{z}}$, and a non-uniform electric field $\bm{E}$ resulting in anisotropic drift motion of particles along periodic orbits. Specifically, the test problems are set up with electric fields characterized by electric scalar potential wells and hills (in the $xy$ plane) of quadratic, cubic, and quartic forms. To enable comparison across the different potentials, the configurations of the potentials are such that the largest absolute eigenvalue of the Hessian matrix for each potential form (quadratic, cubic, and quartic) is set to the same value at the initial particle position. Our reasoning is that the largest absolute eigenvalue, which we denote by $|V''|$, gives a measure of the gradient of the electric field and also is a rough estimate of the electric oscillation frequency. The ratio of this eigenvalue to the magnetic field strength defines an ordering parameter that determines different regimes of particle motion \cite{Ilon}.  Thus, we conduct numerical experiments for several values of this ratio, i.e. $|V''|/B = 1/100$, $|V''|/B = 1/10$, and $|V''|/B = 1$.  The equations were solved over the time interval of $[0,100]$ which is equivalent to just nearly 1,600 gyroperiods.  

To get an estimate of the error in our numerical experiments, we computed approximations to the solutions of test problems using the MATLAB \texttt{ode113} integrator with error tolerances set to $10^{-12}$ for \texttt{RelTol} (relative error tolerance) and $10^{-12}$ for \texttt{AbsTol} (absolute error tolerance) and designated it as the reference solution. The relative error of the numerical solution is defined by
\[
\text{error } = \frac{\|\bm{x}^* - \bm{x}\|}{\|\bm{x}^*\|}, 
\]
where $\bm{x}^*$ is the particle position of the reference solution, $\bm{x}$ is the particle position of the solution of the particle pusher, and $\|\cdot\|$ denotes the Euclidean norm.

We also examined the particle energy over the longer time interval of $[0, 20000]$ corresponding to over $3.18 \times 10^{5}$ gyroperiods. Each electric scalar potential is configured such that at the initial condition the particle energy is unity. Since the particle pushing problem is a Hamiltonian system, energy is a conserved quantity and, therefore, any error in the computed energy gives an indication of the long term accuracy of the particle pusher under examination. The experiments were first performed with all particle pushers using the same step size set to the minimum of $h = 0.01$ or the largest step size such that the energy error is within 10\% of the true value. That is,
\[
h = \min\{0.01, \underline{h}\},
\]
where
\[
\underline{h} \coloneqq \max\left\{h > 0 \colon\quad 0.9 \leq \text{energy} \leq 1.1 \quad\text{and}\quad \frac{20000}{h}\in\mathbb{N}\right\}.
\]
The experiments were then repeated for the exponential integrators using the largest step such that relative energy error of the EP2 solution is within 10\% of the true energy.

For the two dimensional models, we also included two additional experiments. The first experiment examines the performances of the particle pushers for a simple non-uniform magnetic field problem called the grad-$B$ drift problem. The second experiment examines the computed gyroradius of each particle pusher for a linear $\bm{E}\times\bm{B}$ drift problem.

All experiments in this section were run on a PC with an Intel Core i7-1255U processor at clock speed 1.7 GHz and 16 GB of RAM and implemented in C++ using the Eigen C++ template library for linear algebra \cite{Eigen} with the exception of the grad-B drift problem and the gyroradius experiments, which were implemented in MATLAB. All computations were calculated with double precision floating point operations.

\subsection{Two Dimensional Model}\label{Results2DModel}
All two-dimensional test problems set initial particle position and velocity at $\bm{x}_0 = (1, 0)$ and $\bm{v}_0 = (0, -1)$, respectively. Table \ref{2DWellConfig} lists the configurations for the electric scalar potential wells and their corresponding electric fields. Configurations for electric scalar potential hills and the corresponding electric fields are shown in table \ref{2DHillConfig}.

\begin{figure}[h!]
\centering
\begin{tabular}{ccc}
\multicolumn{3}{c}{Potential Wells $\frac{|V''|}{B} = \frac{1}{100}$} \\
Quadratic & Cubic & Quartic \\
\includegraphics[scale=0.325]{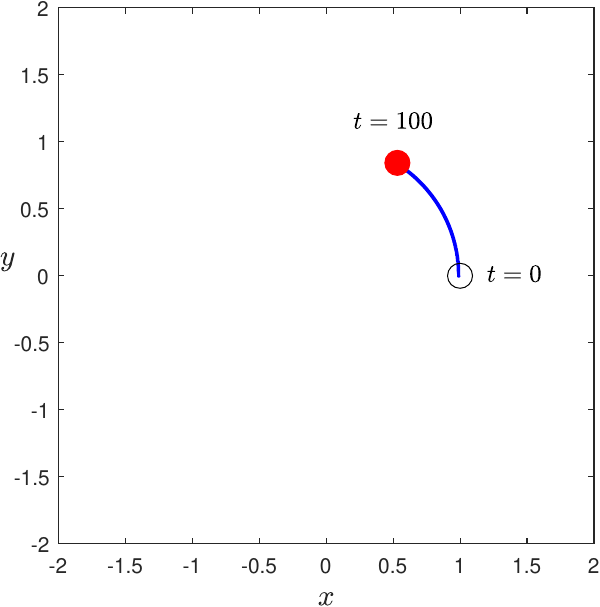} &
\includegraphics[scale=0.325]{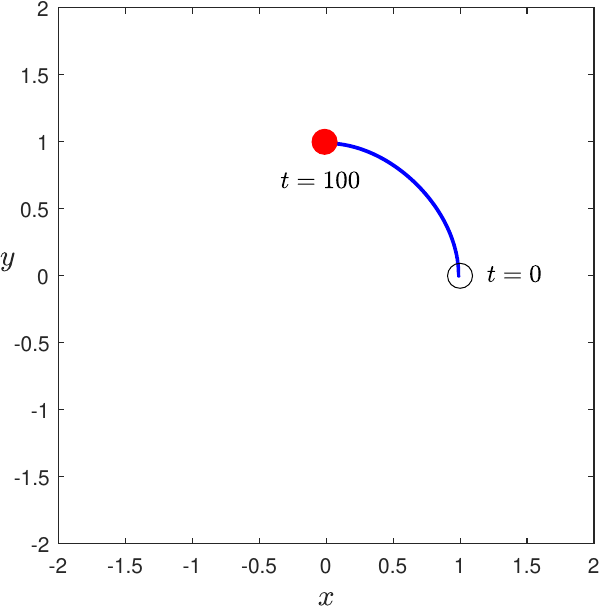} &
\includegraphics[scale=0.325]{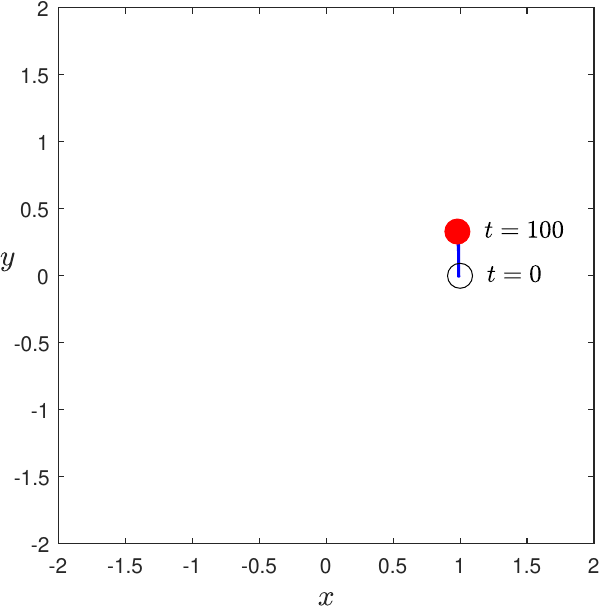} \\
\includegraphics[scale=0.325]{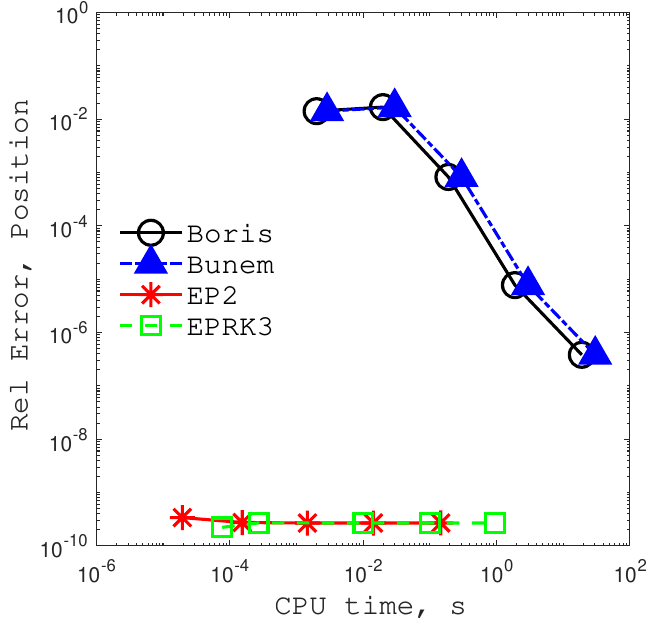} &
\includegraphics[scale=0.325]{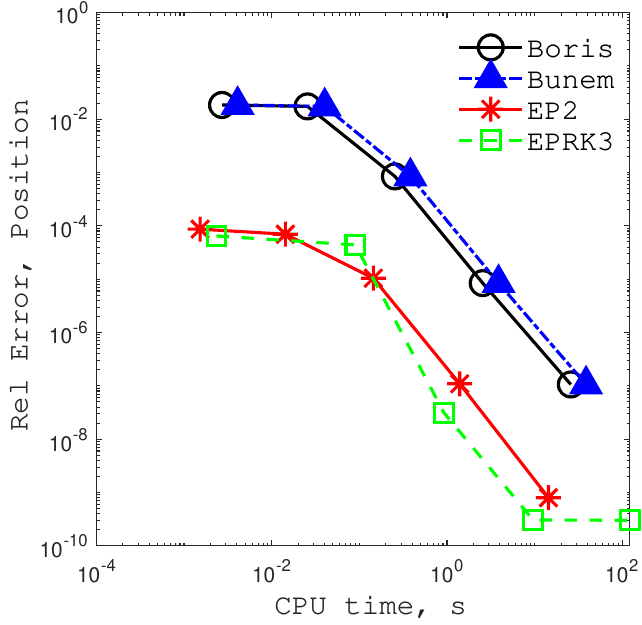} &
\includegraphics[scale=0.325]{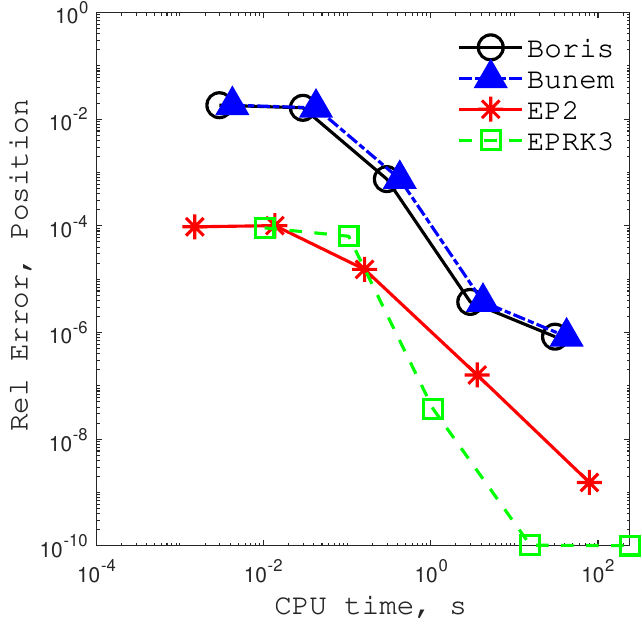} \\[0.5em]
\multicolumn{3}{c}{Potential Hills $\frac{|V''|}{B} = \frac{1}{100}$} \\
Quadratic & Cubic & Quartic \\
\includegraphics[scale=0.325]{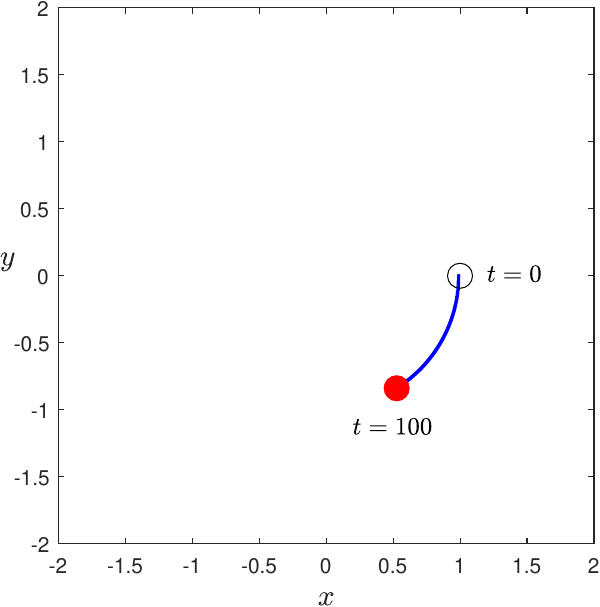} &
\includegraphics[scale=0.325]{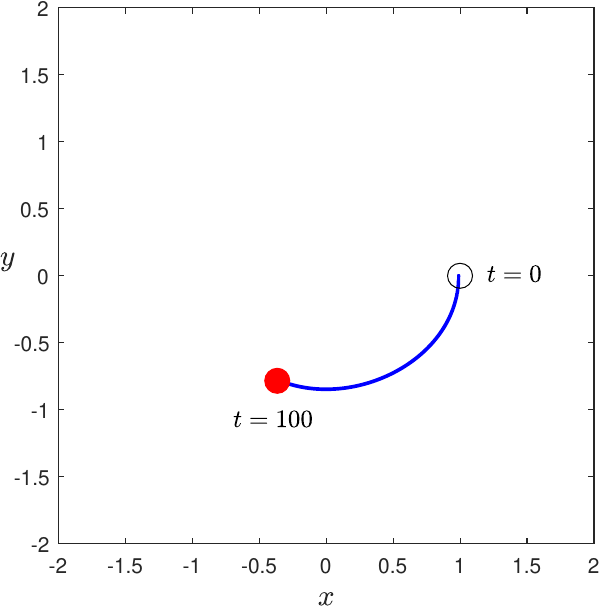} &
\includegraphics[scale=0.325]{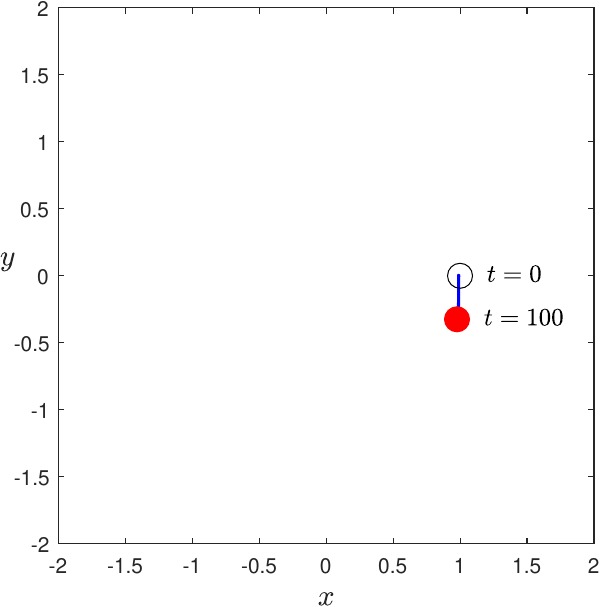} \\
\includegraphics[scale=0.325]{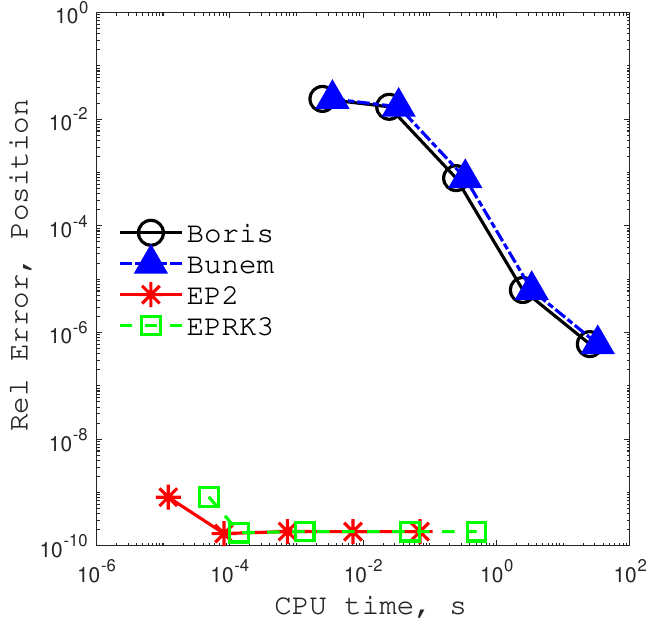} &
\includegraphics[scale=0.325]{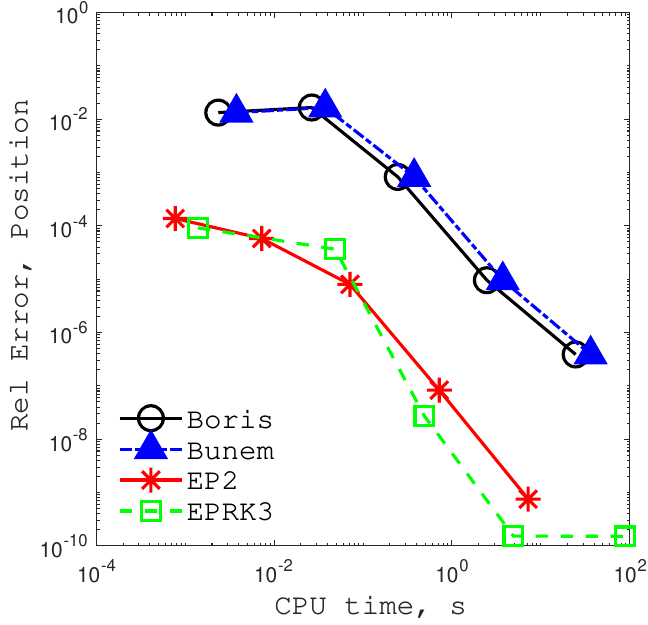} &
\includegraphics[scale=0.325]{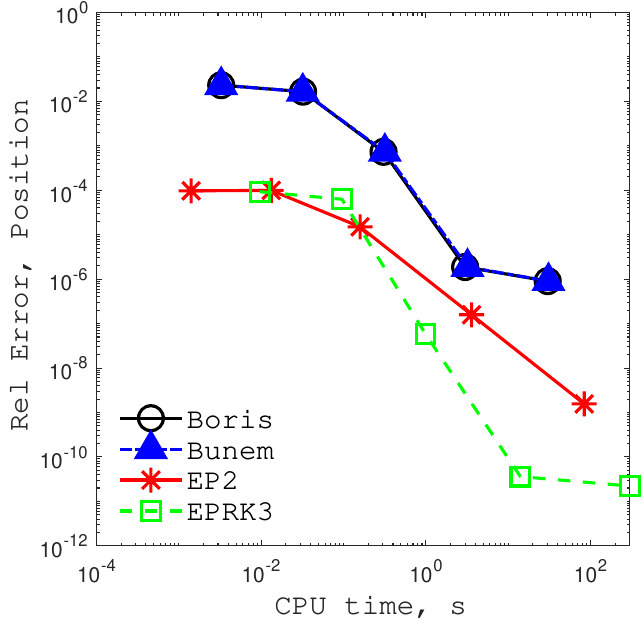}
\end{tabular}
\caption{Results for 2D test problems with $|V''|/B = 1/100$: potential well reference solution orbits (first row), potential well precision diagrams (second row), potential hill reference solution orbits (third row), and potential hill precision diagrams (fourth row). Boris/Buneman step sizes are $h = 10^{-2}, 10^{-3}, 10^{-4}, 10^{-5}, 10^{-6}$ for quadratic potential problems and $h = 10^{-3}, 10^{-4}, 10^{-5}, 10^{-6}, 10^{-7}$ for cubic/quartic potential problems. EP2/EPRK3 step sizes are $h = 100, 10, 1, 10^{-1}, 10^{-2}$ for quadratic potential problems and $h = 10^{-1}, 10^{-2}, 10^{-3}, 10^{-4}, 10^{-5}$ for cubic/quartic potential problems.}\label{2Dpot1}
\end{figure}

\begin{figure}[h!]
\centering
\begin{tabular}{ccc}
\multicolumn{3}{c}{Potential Wells $\frac{|V''|}{B} = \frac{1}{10}$} \\
Quadratic & Cubic & Quartic \\
\includegraphics[scale=0.325]{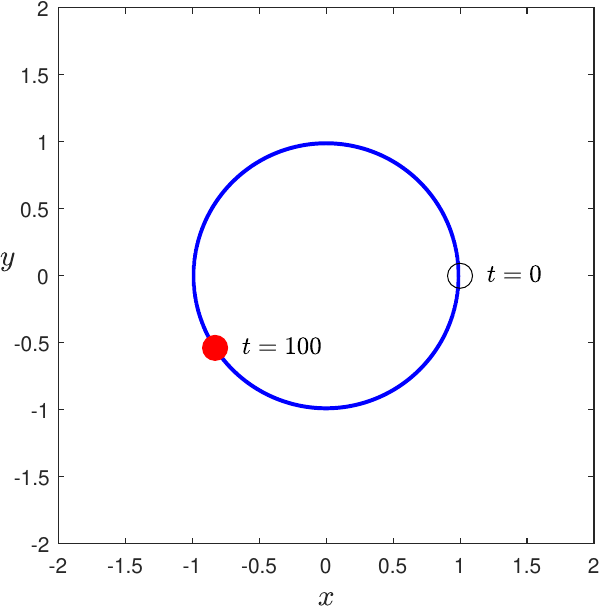} &
\includegraphics[scale=0.325]{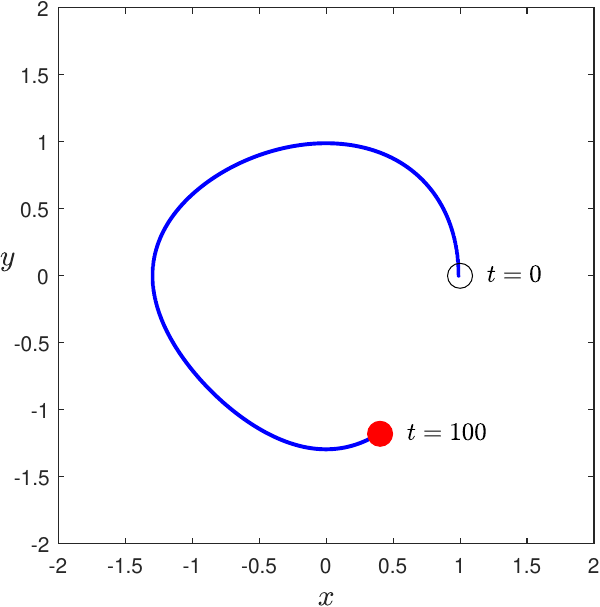} &
\includegraphics[scale=0.325]{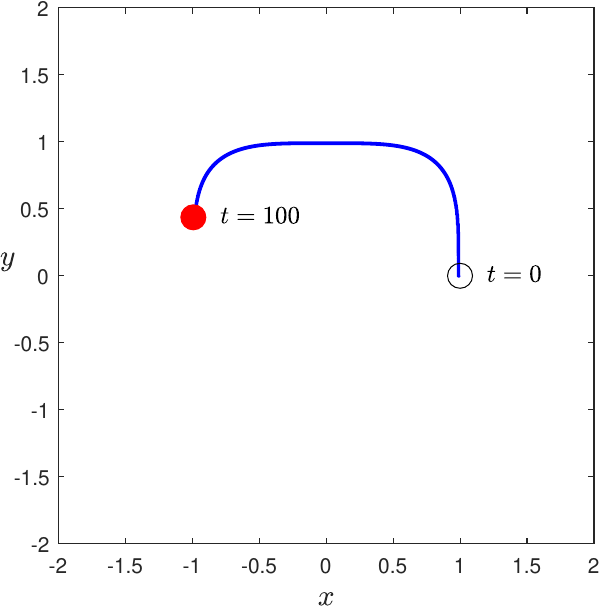} \\
\includegraphics[scale=0.325]{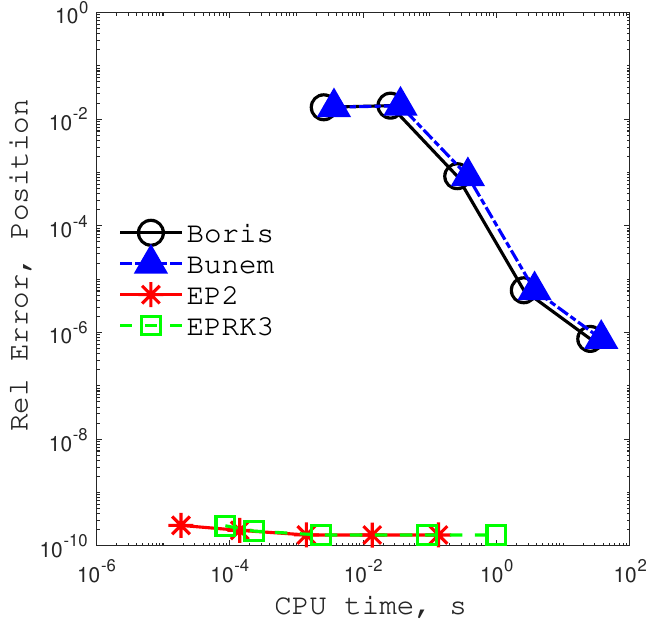} &
\includegraphics[scale=0.325]{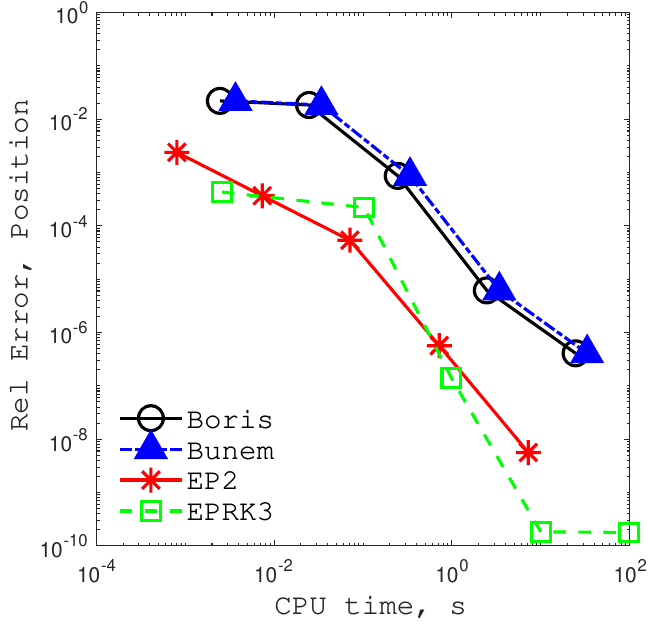} &
\includegraphics[scale=0.325]{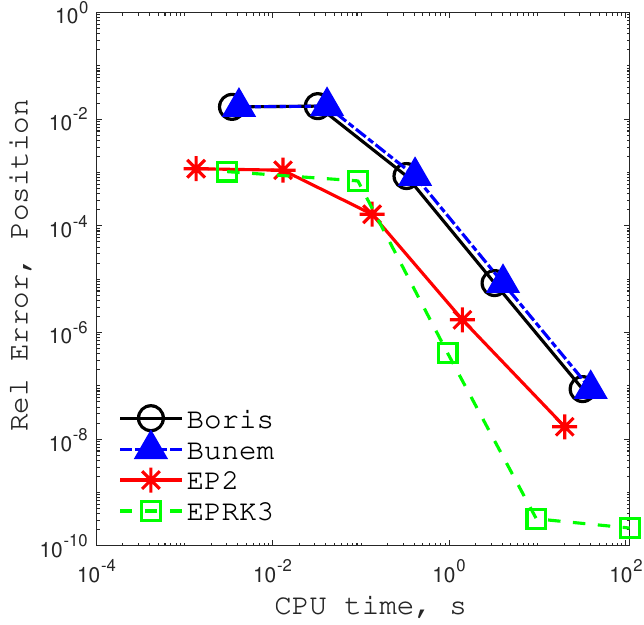} \\[0.5em]
\multicolumn{3}{c}{Potential Hills $\frac{|V''|}{B} = \frac{1}{10}$} \\
Quadratic & Cubic & Quartic \\
\includegraphics[scale=0.325]{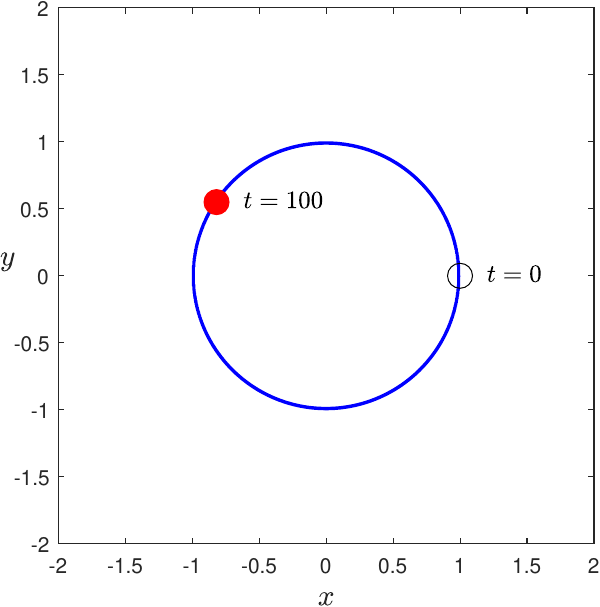} &
\includegraphics[scale=0.325]{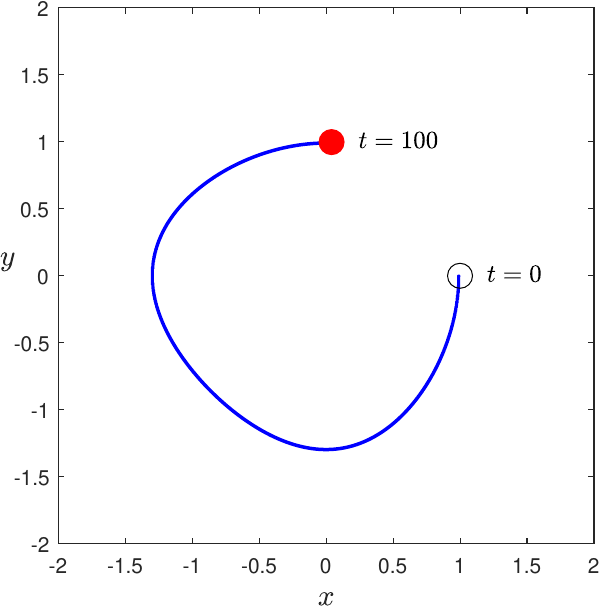} &
\includegraphics[scale=0.325]{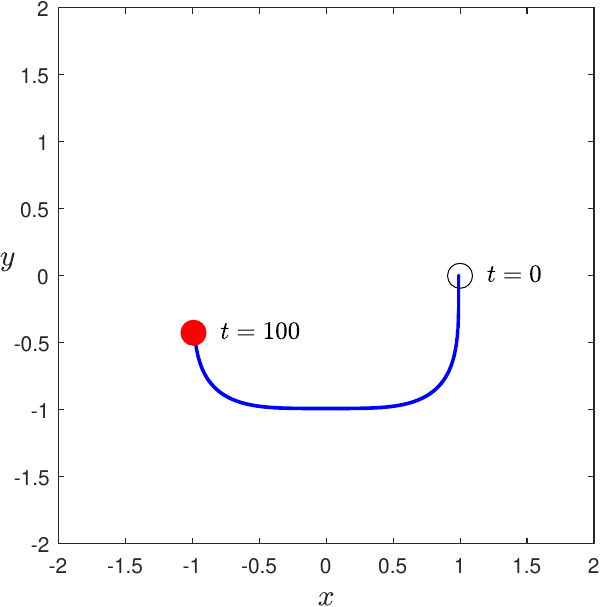} \\
\includegraphics[scale=0.325]{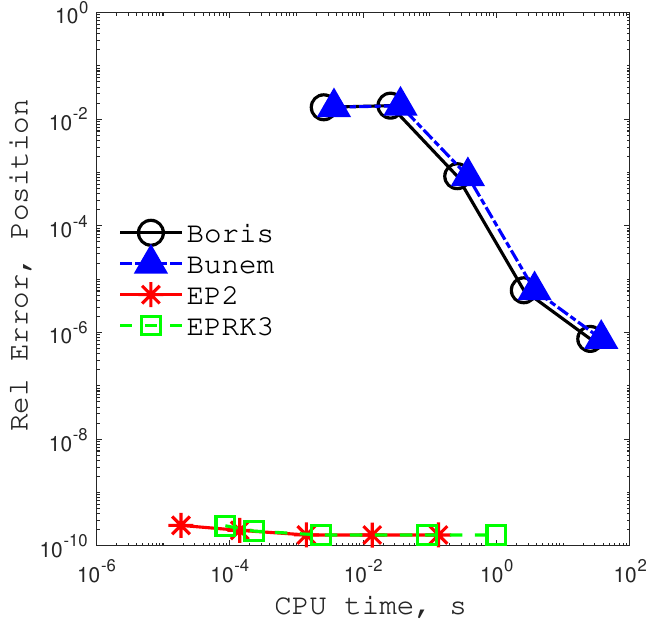} &
\includegraphics[scale=0.325]{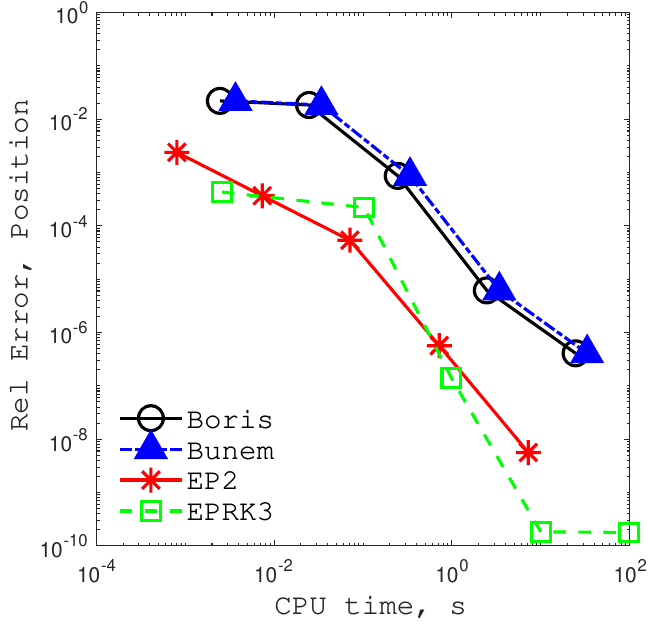} &
\includegraphics[scale=0.325]{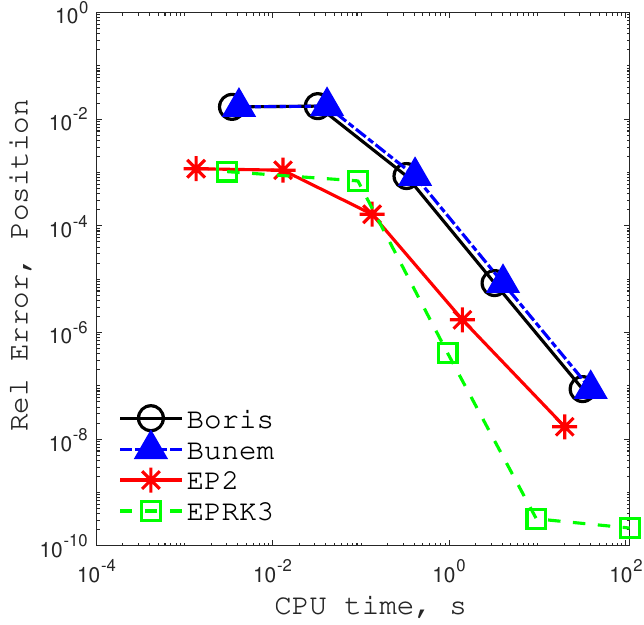}
\end{tabular}
\caption{Results for 2D test problems with $|V''|/B = 1/10$: potential well reference solution orbits (first row), potential well precision diagrams (second row), potential hill reference solution orbits (third row), and potential hill precision diagrams (fourth row). Boris/Buneman step sizes are $h = 10^{-2}, 10^{-3}, 10^{-4}, 10^{-5}, 10^{-6}$ for quadratic potential problems and $h = 10^{-3}, 10^{-4}, 10^{-5}, 10^{-6}, 10^{-7}$ for cubic/quartic potential problems. EP2/EPRK3 step sizes are $h = 100, 10, 1, 10^{-1}, 10^{-2}$ for quadratic potential problems and $h = 10^{-1}, 10^{-2}, 10^{-3}, 10^{-4}, 10^{-5}$ for cubic/quartic potential problems.}\label{2Dpot2}
\end{figure}

\begin{figure}[h!]
\centering
\begin{tabular}{ccc}
\multicolumn{3}{c}{Potential Wells $\frac{|V''|}{B} = 1$} \\
Quadratic & Cubic & Quartic \\
\includegraphics[scale=0.325]{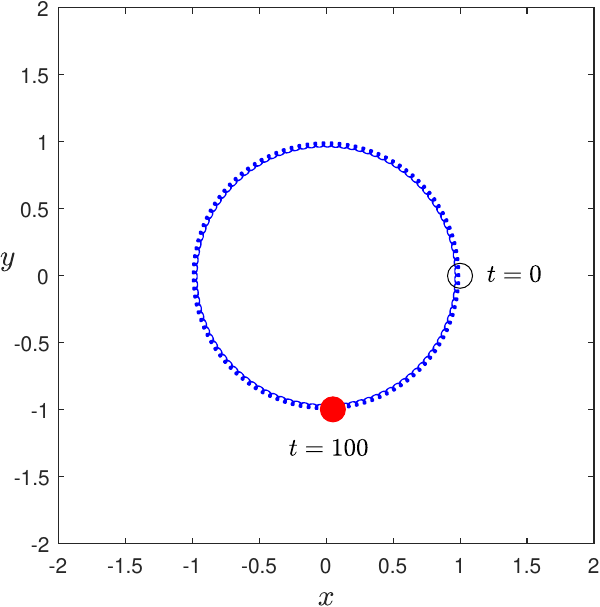} &
\includegraphics[scale=0.325]{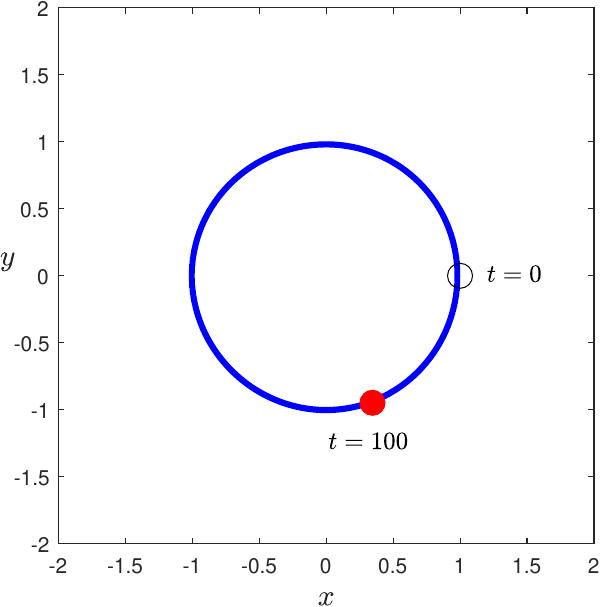} &
\includegraphics[scale=0.325]{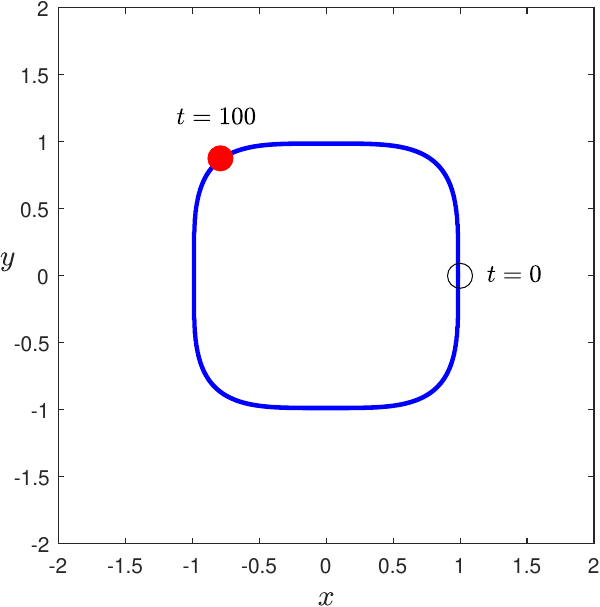} \\
\includegraphics[scale=0.325]{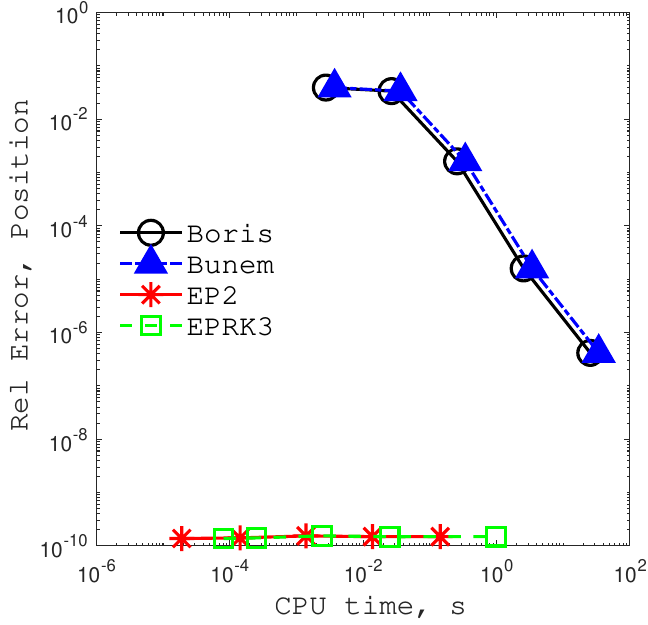} &
\includegraphics[scale=0.325]{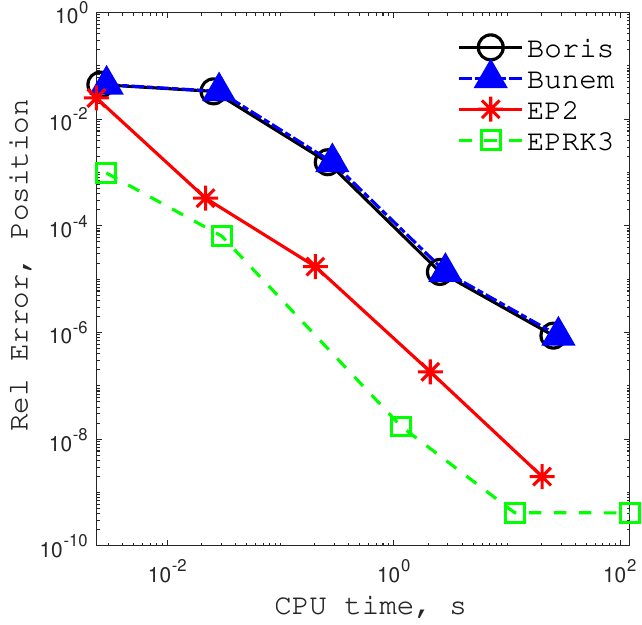} &
\includegraphics[scale=0.325]{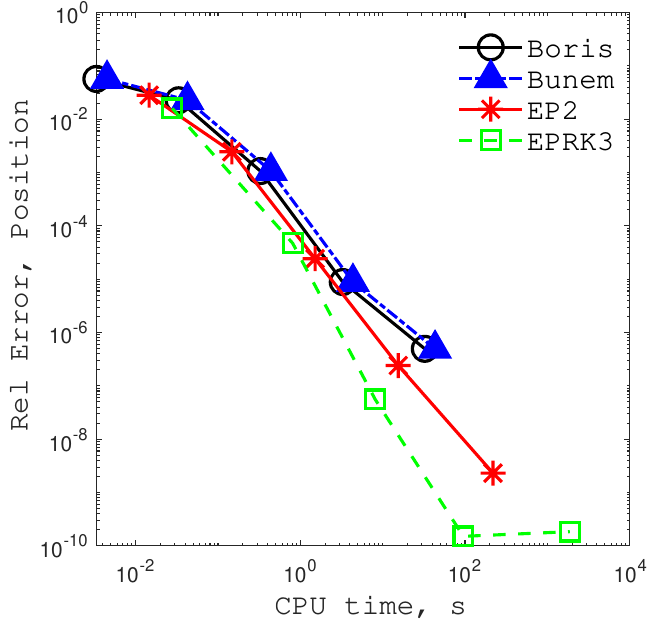} \\[0.5em]
\multicolumn{3}{c}{Potential Hills $\frac{|V''|}{B} = 1$} \\
Quadratic & Cubic & Quartic \\
\includegraphics[scale=0.325]{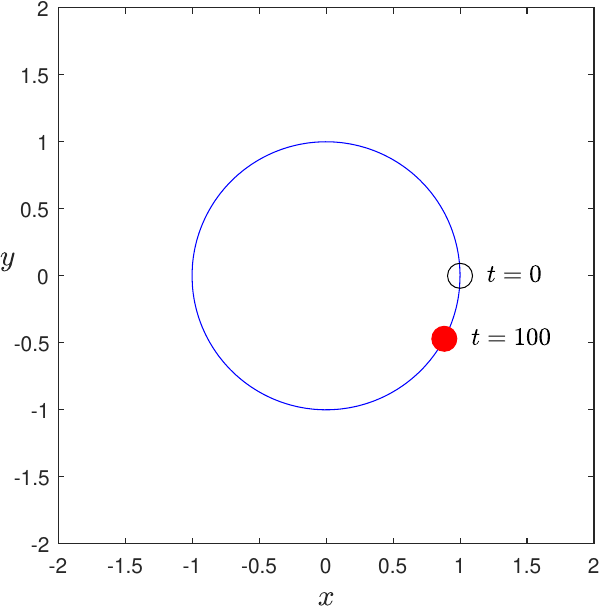} &
\includegraphics[scale=0.325]{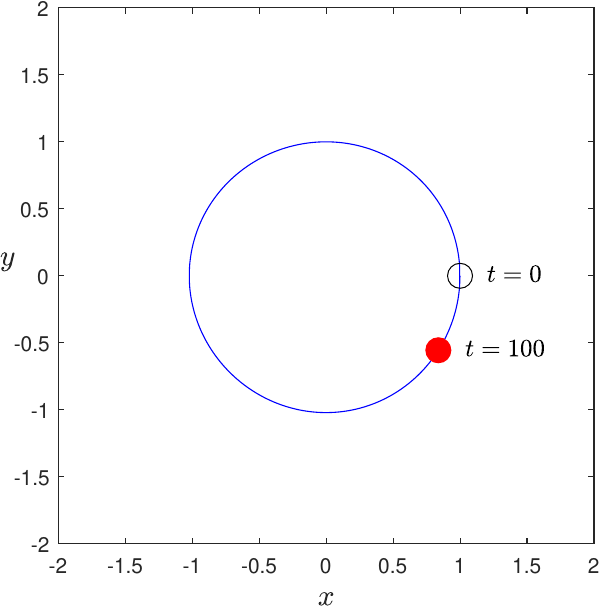} &
\includegraphics[scale=0.325]{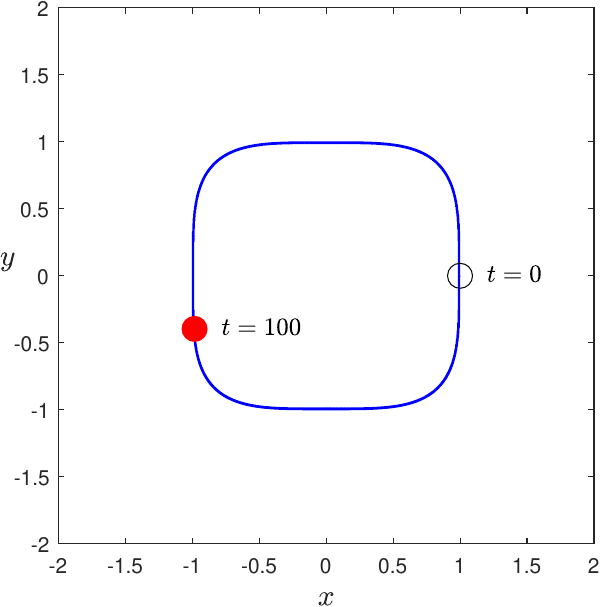} \\
\includegraphics[scale=0.325]{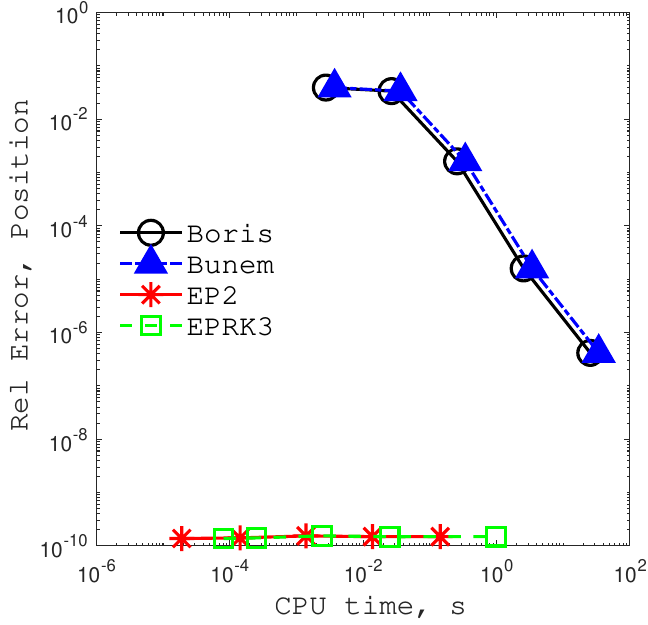} &
\includegraphics[scale=0.325]{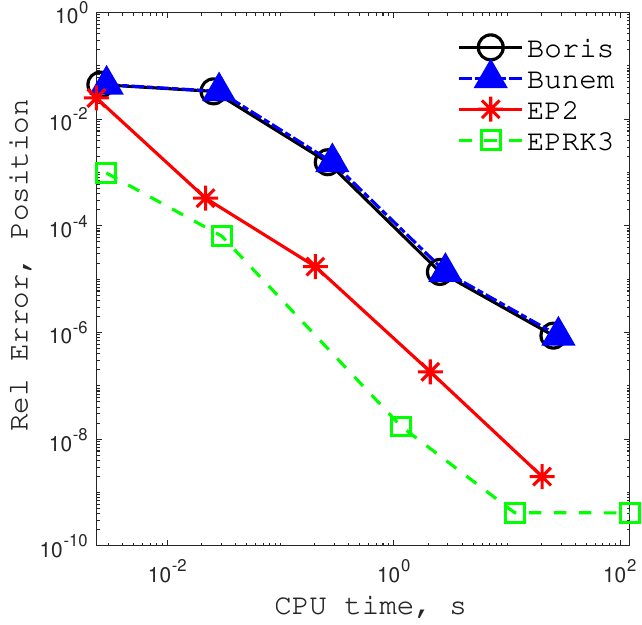} &
\includegraphics[scale=0.325]{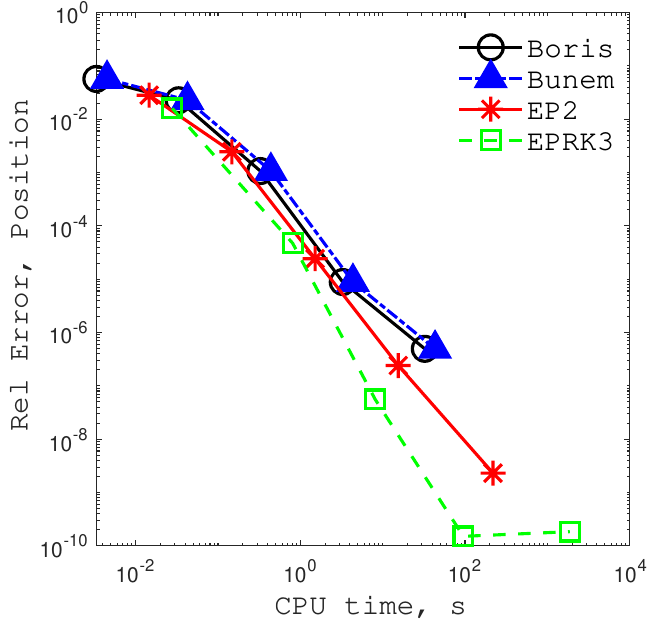}
\end{tabular}
\caption{Results for 2D test problems with $|V''|/B = 1$: potential well reference solution orbits (first row), potential well precision diagrams (second row), potential hill reference solution orbits (third row), and potential hill precision diagrams (fourth row). Boris/Buneman step sizes are $h = 10^{-2}, 10^{-3}, 10^{-4}, 10^{-5}, 10^{-6}$ for quadratic potential problems and $h = 10^{-3}, 10^{-4}, 10^{-5}, 10^{-6}, 10^{-7}$ for cubic/quartic potential problems. EP2/EPRK3 step sizes are $100, 10, 1, 10^{-1}, 10^{-2}$ for quadratic potential problems and $10^{-1}, 10^{-2}, 10^{-3}, 10^{-4}, 10^{-5}$ for cubic/quartic potential problems.}\label{2Dpot3}
\end{figure}

Plots of the reference solution orbits and precision diagrams for test problems with $|V''|/B = 1/100$, $|V''|/B = 1/10$, and $|V''|/B = 1$ are shown in figures \ref{2Dpot1}, \ref{2Dpot2}, and \ref{2Dpot3}, respectively. They show that the performances of the particle pushers are roughly similar between the potential well problems and the potential hill problems. For the test problems with quadratic potentials, the exponential integrators exhibit superior performance as expected, because the problems are linear for which exponential integrators solve exactly. For the nonlinear test problems with cubic and quartic potentials the computational advantage of the exponential methods is not as dramatic but they are still competitive with the Boris and Buneman particle pushers.

\begin{figure}[h!]
\centering
\begin{tabular}{ccc}
\multicolumn{3}{c}{Potential Wells, $\frac{|V''|}{B} = \frac{1}{100}$} \\[0.5em]
Quadratic & Cubic & Quartic \\
\includegraphics[scale=0.3]{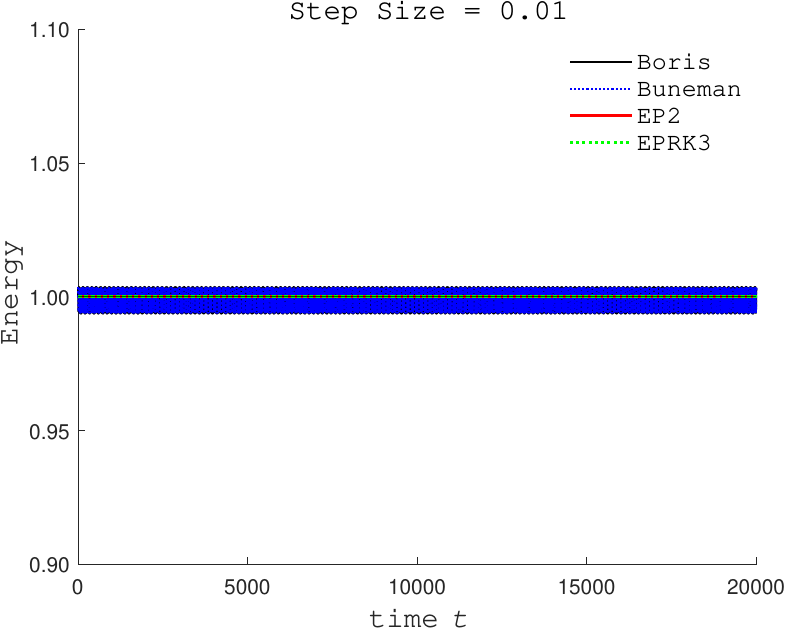} & \includegraphics[scale=0.3]{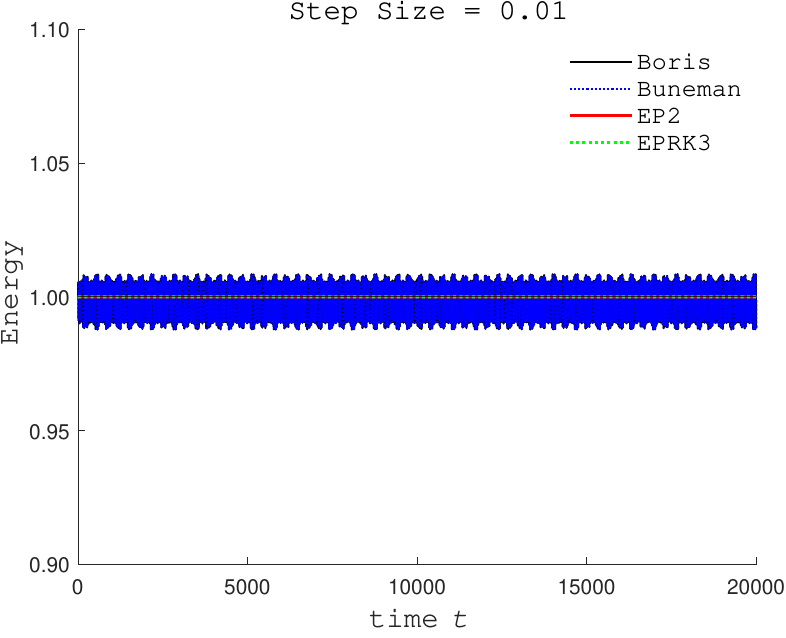} & \includegraphics[scale=0.3]{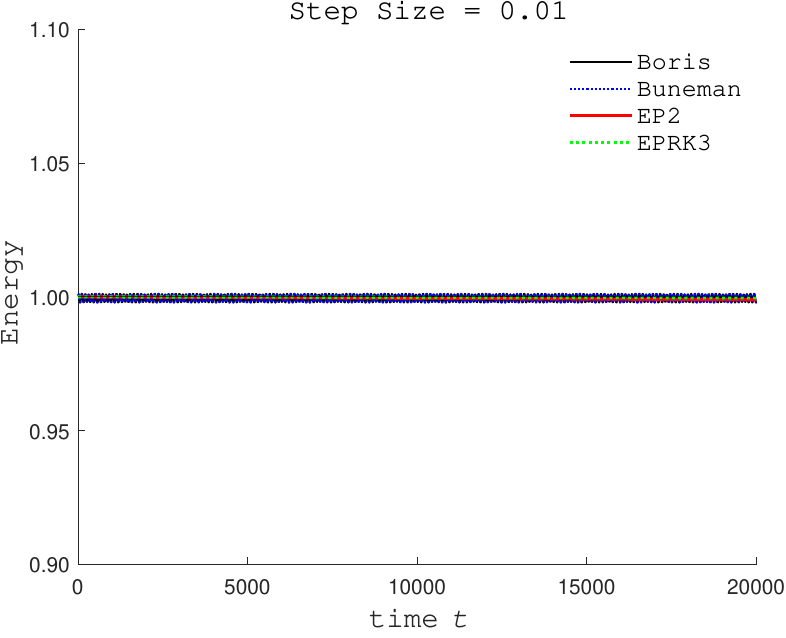} \\
\includegraphics[scale=0.3]{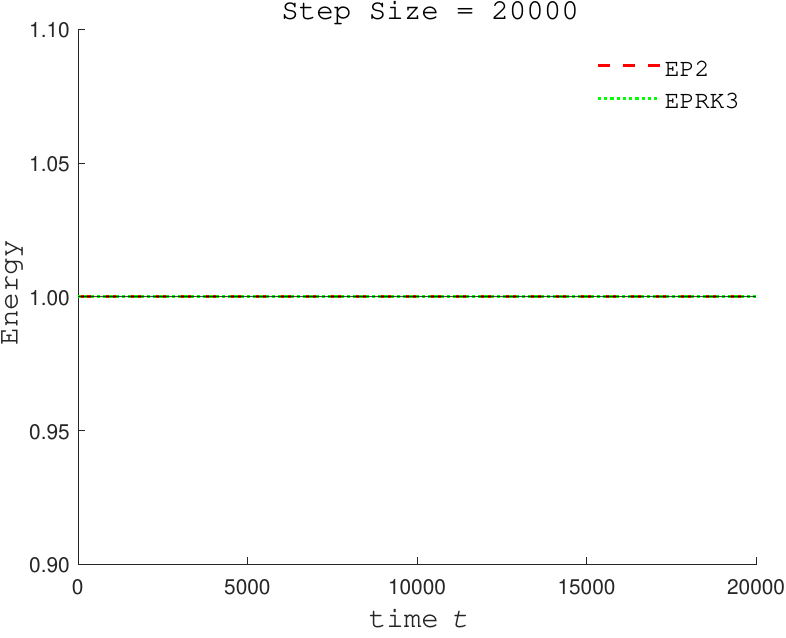} & \includegraphics[scale=0.3]{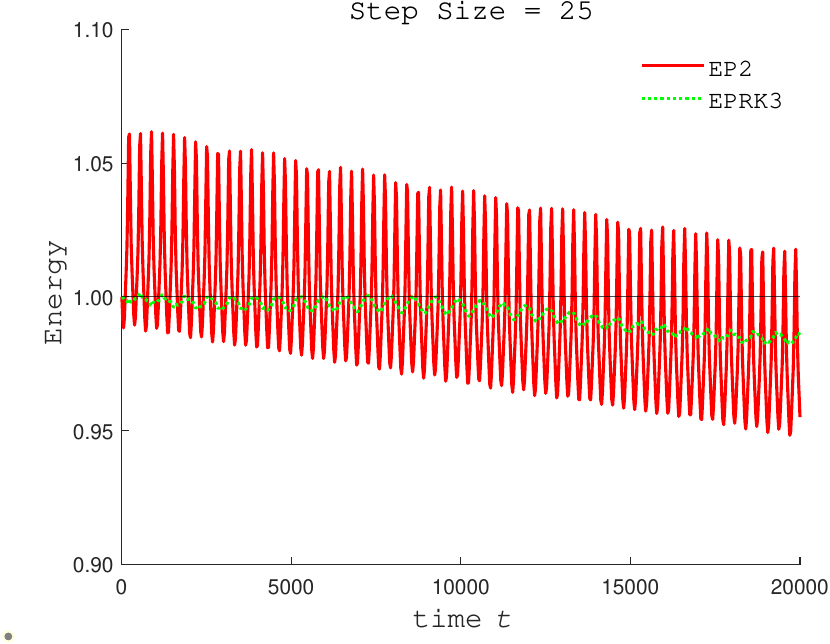} & \includegraphics[scale=0.3]{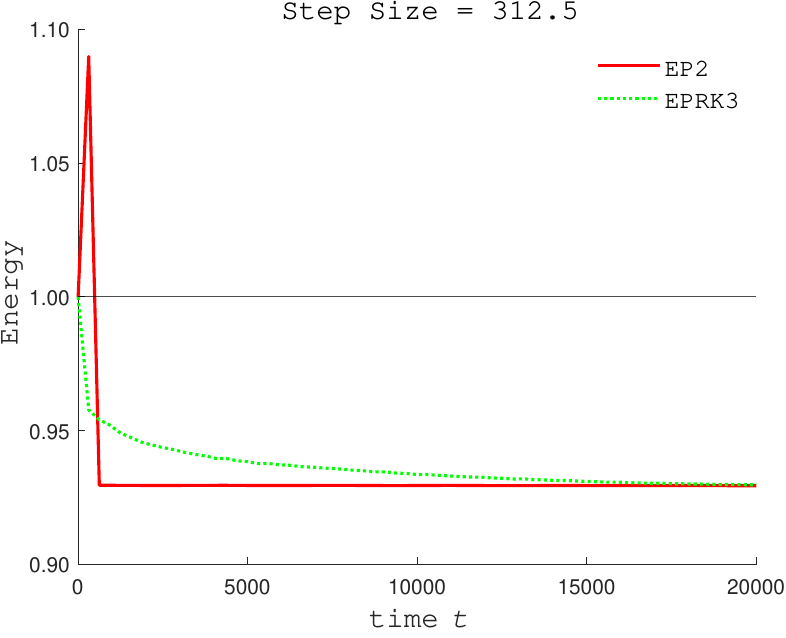} \\[1.5em]
\multicolumn{3}{c}{Potential Hills, $\frac{|V''|}{B} = \frac{1}{100}$} \\[0.5em]
Quadratic & Cubic & Quartic \\
\includegraphics[scale=0.3]{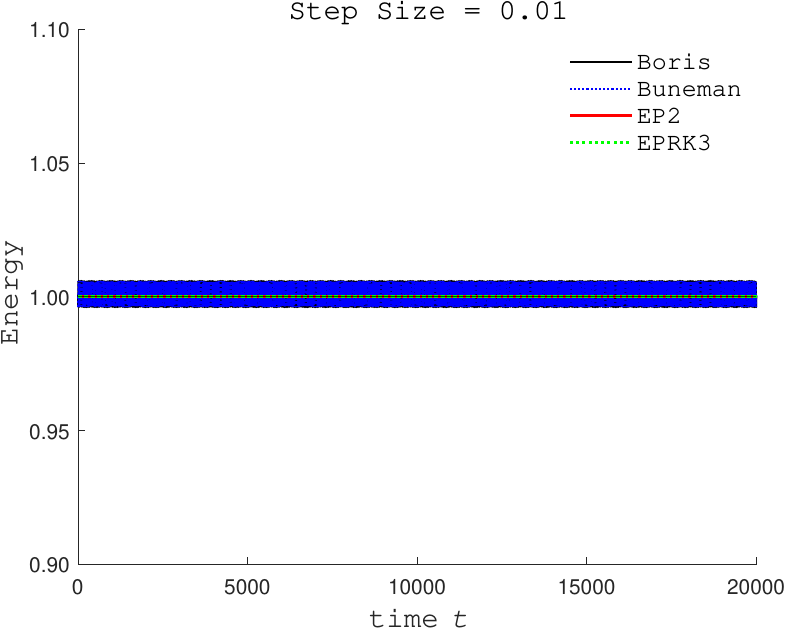} & \includegraphics[scale=0.3]{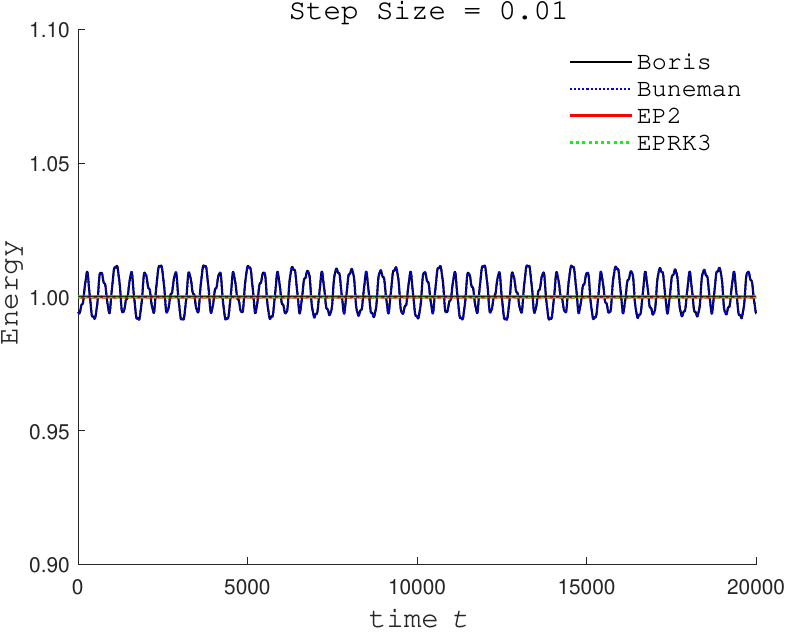} & \includegraphics[scale=0.3]{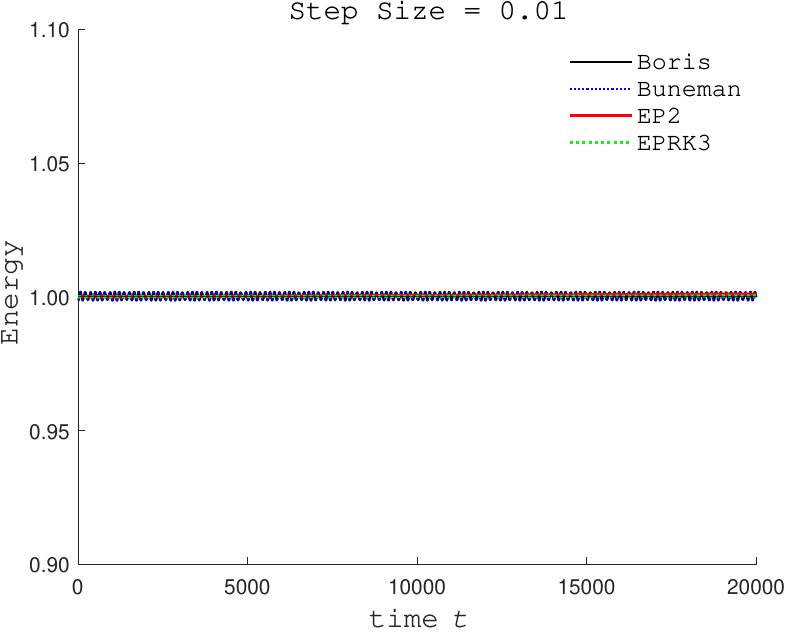} \\
\includegraphics[scale=0.3]{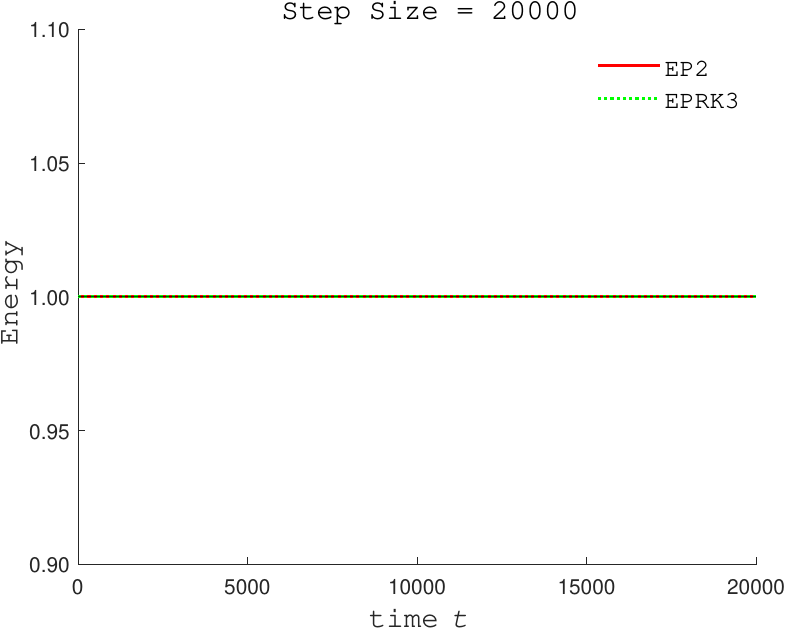} & \includegraphics[scale=0.3]{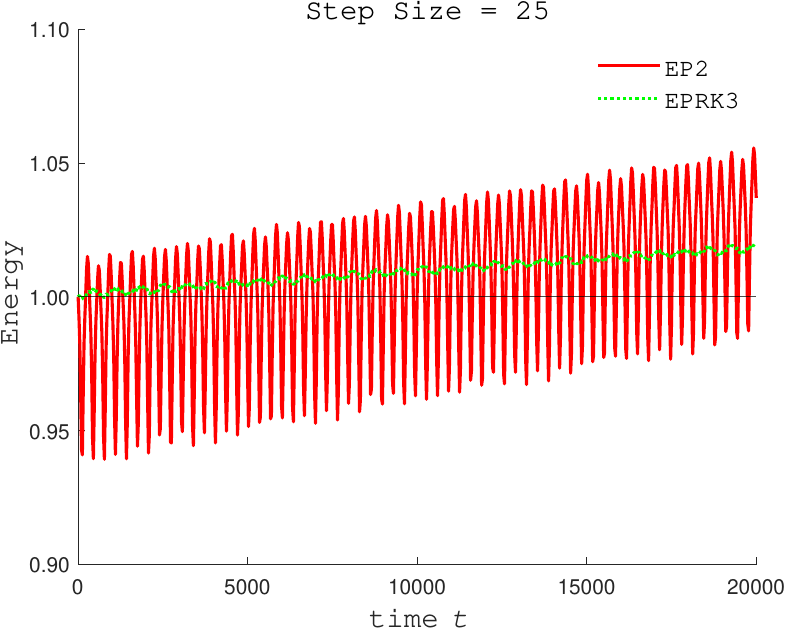} & \includegraphics[scale=0.3]{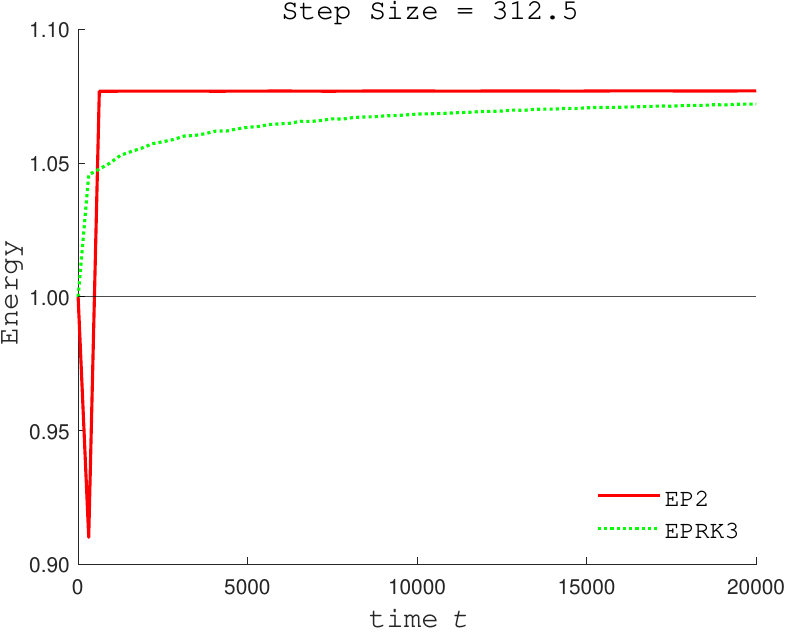}
\end{tabular}
\caption{Energy of 2D test problems with $|V''|/B = 1/100$}\label{2DEnergy1}
\end{figure}

\begin{figure}[h!]
\centering
\begin{tabular}{ccc}
\multicolumn{3}{c}{Potential Wells, $\frac{|V''|}{B} = \frac{1}{10}$} \\[0.5em]
Quadratic & Cubic & Quartic \\
\includegraphics[scale=0.3]{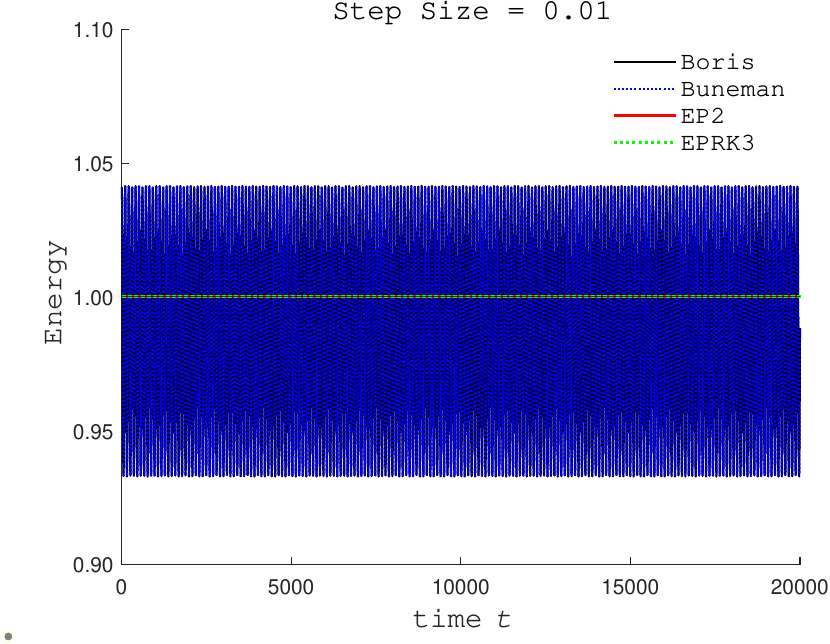} & \includegraphics[scale=0.3]{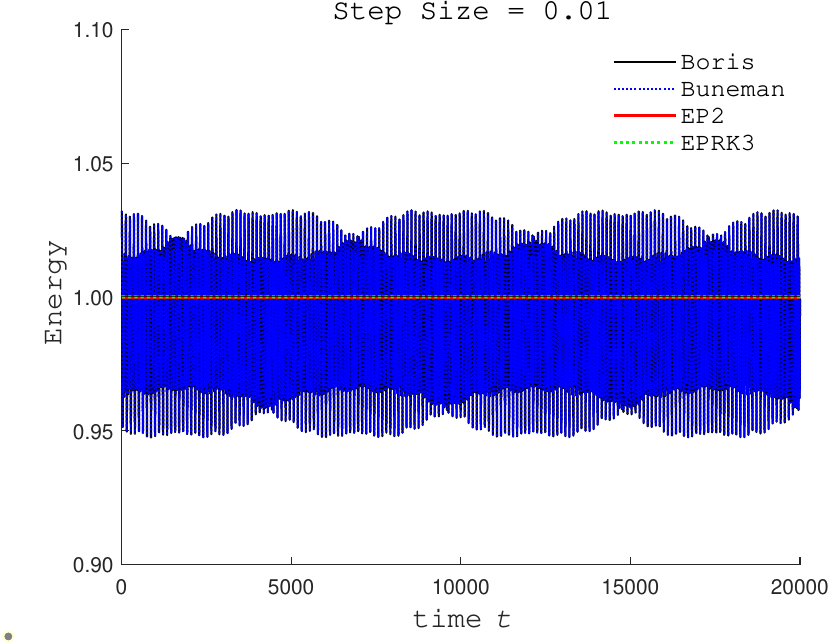} & \includegraphics[scale=0.3]{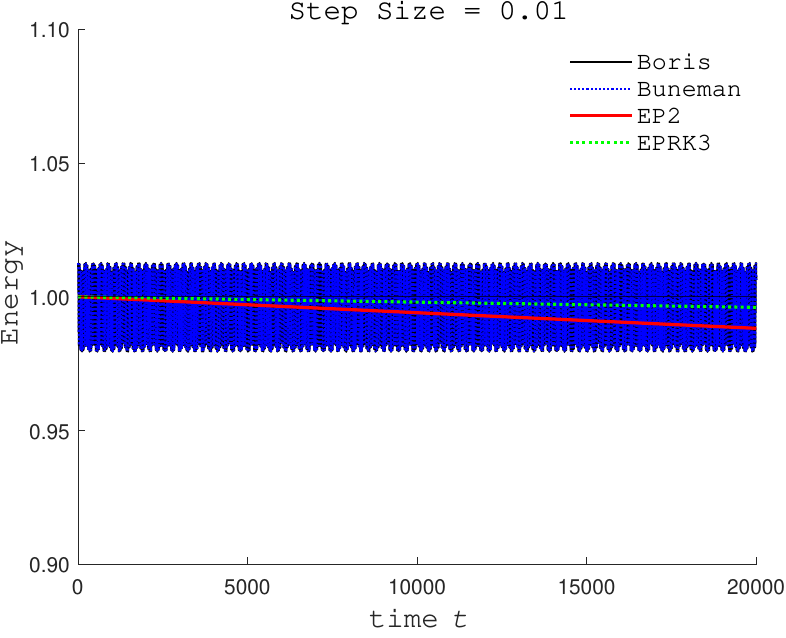} \\
\includegraphics[scale=0.3]{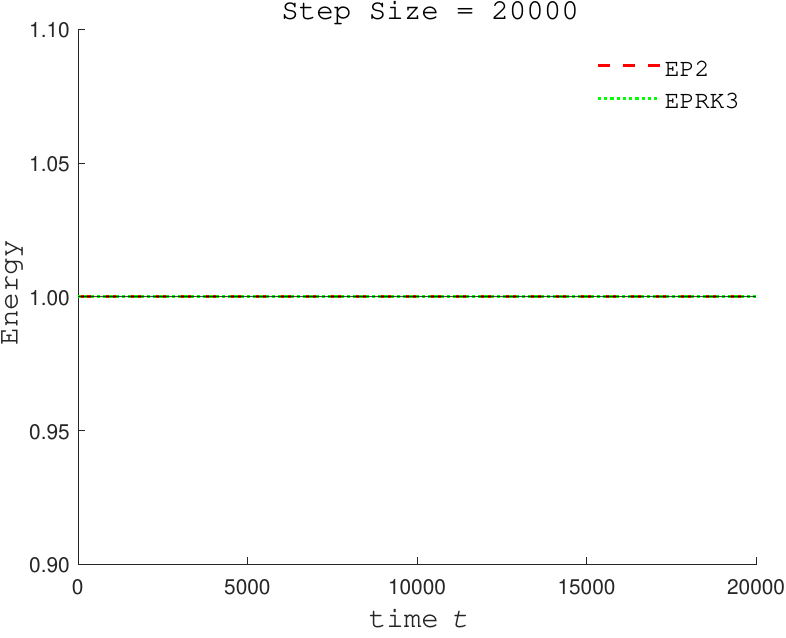} & \includegraphics[scale=0.3]{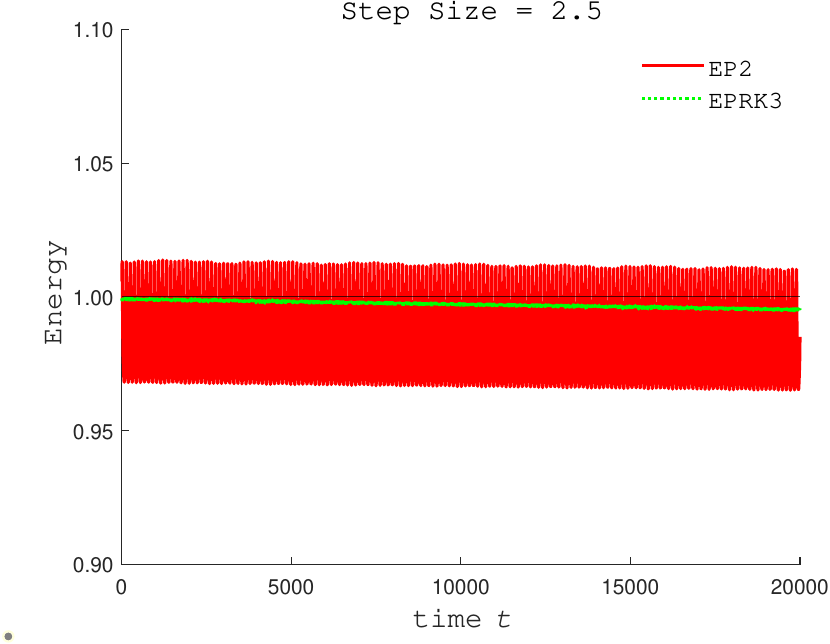} & \includegraphics[scale=0.3]{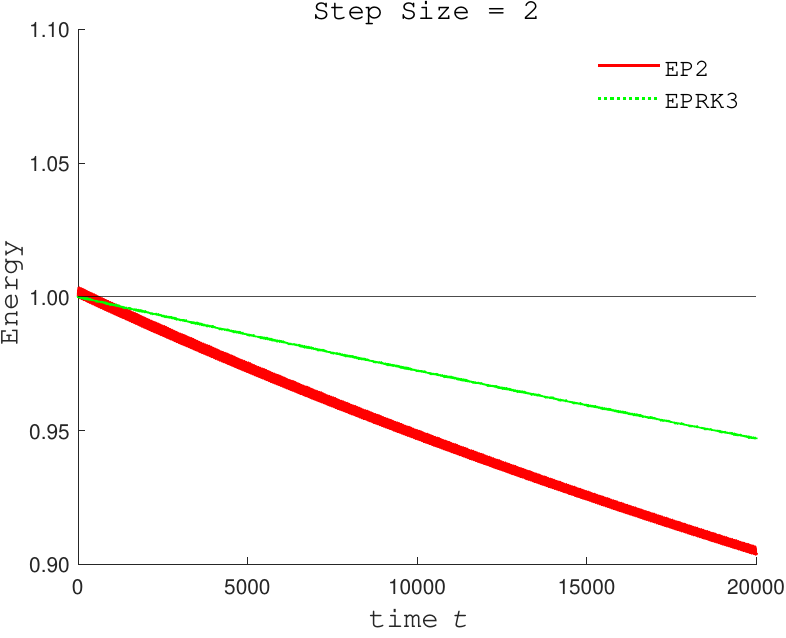} \\[1.5em]
\multicolumn{3}{c}{Potential Hills, $\frac{|V''|}{B} = \frac{1}{10}$} \\[0.5em]
Quadratic & Cubic & Quartic \\
\includegraphics[scale=0.3]{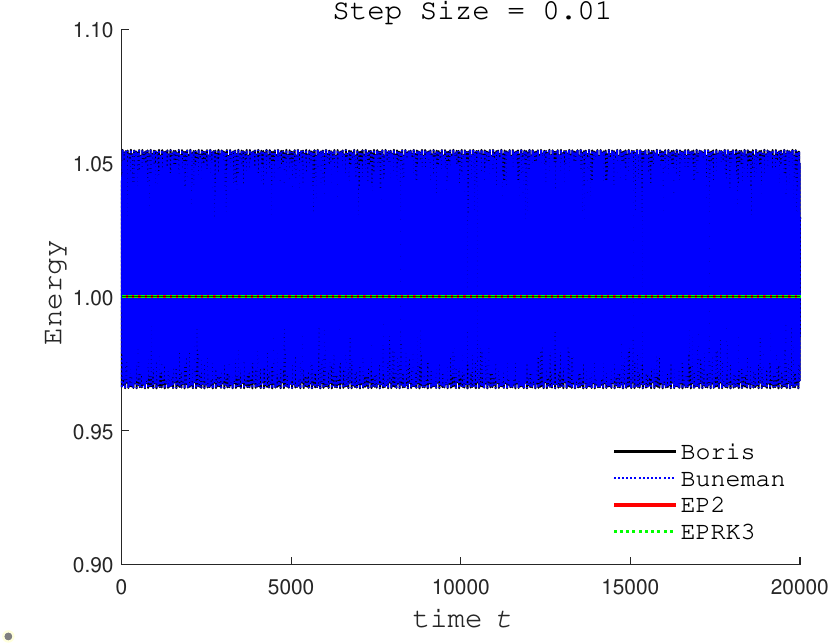} & \includegraphics[scale=0.3]{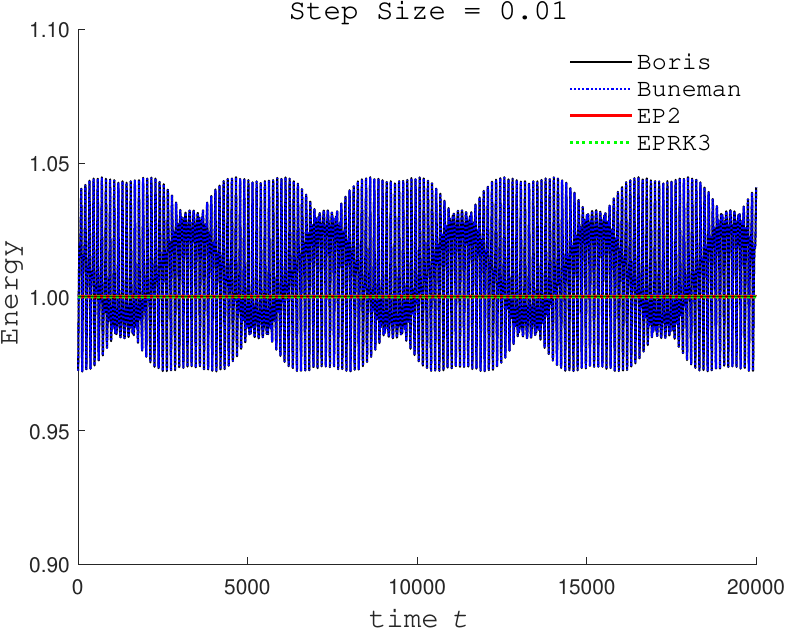} & \includegraphics[scale=0.3]{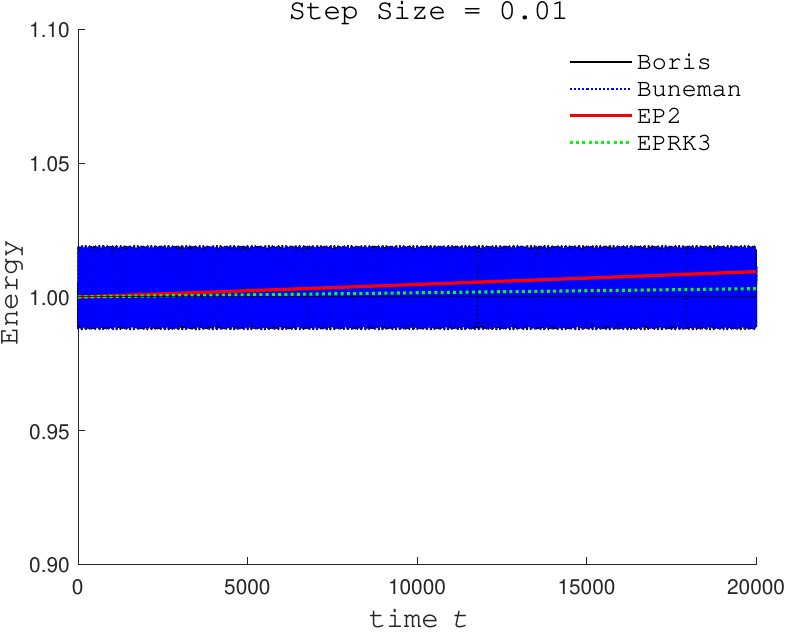} \\
\includegraphics[scale=0.3]{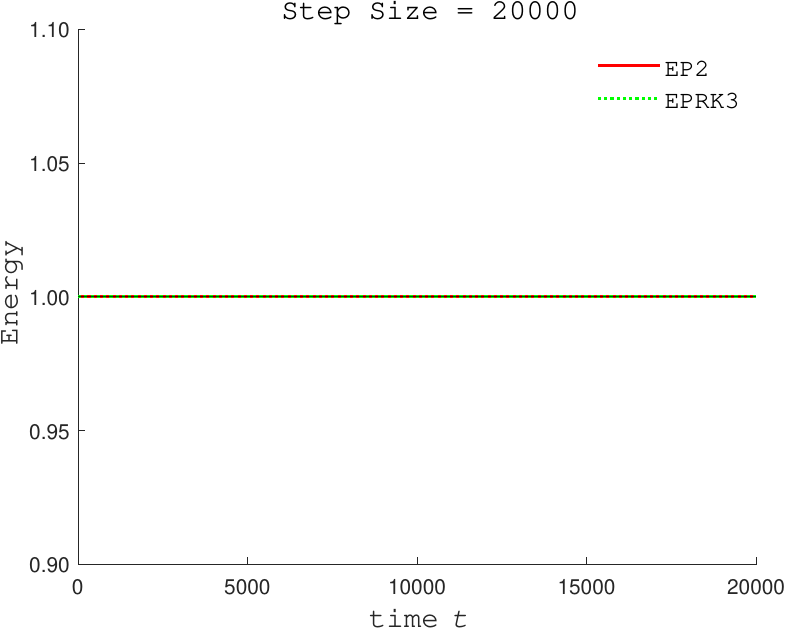} & \includegraphics[scale=0.3]{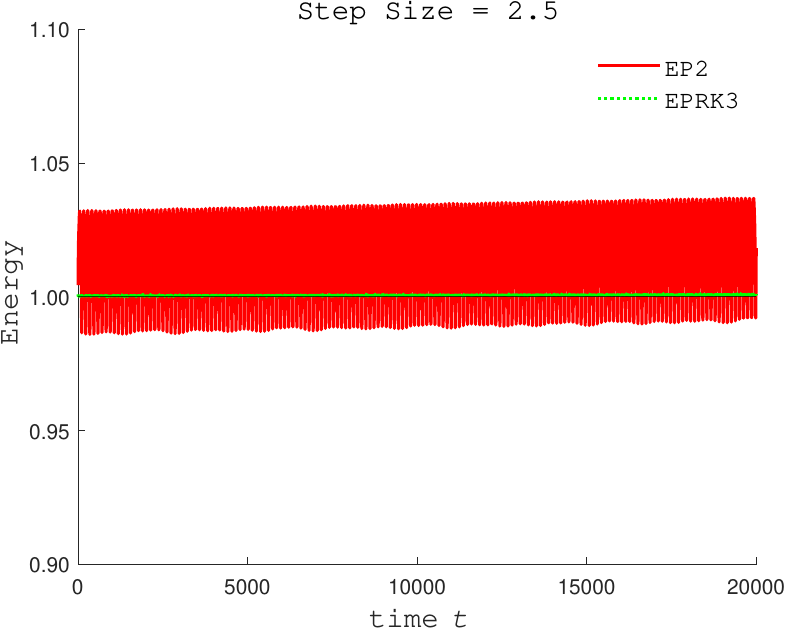} & \includegraphics[scale=0.3]{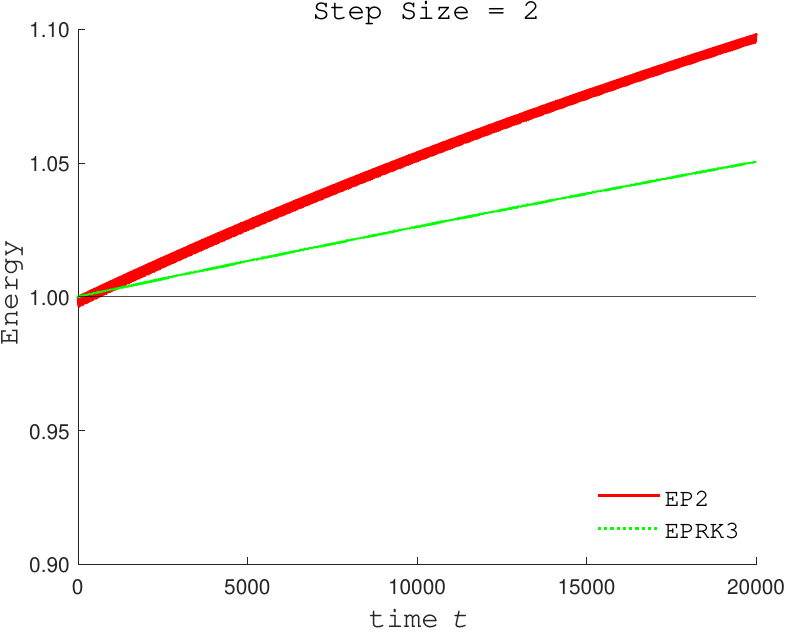}
\end{tabular}
\caption{Energy of 2D test problems with $|V''|/B = 1/10$}\label{2DEnergy2}
\end{figure}

\begin{figure}[h!]
\centering
\begin{tabular}{ccc}
\multicolumn{3}{c}{Potential Wells, $\frac{|V''|}{B} = 1$} \\[0.5em]
Quadratic & Cubic & Quartic \\
\includegraphics[scale=0.3]{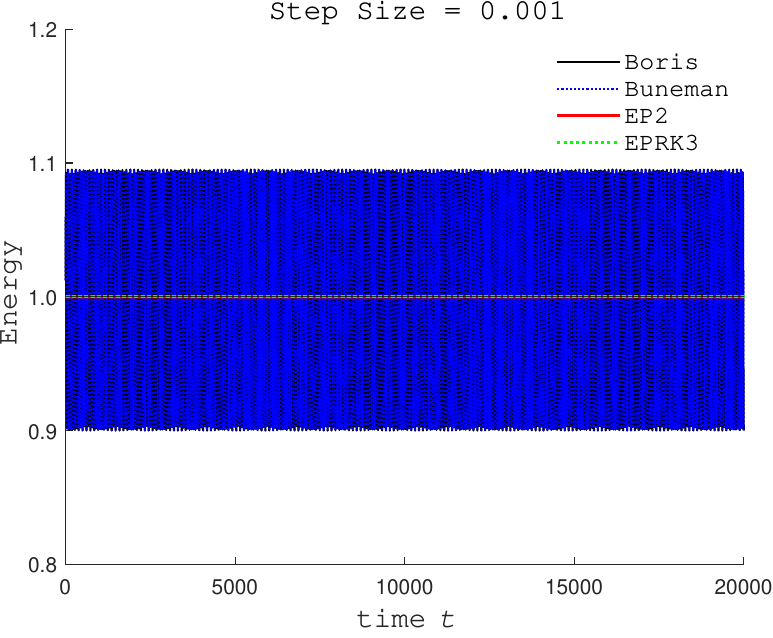} & \includegraphics[scale=0.3]{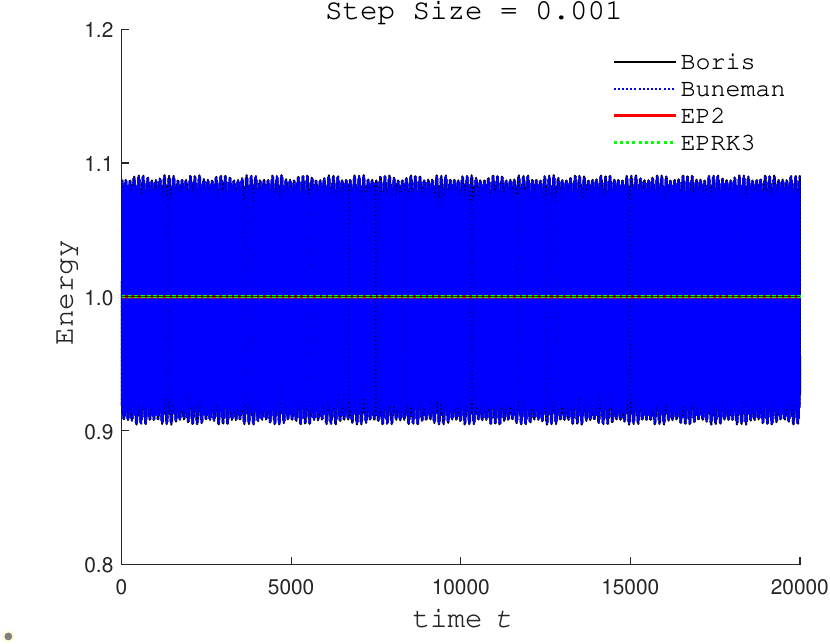} & \includegraphics[scale=0.3]{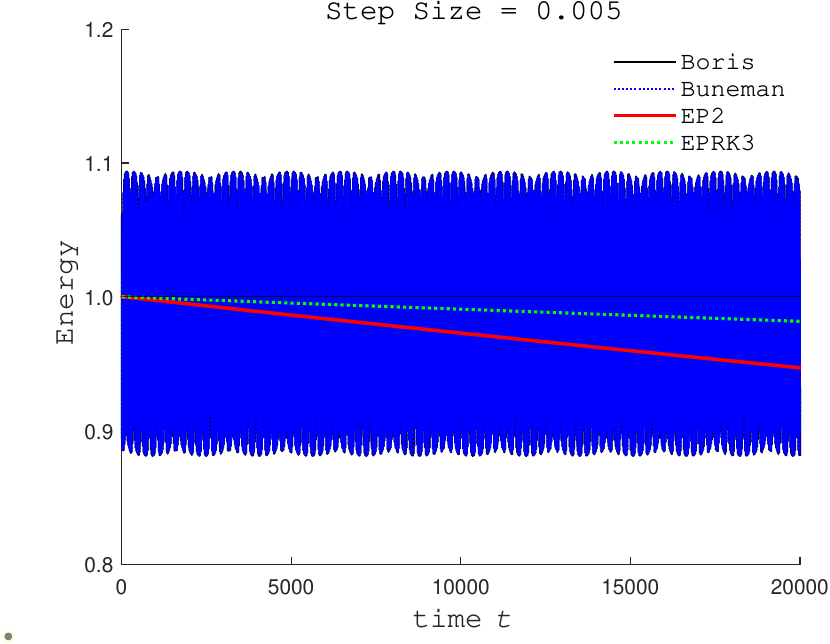} \\
\includegraphics[scale=0.3]{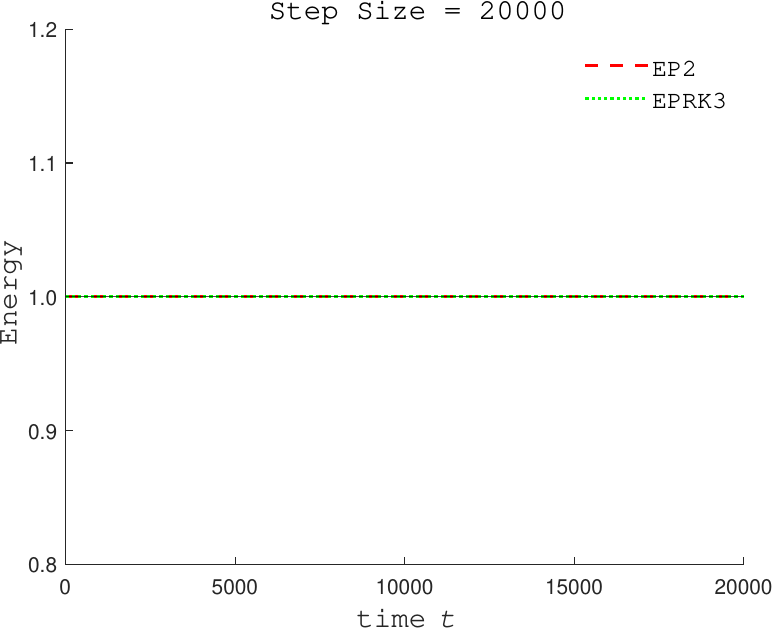} & \includegraphics[scale=0.3]{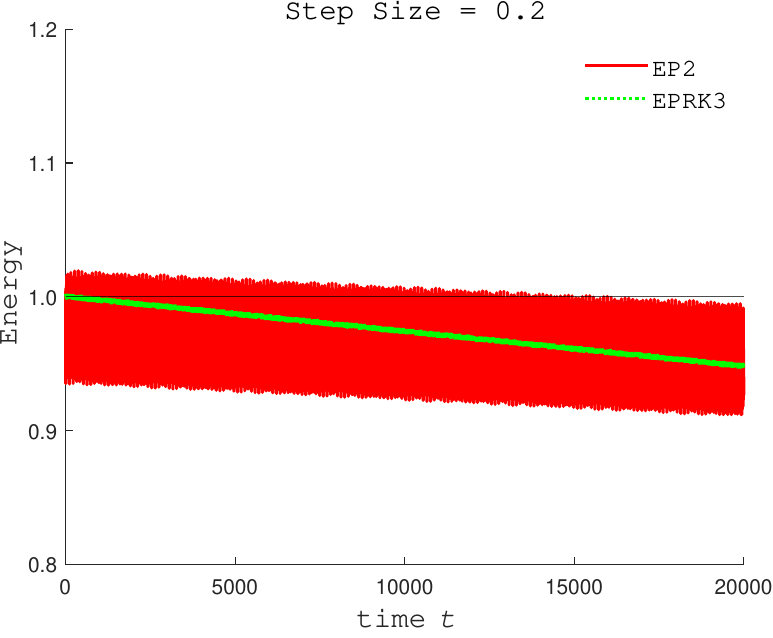} & \includegraphics[scale=0.3]{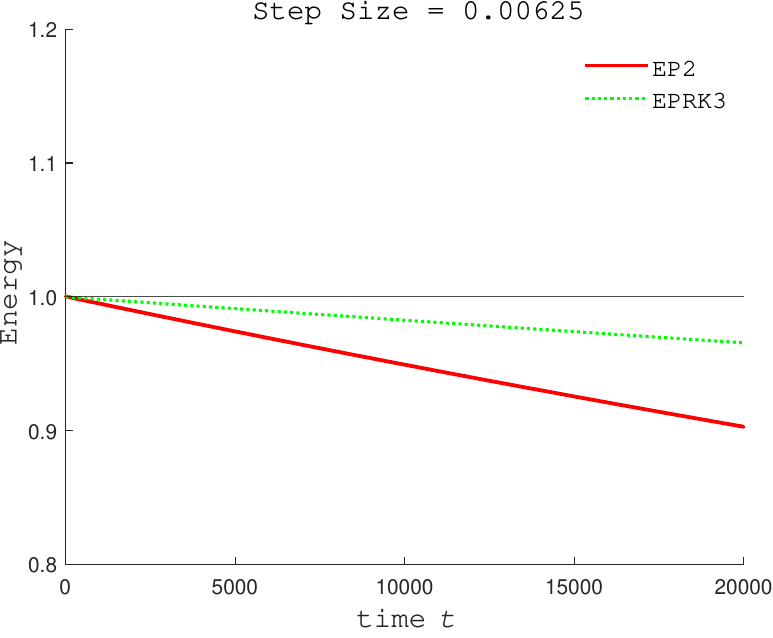} \\[1.5em]
\multicolumn{3}{c}{Potential Hills, $\frac{|V''|}{B} = 1$} \\[0.5em]
Quadratic & Cubic & Quartic \\
\includegraphics[scale=0.3]{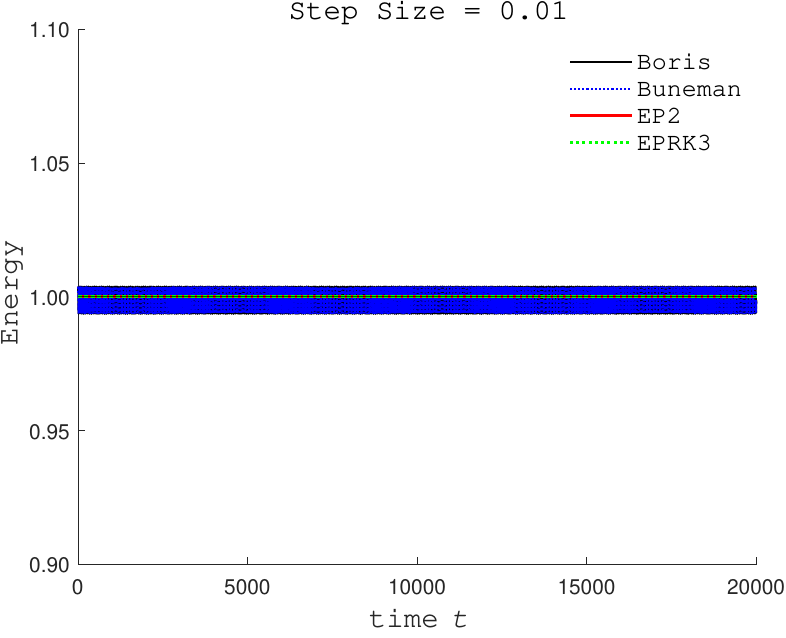} & \includegraphics[scale=0.3]{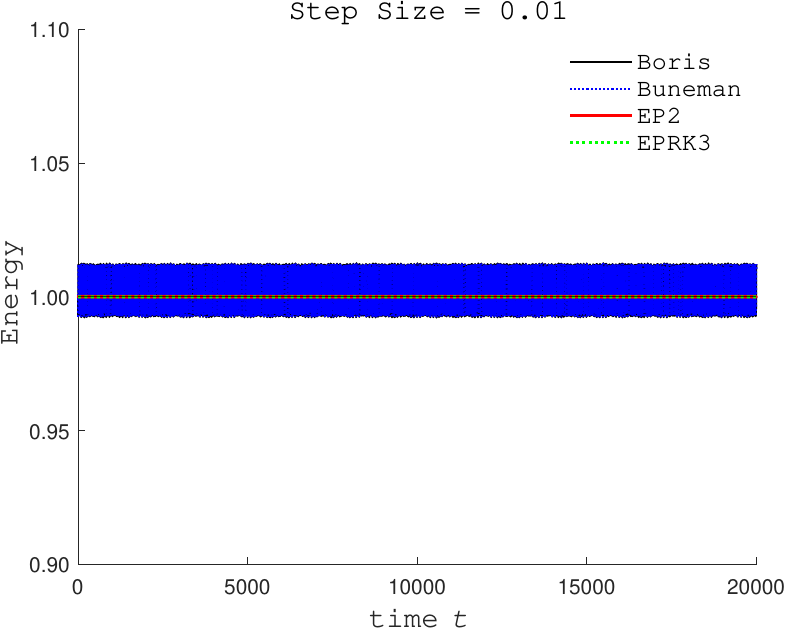} & \includegraphics[scale=0.3]{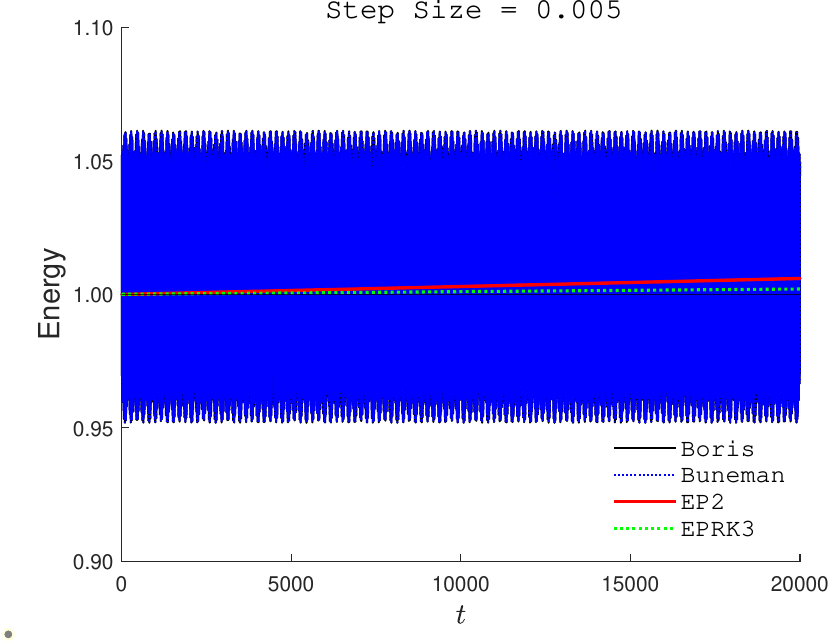} \\
\includegraphics[scale=0.3]{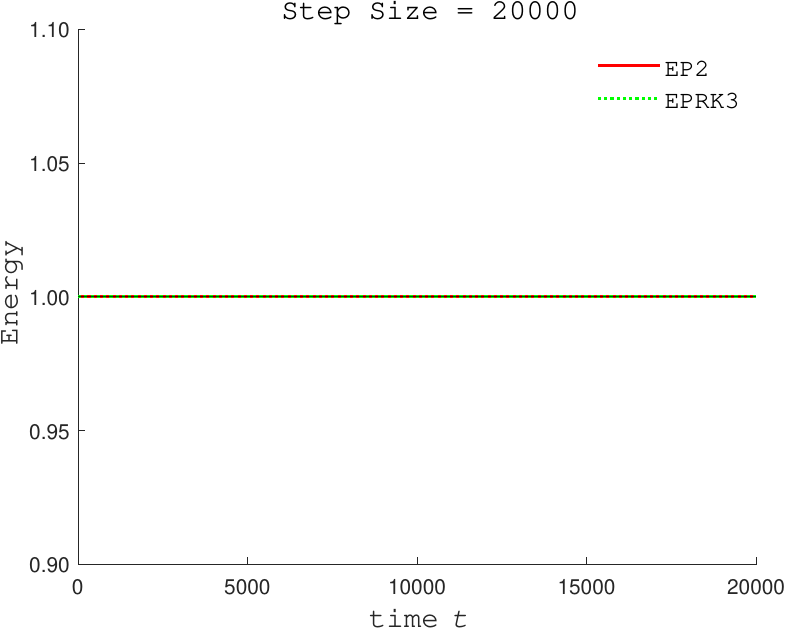} & \includegraphics[scale=0.3]{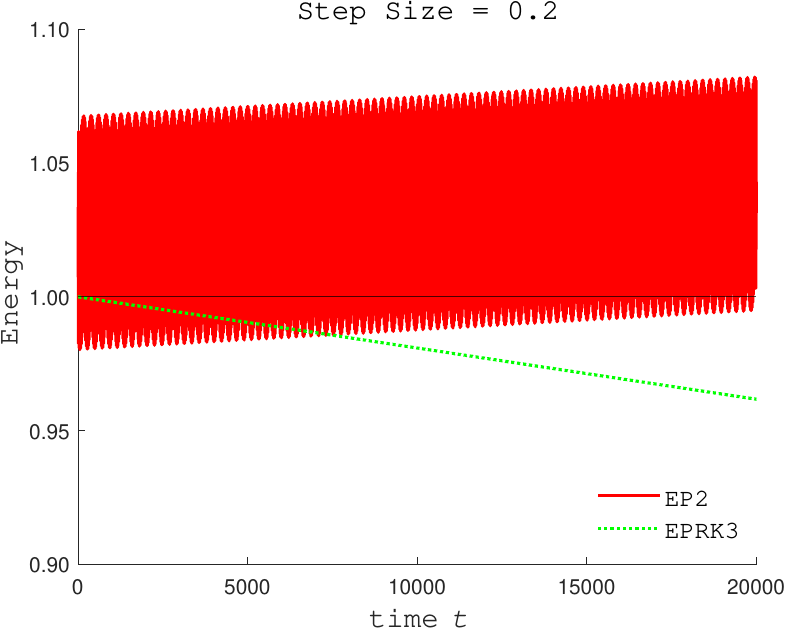} & \includegraphics[scale=0.3]{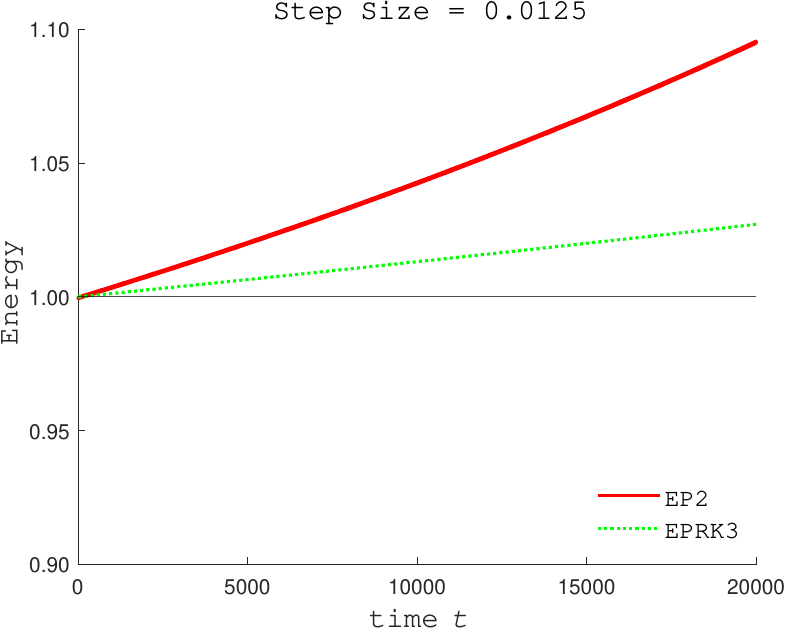}
\end{tabular}
\caption{Energy of 2D test problems with $|V''|/B = 1$}\label{2DEnergy3}
\end{figure}

Figures \ref{2DEnergy1}, \ref{2DEnergy2}, and \ref{2DEnergy3} show the energy plots for test problems with $|V''|/B = 1/100$, $|V''|/B = 1/10$, and $|V''|/B = 1$, respectively. Note that the exponential integrators compute the exact energy in one single time step for linear test problems with quadratic potentials. For the nonlinear test problems with cubic and quartic potentials, we point out several key observations. Since these exponential integrators have not been designed to preserve energy exactly, their computed energies are expected to drift over the time interval. However, the errors in energy of the exponential methods remains within the same bounds of the errors for the Boris and Buneman algorithms for comparable time step sizes. For large step sizes, the drift causes the energy to eventually exceed those bounds. It is also important to note that the EPRK3 integrator performs better than the EP2 integrator in two respects: the EP2 energy drifts are larger than the EPRK3 energy drifts and there is wider variation in the EP2 energies compared to the EPRK3 energies.  These results indicate a possibility of construction of higher order exponential methods that can yield sufficient accuracy within the time interval of interest and, if they are carefully designed, could still remain competitive from the efficiency standpoint.  We will pursue development of such techniques in our future publications.

\vfill
\begin{table}[h!]
\centering
\begin{tabular}{ll|c|c|c|}
& & \multicolumn{1}{|c|}{} & \multicolumn{1}{|c|}{} & \multicolumn{1}{|c|}{} \\[-0.75em]
\cline{3-5}
& & \multicolumn{1}{|c|}{} & \multicolumn{1}{|c|}{} & \multicolumn{1}{|c|}{} \\[-0.75em]
& & \multicolumn{1}{|c|}{$|V''|=1$} & \multicolumn{1}{|c|}{$|V''|=10$} & \multicolumn{1}{|c|}{$|V''|=100$} \\[0.5em]
\hline
\multicolumn{1}{|c|}{} & \multicolumn{1}{|c|}{} & \multicolumn{1}{|c|}{} & \multicolumn{1}{|c|}{} & \\[-0.5em]
\multicolumn{1}{|c|}{\multirow{3}{4.5em}{Quadratic Well}} & \multicolumn{1}{|c|}{$V$} & \multicolumn{1}{|c|}{$\frac{1}{2}(x^2 + y^2)$} & \multicolumn{1}{|c|}{$-\frac{9}{2} + 5(x^2 + y^2)$} & $-\frac{99}{2} + 50(x^2 + y^2)$ \\[0.75em]
\cline{2-5}
\multicolumn{1}{|c|}{} & \multicolumn{1}{|c|}{} & \multicolumn{1}{|c|}{} & \multicolumn{1}{|c|}{} & \multicolumn{1}{|c|}{} \\[-0.5em]
\multicolumn{1}{|c|}{} & \multicolumn{1}{|c|}{$\bm{E}$} & $-\begin{bmatrix}
x \\ y
\end{bmatrix}$ & $-10\begin{bmatrix}
x \\ y
\end{bmatrix}$ & $-100\begin{bmatrix}
x \\ y
\end{bmatrix}$ \\[1.25em]
\hline
\multicolumn{1}{|c|}{} & \multicolumn{1}{|c|}{} & \multicolumn{1}{|c|}{} & \multicolumn{1}{|c|}{} & \multicolumn{1}{|c|}{} \\[-0.5em]
\multicolumn{1}{|c|}{\multirow{5}{4.25em}{Cubic Well}} & \multicolumn{1}{|c|}{\multirow{2}{*}{$V$}} & $-\frac{2}{3} + x^2 + y^2$ & $-\frac{19}{6} + 3(x^2 + y^2)$ & $-\frac{95}{2} + 47(x^2 + y^2)$ \\[0.25em]
\multicolumn{1}{|c|}{} & & $+ \frac{1}{6}(x^3 + y^3)$ & $+ \frac{2}{3}(x^3 + y^3)$ & $+ x^3 + y^3$ \\[0.75em]
\cline{2-5}
\multicolumn{1}{|c|}{} & \multicolumn{1}{|c|}{} & \multicolumn{1}{|c|}{} & \multicolumn{1}{|c|}{} & \multicolumn{1}{|c|}{} \\[-0.5em]
\multicolumn{1}{|c|}{} & \multicolumn{1}{|c|}{$\bm{E}$} & $-\begin{bmatrix}
2x - \frac{1}{2}x^2 \\[0.25em] 2y - \frac{1}{2}y^2
\end{bmatrix}$ & $-\begin{bmatrix}
6x - 2x^2 \\[0.25em] 6y - 2y^2
\end{bmatrix}$ & $-\begin{bmatrix}
94x + 3x^2 \\[0.25em] 94y + 3y^2
\end{bmatrix}$ \\[1.25em]
\hline
\multicolumn{1}{|c|}{} & \multicolumn{1}{|c|}{} & \multicolumn{1}{|c|}{} & \multicolumn{1}{|c|}{} & \multicolumn{1}{|c|}{} \\[-0.5em]
\multicolumn{1}{|c|}{\multirow{3}{4em}{Quartic Well}} & \multicolumn{1}{|c|}{$V$} & $\frac{5}{12} + \frac{1}{12}(x^4 + y^4)$ & $-\frac{1}{3} + \frac{5}{6}(x^4 + y^4)$ & $-\frac{47}{6} +\frac{25}{3}(x^4 + y^4)$ \\[0.75em]
\cline{2-5}
\multicolumn{1}{|c|}{} & \multicolumn{1}{|c|}{} & \multicolumn{1}{|c|}{} & \multicolumn{1}{|c|}{} & \multicolumn{1}{|c|}{} \\[-0.5em]
\multicolumn{1}{|c|}{} & \multicolumn{1}{|c|}{$\bm{E}$} & $-\frac{1}{3}\begin{bmatrix}
x^3 \\ y^3
\end{bmatrix}$ & $-\frac{10}{3}\begin{bmatrix}
x^3 \\ y^3
\end{bmatrix}$ & $-\frac{100}{3}\begin{bmatrix}
x^3 \\ y^3
\end{bmatrix}$ \\[1.25em]
\hline
\end{tabular}
\caption{Electric scalar potential wells and corresponding electric fields for 2D model test problems}\label{2DWellConfig}
\end{table}
\vfill

\begin{table}[h!]
\centering
\begin{tabular}{ll|c|c|c|}
& & \multicolumn{1}{|c|}{} & \multicolumn{1}{|c|}{} & \multicolumn{1}{|c|}{} \\[-0.75em]
\cline{3-5}
& & \multicolumn{1}{|c|}{} & \multicolumn{1}{|c|}{} & \multicolumn{1}{|c|}{} \\[-0.75em]
& & \multicolumn{1}{|c|}{$|V''|=1$} & \multicolumn{1}{|c|}{$|V''|=10$} & \multicolumn{1}{|c|}{$|V''|=100$} \\[0.5em]
\hline
\multicolumn{1}{|c|}{} & \multicolumn{1}{|c|}{} & \multicolumn{1}{|c|}{} & \multicolumn{1}{|c|}{} & \\[-0.5em]
\multicolumn{1}{|c|}{\multirow{3}{4.5em}{Quadratic Hill}} & \multicolumn{1}{|c|}{$V$} & \multicolumn{1}{|c|}{$1 - \frac{1}{2}(x^2 + y^2)$} & \multicolumn{1}{|c|}{$\frac{11}{2} - 5(x^2 + y^2)$} & \multicolumn{1}{|c|}{$\frac{101}{2} - 50(x^2 + y^2)$} \\[0.75em]
\cline{2-5}
\multicolumn{1}{|c|}{} & \multicolumn{1}{|c|}{} & \multicolumn{1}{|c|}{} & \multicolumn{1}{|c|}{} & \multicolumn{1}{|c|}{} \\[-0.5em]
\multicolumn{1}{|c|}{} & \multicolumn{1}{|c|}{$\bm{E}$} & \multicolumn{1}{|c|}{$\begin{bmatrix}
x \\ y
\end{bmatrix}$} & \multicolumn{1}{|c|}{$10\begin{bmatrix}
x \\ y
\end{bmatrix}$} & \multicolumn{1}{|c|}{$100\begin{bmatrix}
x \\ y
\end{bmatrix}$} \\[1.25em]
\hline
\multicolumn{1}{|c|}{} & \multicolumn{1}{|c|}{} & \multicolumn{1}{|c|}{} & \multicolumn{1}{|c|}{} & \multicolumn{1}{|c|}{} \\[-0.5em]
\multicolumn{1}{|c|}{\multirow{5}{4.25em}{Cubic Hill}} & \multicolumn{1}{|c|}{\multirow{2}{*}{$V$}} & \multicolumn{1}{|c|}{$\frac{4}{3} - x^2 - y^2$} & \multicolumn{1}{|c|}{$\frac{25}{6} - 3(x^2 + y^2)$} & \multicolumn{1}{c|}{$\frac{97}{2} - 47(x^2 + y^2)$} \\[0.25em]
\multicolumn{1}{|c|}{} & \multicolumn{1}{|c|}{} & $+ \frac{1}{6}(x^3 + y^3)$ & $-\frac{2}{3}(x^3 + y^3)$ & \multicolumn{1}{|c|}{$- x^3 - y^3$} \\[0.75em]
\cline{2-5}
\multicolumn{1}{|c|}{} & \multicolumn{1}{|c|}{} & \multicolumn{1}{|c|}{} & \multicolumn{1}{|c|}{} & \multicolumn{1}{|c|}{} \\[-0.5em]
\multicolumn{1}{|c|}{} & \multicolumn{1}{|c|}{$\bm{E}$} & \multicolumn{1}{|c|}{$\begin{bmatrix}
2x - \frac{1}{2}x^2 \\[0.25em] 2y - \frac{1}{2}y^2
\end{bmatrix}$} & \multicolumn{1}{|c|}{$\begin{bmatrix}
6x - 2x^2 \\[0.25em] 6y - 2y^2
\end{bmatrix}$} & \multicolumn{1}{|c|}{$\begin{bmatrix}
94x + 3x^2 \\[0.25em] 94y + 3y^2
\end{bmatrix}$} \\[1.25em]
\hline
\multicolumn{1}{|c|}{} & \multicolumn{1}{|c|}{} & \multicolumn{1}{|c|}{} & \multicolumn{1}{|c|}{} & \multicolumn{1}{|c|}{} \\[-0.5em]
\multicolumn{1}{|c|}{\multirow{3}{4em}{Quartic Hill}} & \multicolumn{1}{|c|}{$V$} & \multicolumn{1}{|c|}{$\frac{7}{12} - \frac{1}{12}(x^4 + y^4)$} & \multicolumn{1}{|c|}{$\frac{4}{3} - \frac{5}{6}(x^4 + y^4)$} & \multicolumn{1}{c|}{$\frac{53}{6} - \frac{25}{3}(x^4 + y^4)$} \\[0.75em]
\cline{2-5}
\multicolumn{1}{|c|}{} & \multicolumn{1}{|c|}{} & \multicolumn{1}{|c|}{} & \multicolumn{1}{|c|}{} & \multicolumn{1}{|c|}{} \\[-0.5em]
\multicolumn{1}{|c|}{} & \multicolumn{1}{|c|}{$\bm{E}$} & \multicolumn{1}{|c|}{$\frac{1}{3}\begin{bmatrix}
x^3 \\ y^3
\end{bmatrix}$} & \multicolumn{1}{|c|}{$\frac{10}{3}\begin{bmatrix}
x^3 \\ y^3
\end{bmatrix}$} & \multicolumn{1}{|c|}{$\frac{100}{3}\begin{bmatrix}
x^3 \\ y^3
\end{bmatrix}$} \\[1.25em]
\hline
\end{tabular}
\caption{Electric scalar potential hills and corresponding electric fields for 2D model test problems}\label{2DHillConfig}
\end{table}
\vfill

\newpage

\newpage

\newpage

\newpage

\newpage

\newpage

\newpage
\subsubsection{Non-Uniform Magnetic Field, Grad-$B$ Drift Problem, 2D Model}
In this experiment we considered a prototype non-uniform magnetic field configuration with the so-called grad-$B$ drift problem. This problem has non-uniform magnetic field with linear spatial variation in which the length scale of spatial variation is of much longer than the gyroradius. Formally, we assume
\[
\frac{r\|\nabla B\|}{B} \ll 1,
\]
where $r$ is the gyroradius, $B = \|\bm{B}\|$, and $\nabla B$ is the magnetic field gradient. Under this assumption, the particle experiences a drift velocity \cite{Chen,Nicholson} approximately given by
\[
\bm{v}_{\nabla B} = \frac{1}{2}\frac{v_\perp^2}{\omega}\frac{\bm{B}\times\nabla B}{B^2},
\]
where $v_\perp$ is the particle speed in the plane perpendicular to the magnetic field and $\omega = qB/m$ is the gyrofrequency.

\begin{figure}[h!]
\centering
\begin{tabular}{ccc}
$\delta B = 0.1$ & $\delta B = 1$ & $\delta B = 10$ \\[0.5em]
\includegraphics[scale=0.35]{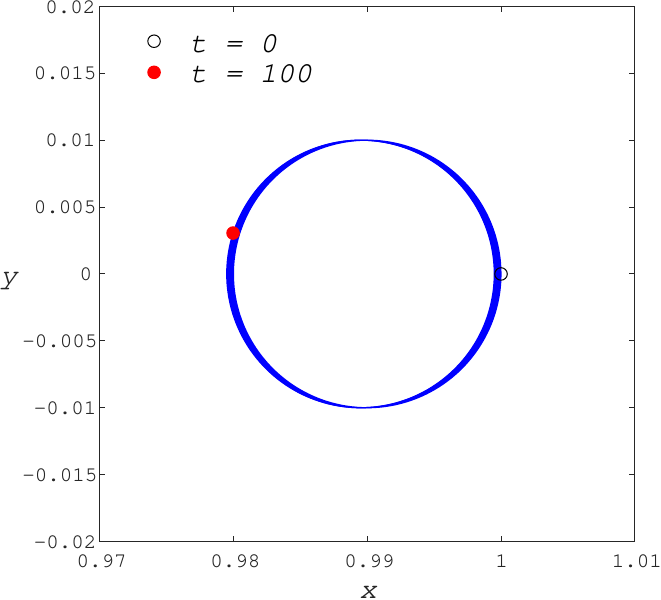} & \includegraphics[scale=0.35]{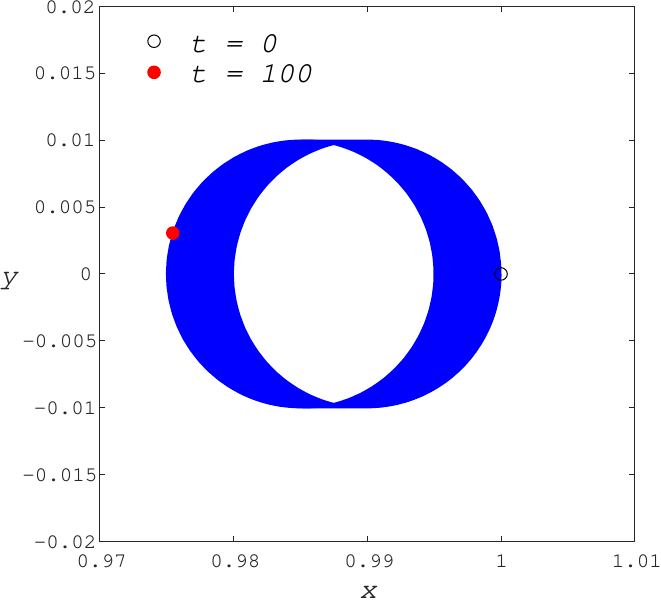} & \includegraphics[scale=0.35]{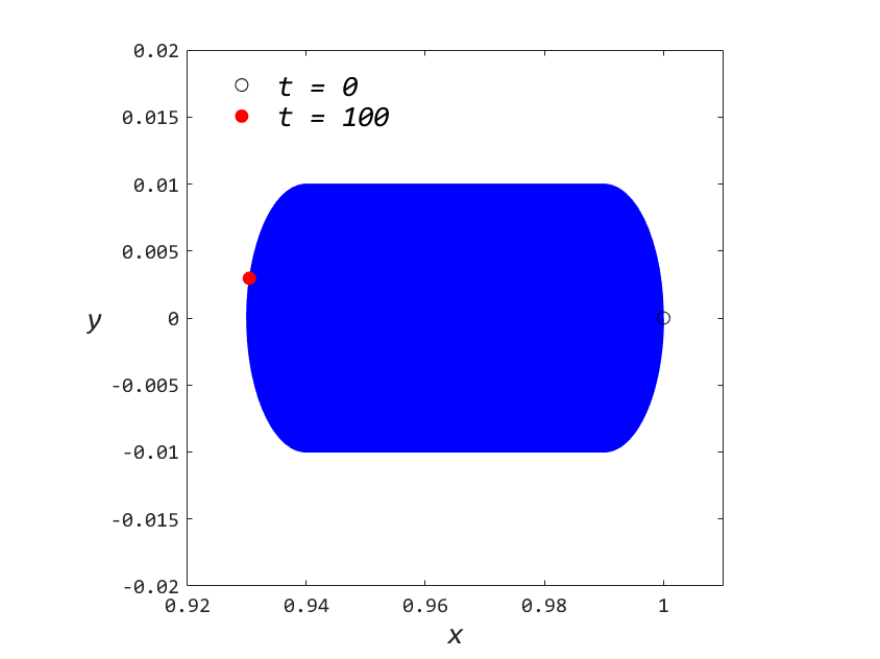} \\
\includegraphics[scale=0.35]{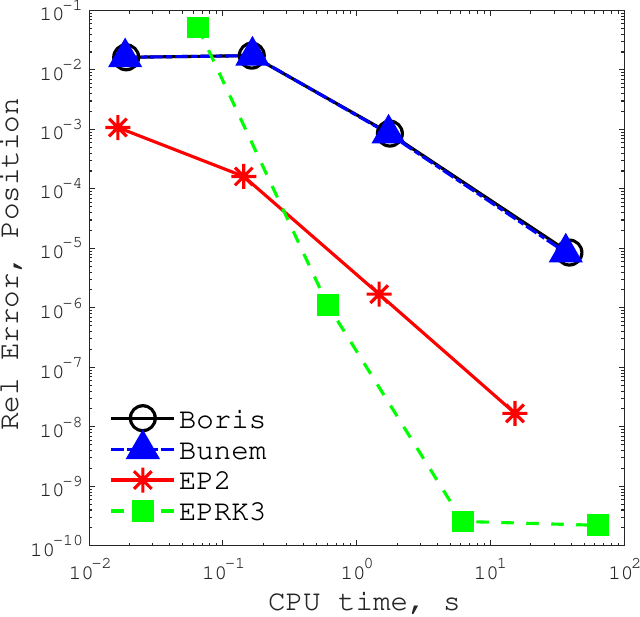} & \includegraphics[scale=0.35]{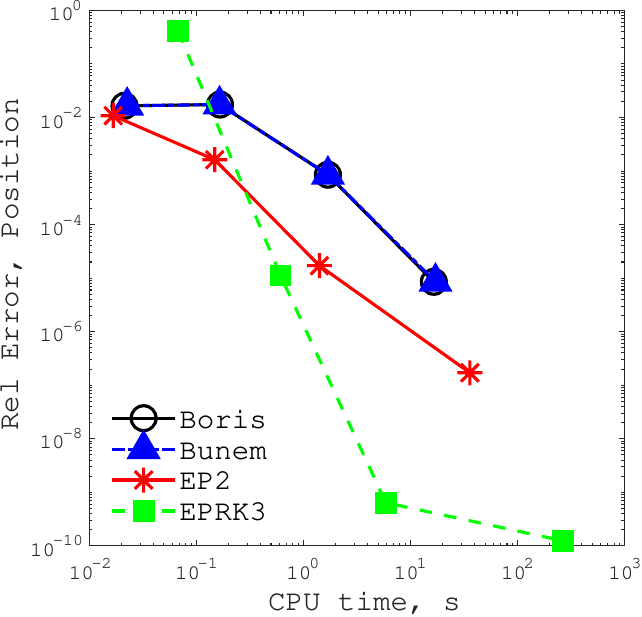} & \includegraphics[scale=0.35]{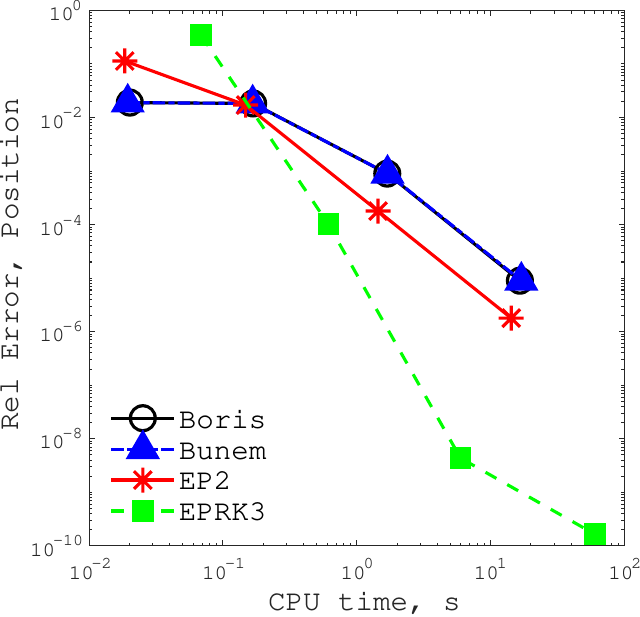} \\
\end{tabular}
\caption{Results for grad-$B$ drift problem: reference solution orbits (top row), and precision diagrams (bottom row). Boris/Buneman step sizes are $h = 10^{-2}, 10^{-3}, 10^{-4}, 10^{-5}$. EP2/EPRK3 step sizes are $h = 10^{-1}, 10^{-2}, 10^{-3}, 10^{-4}$.}\label{GradBDrift}
\end{figure}

The test problem for this experiment was configured with zero electric field and magnetic field set to
\[
\bm{B} = (100 + \delta B\,y)\hat{\bm{z}}.
\]
Similar to the previous experiments in this section, the grad-$B$ drift experiment considers a particle of mass $m = 1$ and charge $q = 1$ with initial conditions $\bm{x}_0 = (1, 0)$ and $\bm{v}_0 = (0, -1)$. Solutions were obtained by integrating the equations of motion over the time interval $[0, 100]$. Figure \ref{GradBDrift} shows plots of the reference solution orbits and the precision diagrams for $\delta B$ = 0.1, 1, 10.

\newpage
\vfill
\vfill

\newpage
\subsubsection{Gyroradius, 2D Model}
\begin{figure}[h!]
\centering
\begin{tabular}{cc}
Boris & Buneman \\
\includegraphics[scale=0.5]{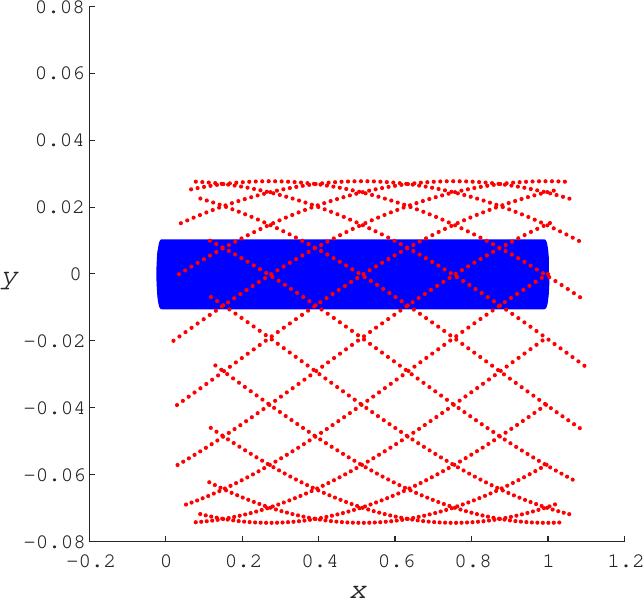} & \includegraphics[scale=0.5]{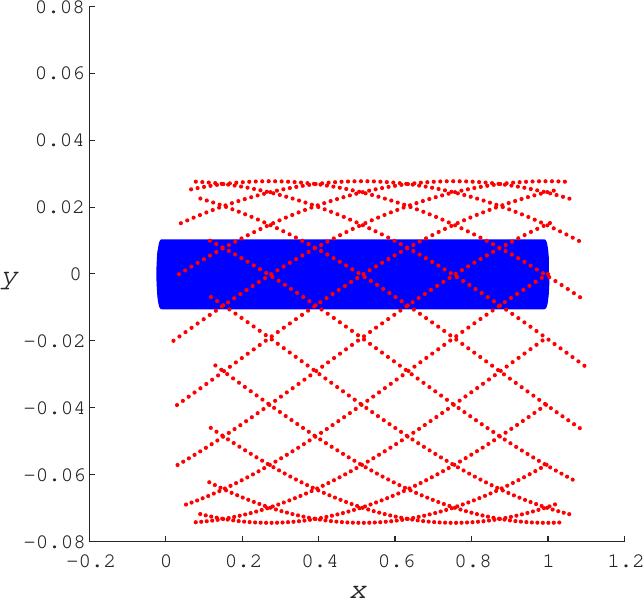} \\[1em]
EP2 & EPRK3 \\
\includegraphics[scale=0.5]{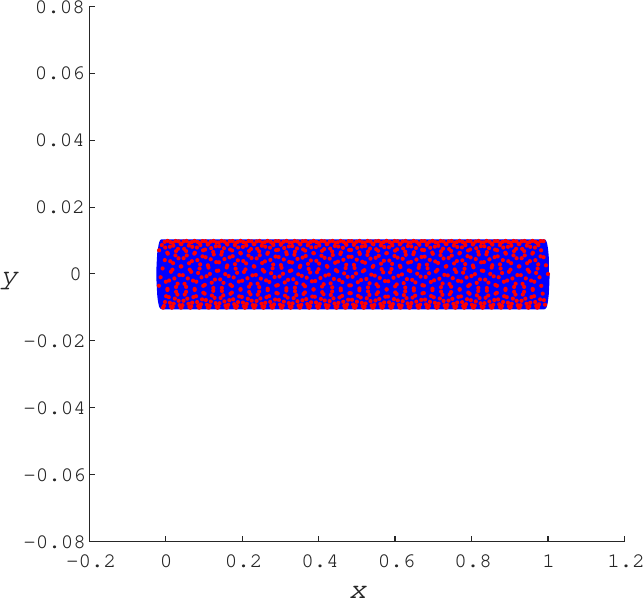} & \includegraphics[scale=0.5]{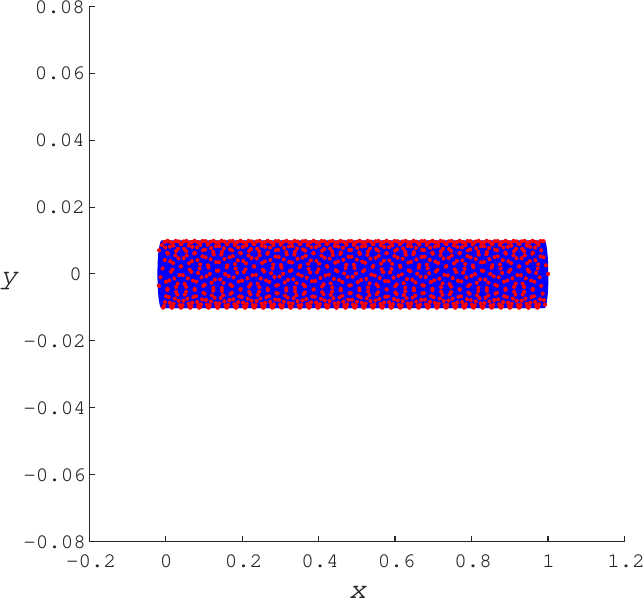}
\end{tabular}
\caption{Plots of computed trajectories for the $\bm{E}\times\bm{B}$ drift problem. Solutions for step size $h = 0.001$ are solid blue and solutions for step size $h = 0.1$ are dotted red.}\label{gyroradius_ExB}
\end{figure}

This experiment examines the gyroradius of the solutions computed by the numerical particle pushers. Since the Boris algorithm is known to compute an artificially enlarged gyroradius when using large step sizes relative to the gyroperiod \cite{Parker}, it is of interest to see how the exponential integrators perform in this regard. Here, the term "large step size relative to the gyroperiod" (or simply "large" step size) is defined by $\omega h \gg 1$, where $\omega = |q|B/m$ is the gyroperiod and $h$ is the time step size. Conversely, the term "small step size relative to the gyroperiod" (or simply "small" step size) is defined by $\omega h \ll 1$. Here we consider a linear $\bm{E}\times\bm{B}$ drift problem with electromagnetic fields
\[
\bm{B} = 100\,\hat{\bm{z}} \quad\text{and}\quad \bm{E} = -\begin{bmatrix}
0 \\
1 + y
\end{bmatrix}.
\]
The gyroradius is $r = 0.01$ for this particular configuration. Using a particle of mass $m = 1$ and charge $q = 1$ with initial conditions $\bm{x}_0 = (1, 0)$ and $\bm{v}_0 = (0, -1)$, we integrated the equations of motion over the time interval [0, 100] using a "small" step size $h = 0.001$ and a "large" step size $h = 0.1$. For the "small" and "large" step sizes, these yield $\omega h = 0.1 < 1$ and $\omega h = 10 > 1$, respectively. Results of the experiment are shown in figure \ref{gyroradius_ExB}. Observe that all the numerical particle pushers accurately computed the correct gyroradius for the "small" step size $h = 0.001$. However, for the "large" step size $h = 0.1$ both Boris and Buneman algorithms compute a drastically enlarged gyroradius while the exponential integrators compute the correct gyroradius.
\vfill
\vfill

\newpage
\subsection{Three Dimensional Model}\label{Results3DModel}
All three-dimensional test problems set the initial particle position at $\bm{x}_0 = (1, 0, 0)$ and the initial particle velocity at $\bm{v}_0 = (0, -1, 1)$. Configurations for the electric scalar potential wells and their corresponding electric fields are shown in table \ref{3DWellConfig}. Configurations for the electric scalar potential hills and the corresponding electric fields are shown in table \ref{3DHillConfig}.

\begin{figure}[h!]
\centering
\begin{tabular}{ccc}
\multicolumn{3}{c}{Potential Wells $\frac{|V''|}{B} = \frac{1}{100}$} \\
Quadratic & Cubic & Quartic \\
\includegraphics[scale=0.3]{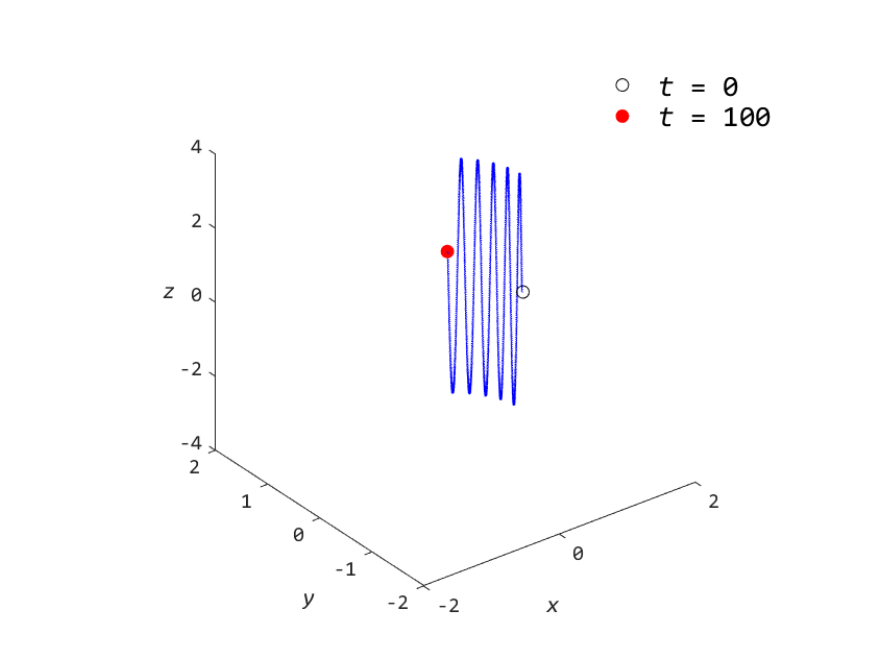} &
\includegraphics[scale=0.3]{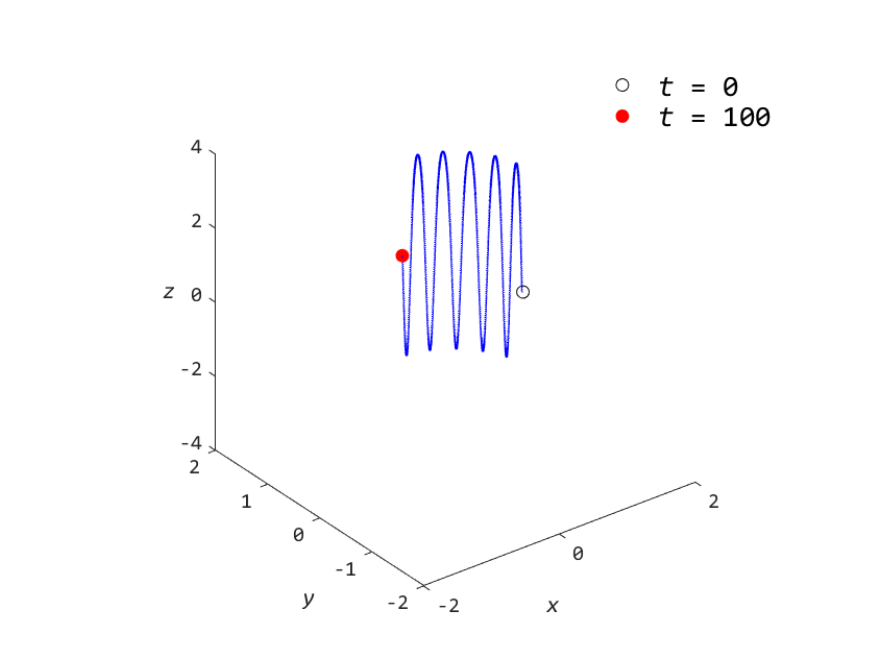} &
\includegraphics[scale=0.3]{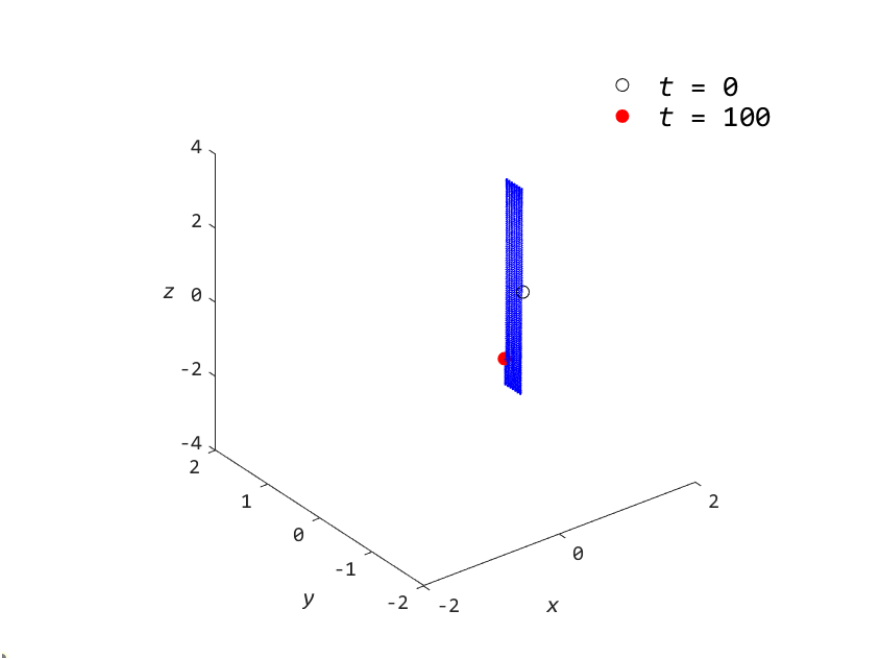} \\
\includegraphics[scale=0.3]{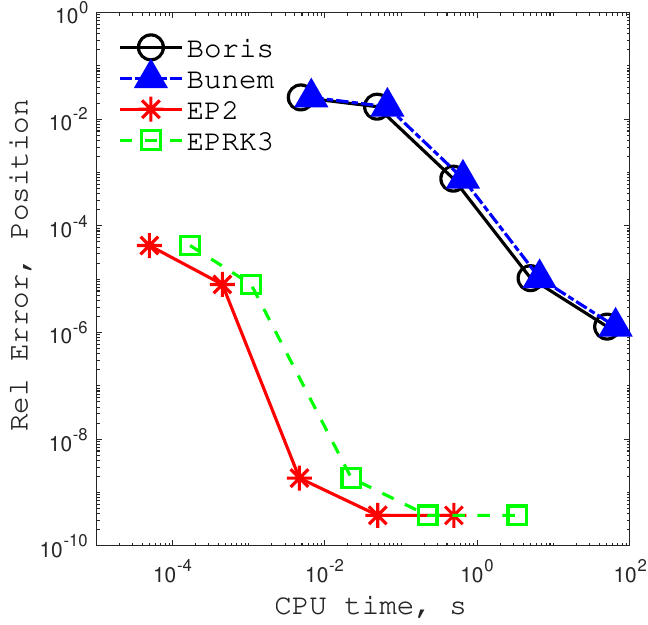} &
\includegraphics[scale=0.3]{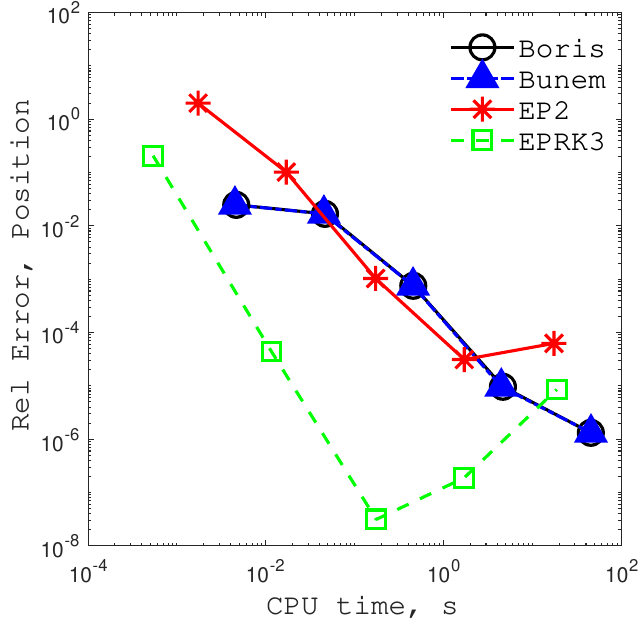} &
\includegraphics[scale=0.3]{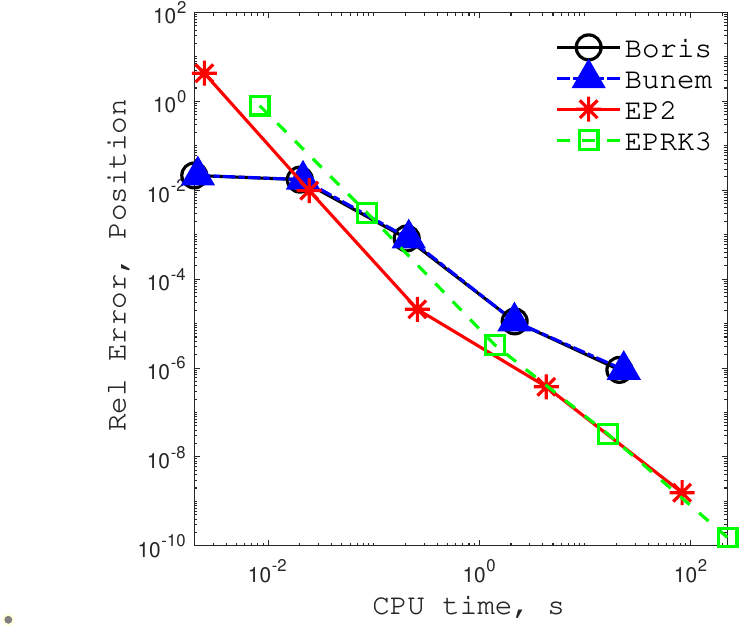} \\[1em]
\multicolumn{3}{c}{Potential Hills $\frac{|V''|}{B} = \frac{1}{100}$} \\
Quadratic & Cubic & Quartic \\
\includegraphics[scale=0.3]{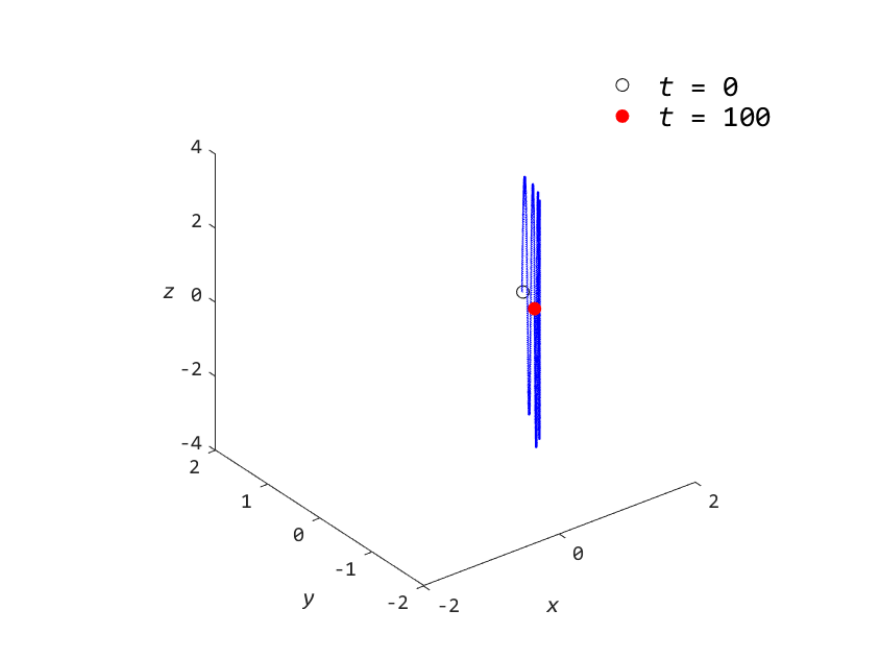} &
\includegraphics[scale=0.3]{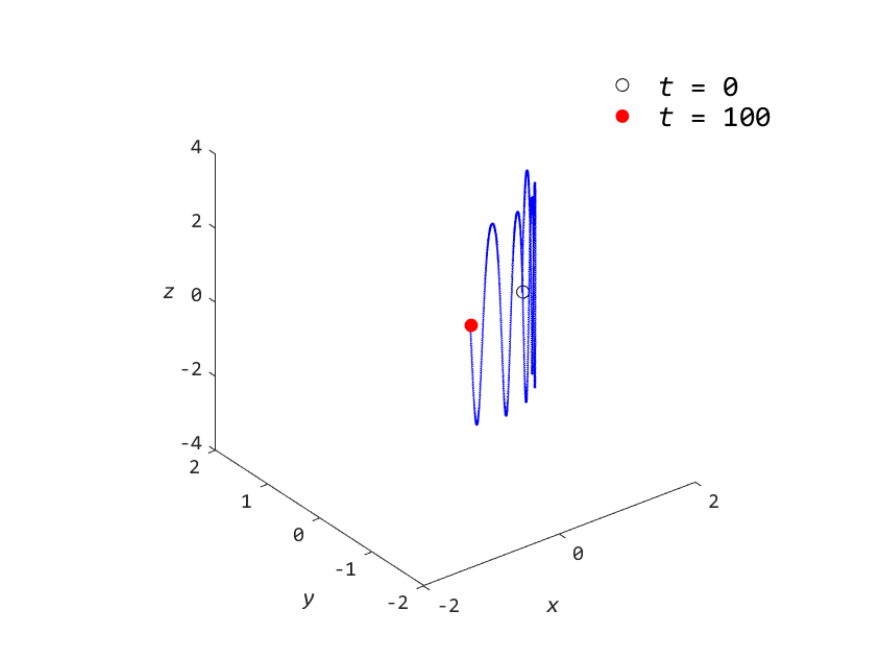} &
\includegraphics[scale=0.3]{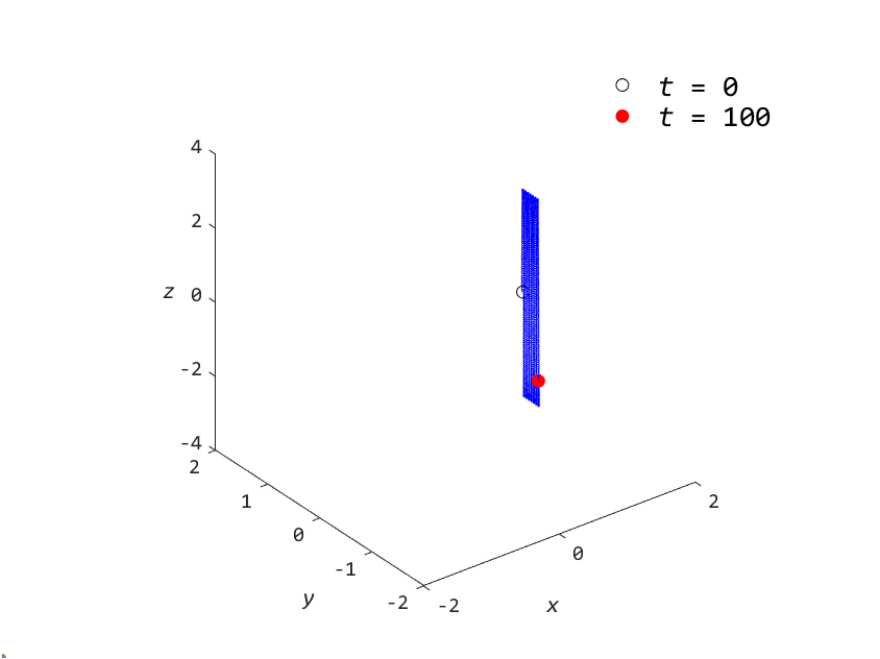} \\
\includegraphics[scale=0.3]{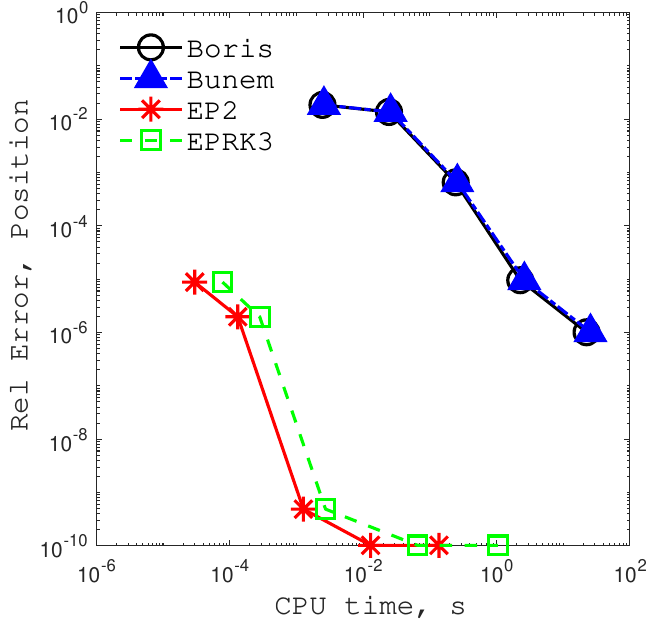} &
\includegraphics[scale=0.3]{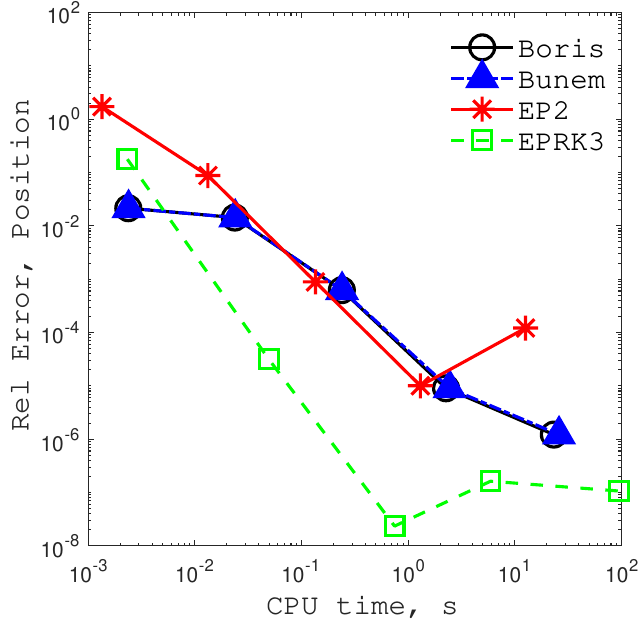} &
\includegraphics[scale=0.3]{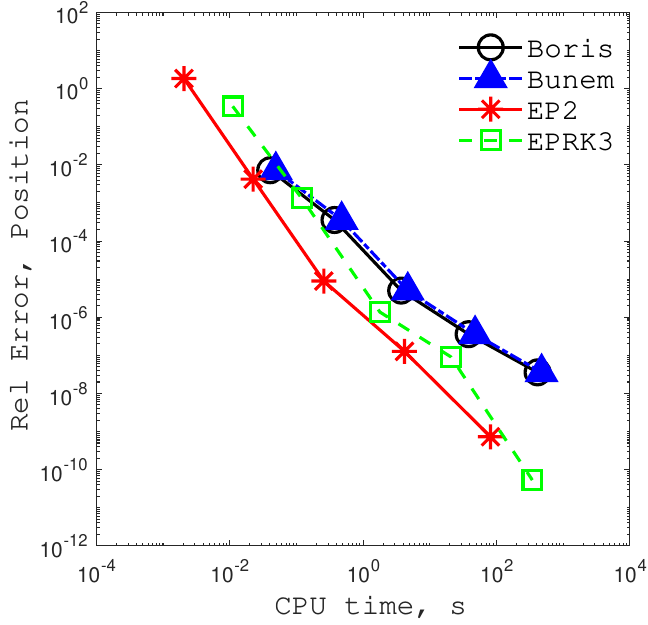}
\end{tabular}
\caption{Results for 3D test problems with $|V''|/B = 1/100$: potential well reference solution orbits (first row), potential well precision diagrams (second row), potential hill reference solution orbits (third row), and potential hill precision diagrams (fourth row). Boris/Buneman step sizes are $h = 10^{-2}, 10^{-3}, 10^{-4}, 10^{-5}, 10^{-6}$ for quadratic potential problems and $h = 10^{-3}, 10^{-4}, 10^{-5}, 10^{-6}, 10^{-7}$ for cubic/quartic potential problems. EP2/EPRK3 step sizes are $h = 100, 10, 1, 10^{-1}, 10^{-2}$ for quadratic potential problems and $h = 10^{-1}, 10^{-2}, 10^{-3}, 10^{-4}, 10^{-5}$ for cubic/quartic potential problems.}\label{3Dpot1}
\end{figure}

\begin{figure}[h!]
\centering
\begin{tabular}{ccc}
\multicolumn{3}{c}{Potential Wells $\frac{|V''|}{B} = \frac{1}{10}$} \\
Quadratic & Cubic & Quartic \\
\includegraphics[scale=0.3]{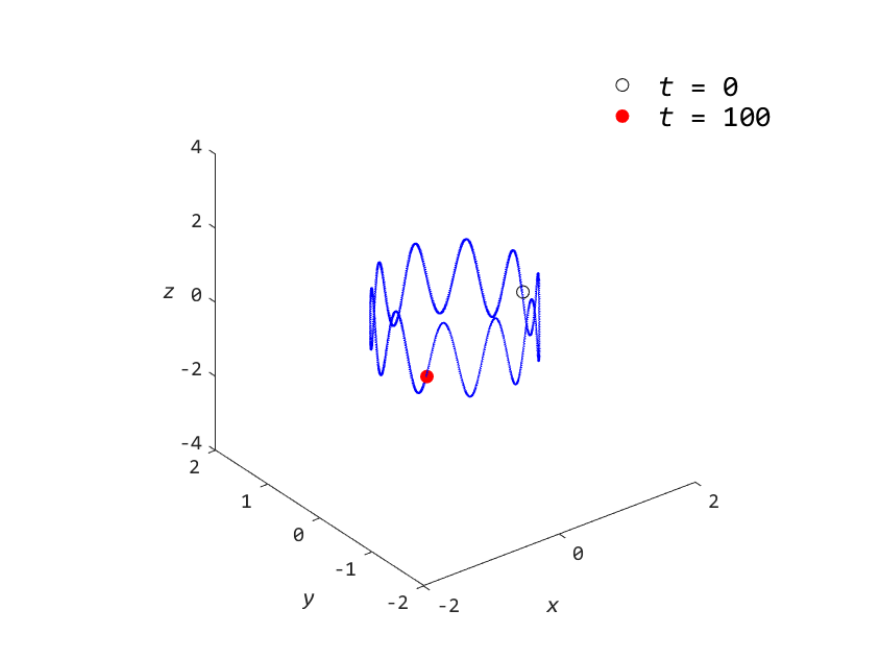} &
\includegraphics[scale=0.3]{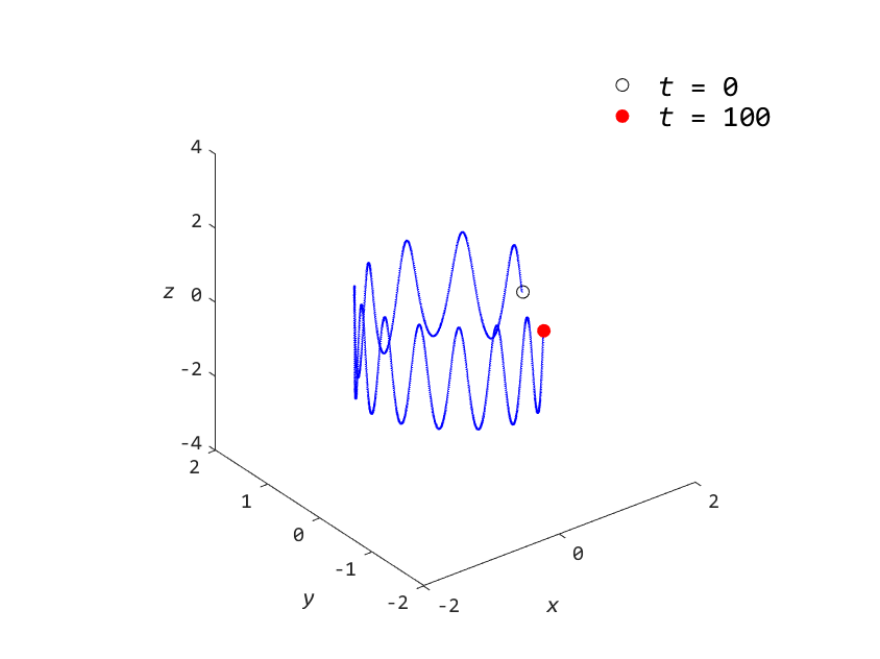} &
\includegraphics[scale=0.3]{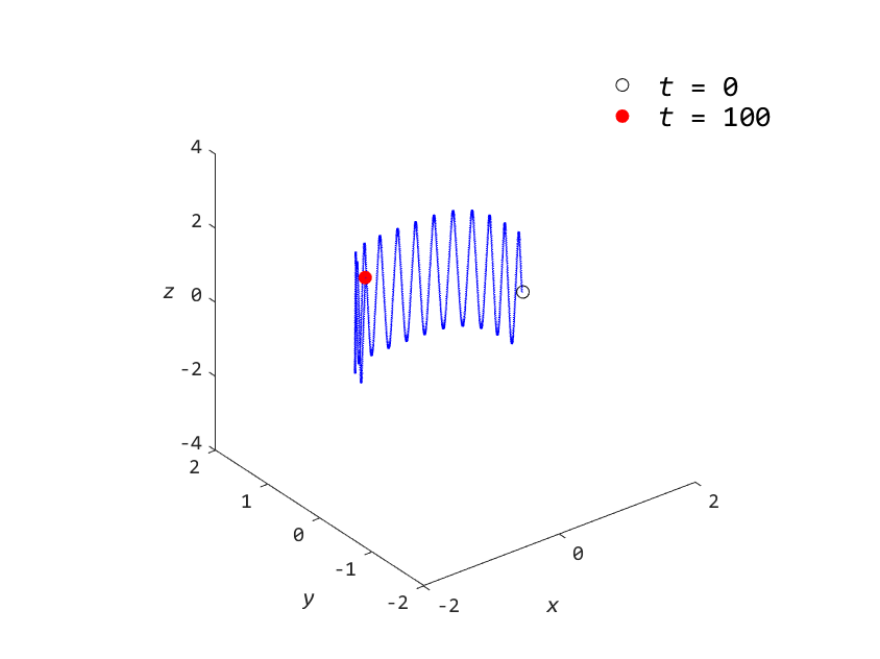} \\
\includegraphics[scale=0.3]{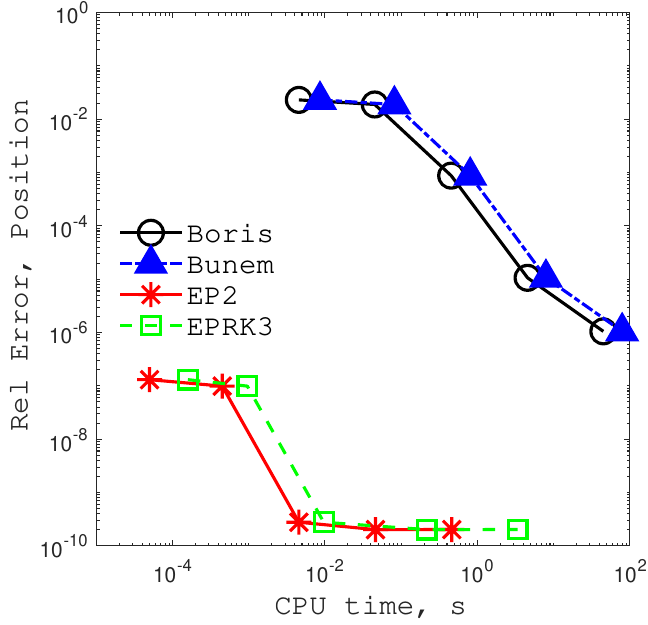} &
\includegraphics[scale=0.3]{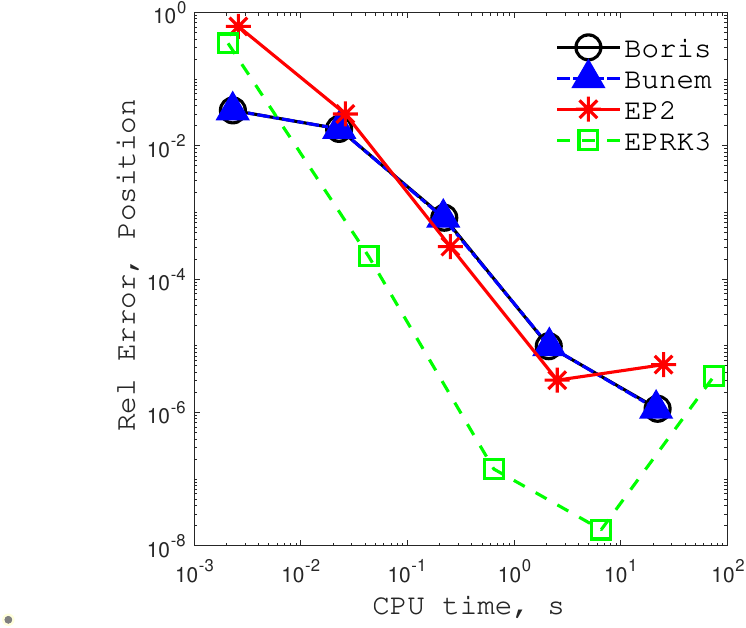} &
\includegraphics[scale=0.3]{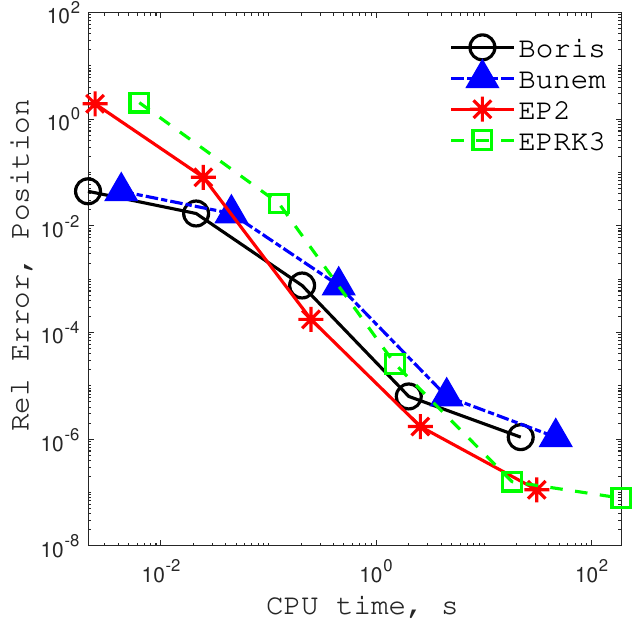} \\[1em]
\multicolumn{3}{c}{Potential Hills $\frac{|V''|}{B} = \frac{1}{10}$} \\
Quadratic & Cubic & Quartic \\
\includegraphics[scale=0.3]{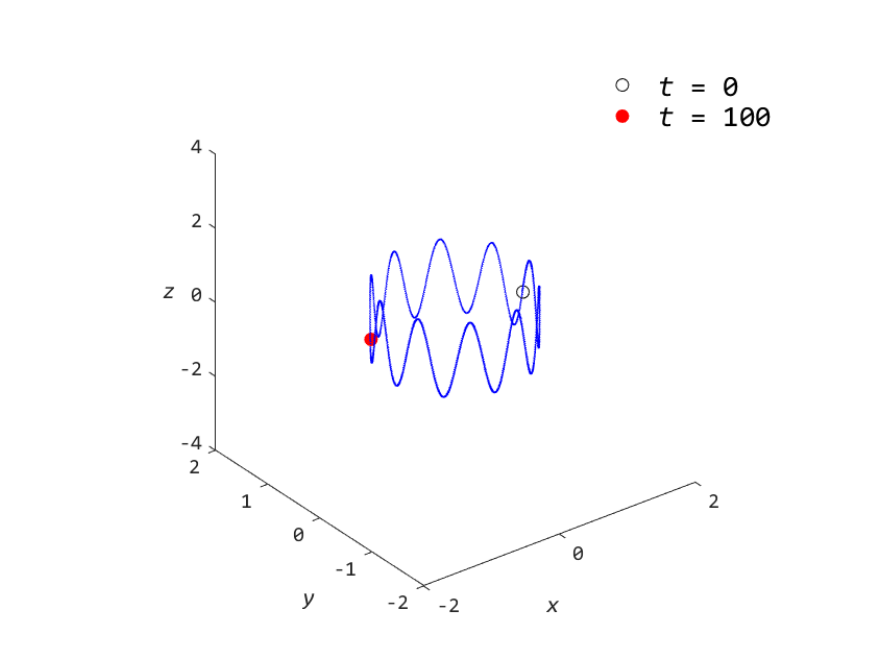} &
\includegraphics[scale=0.3]{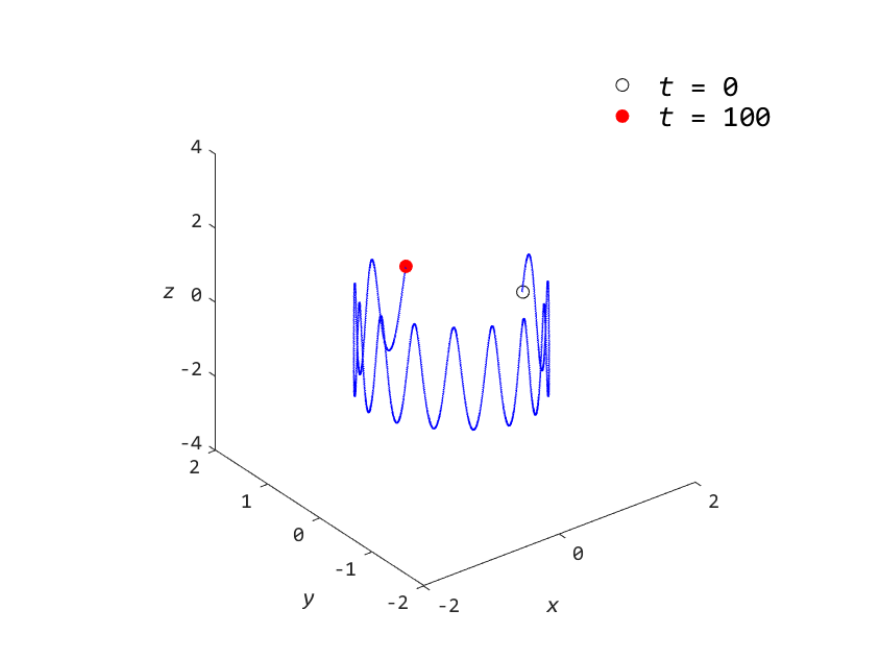} &
\includegraphics[scale=0.3]{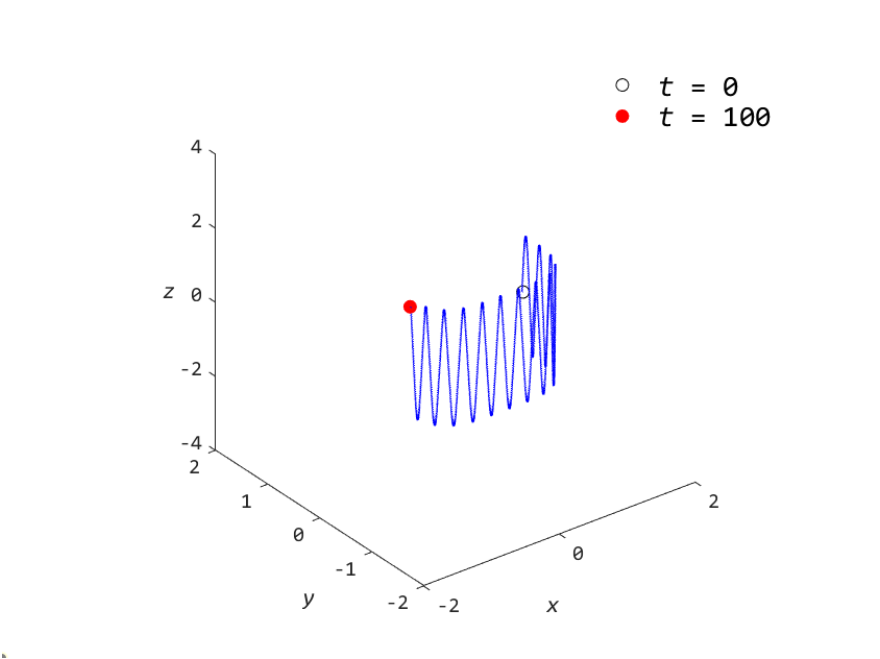} \\
\includegraphics[scale=0.3]{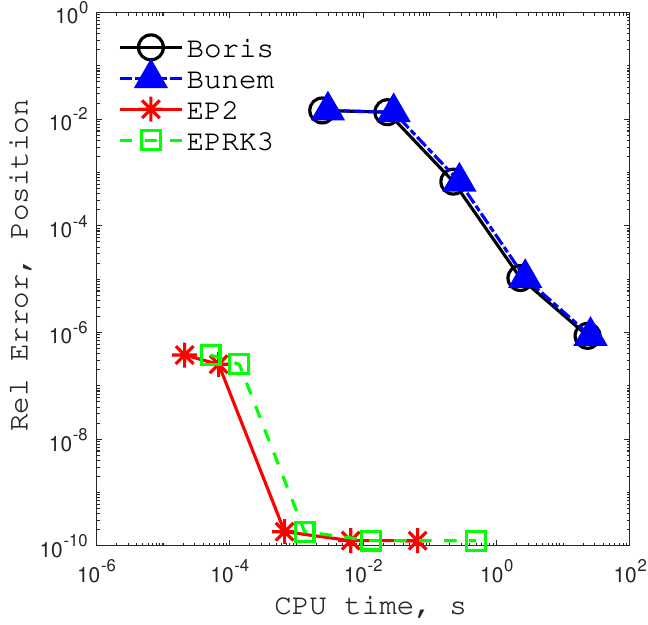} &
\includegraphics[scale=0.3]{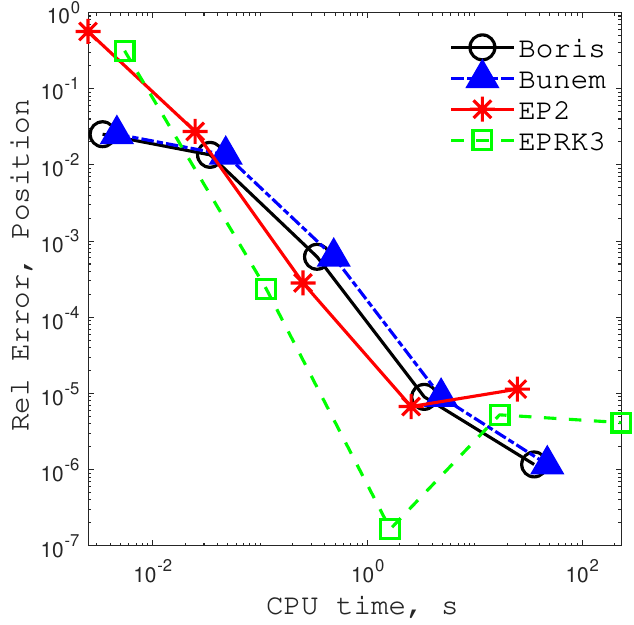} &
\includegraphics[scale=0.3]{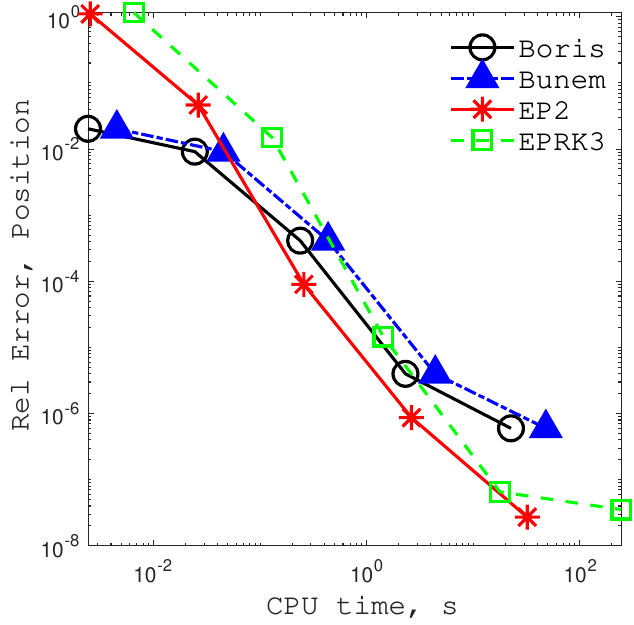}
\end{tabular}
\caption{Results for 3D test problems with $|V''|/B = 1/10$: potential well reference solution orbits (first row), potential well precision diagrams (second row), potential hill reference solution orbits (third row), and potential hill precision diagrams (fourth row). Boris/Buneman step sizes are $h = 10^{-2}, 10^{-3}, 10^{-4}, 10^{-5}, 10^{-6}$ for quadratic potential problems and $h = 10^{-3}, 10^{-4}, 10^{-5}, 10^{-6}, 10^{-7}$ for cubic/quartic potential problems. EP2/EPRK3 step sizes are $h = 100, 10, 1, 10^{-1}, 10^{-2}$ for quadratic potential problems and $h = 10^{-1}, 10^{-2}, 10^{-3}, 10^{-4}, 10^{-5}$ for cubic/quartic potential problems.}\label{3Dpot2}
\end{figure}

\begin{figure}[h!]
\centering
\begin{tabular}{ccc}
\multicolumn{3}{c}{Potential Wells $\frac{|V''|}{B} = 1$} \\
Quadratic & Cubic & Quartic \\
\includegraphics[scale=0.3]{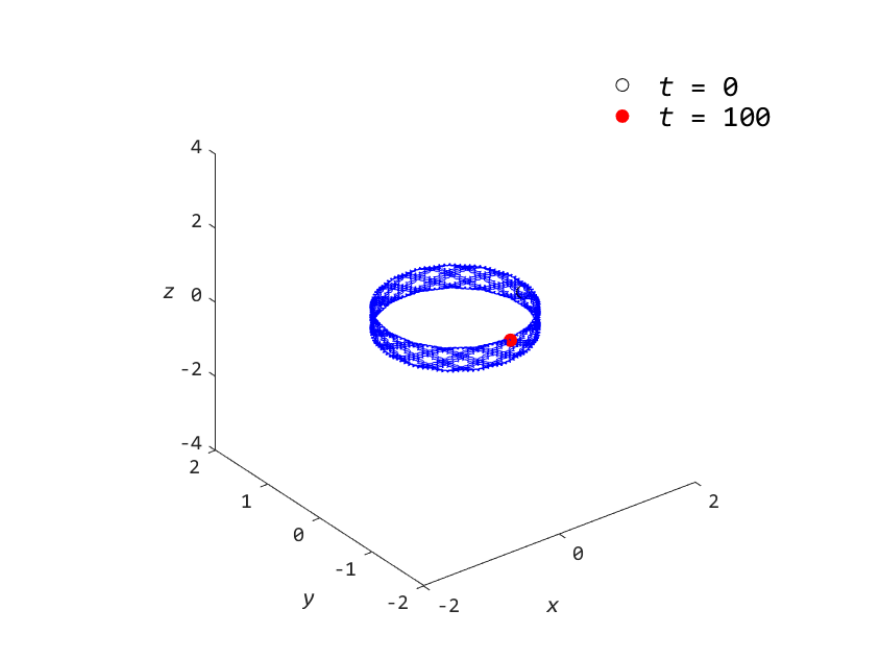} &
\includegraphics[scale=0.3]{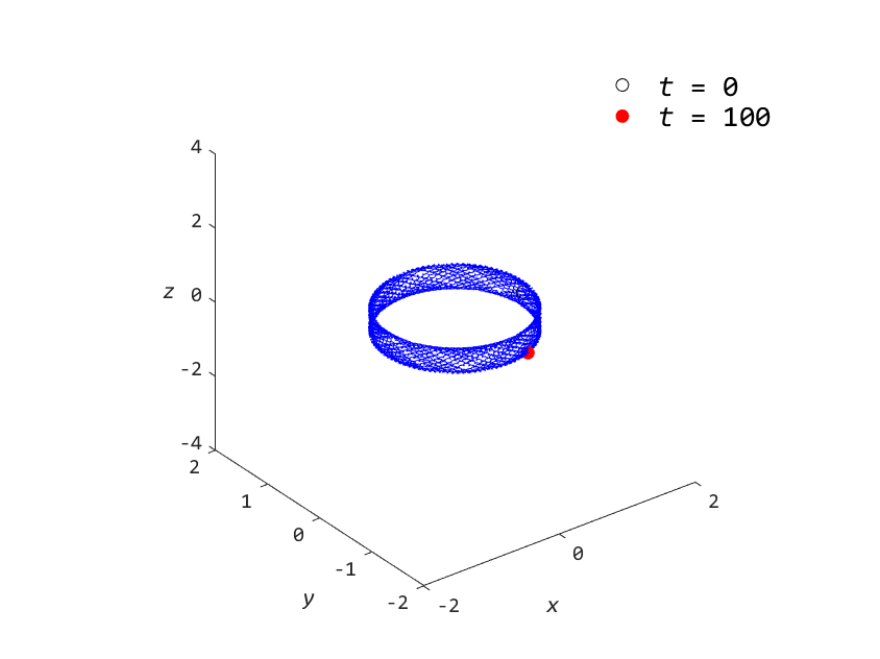} &
\includegraphics[scale=0.3]{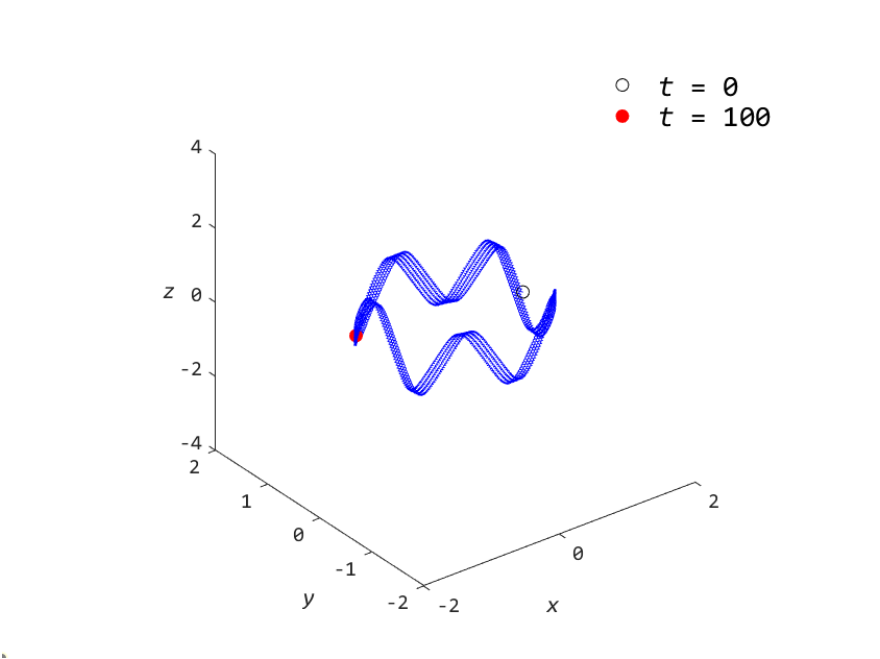} \\
\includegraphics[scale=0.3]{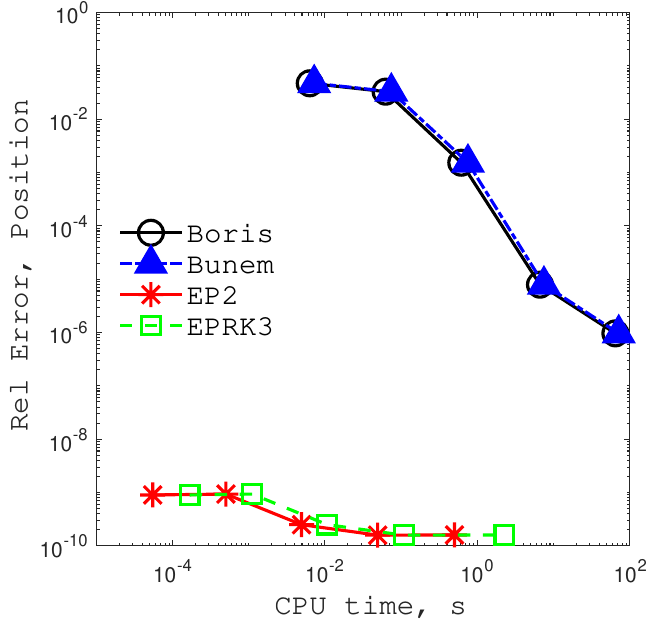} &
\includegraphics[scale=0.3]{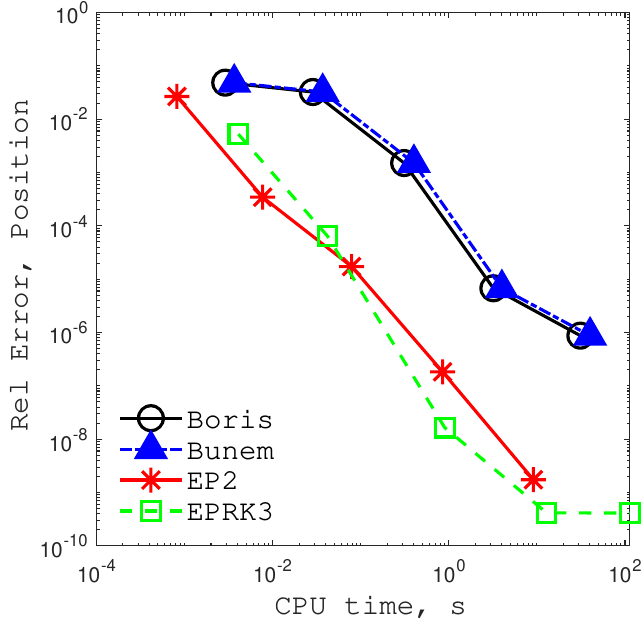} &
\includegraphics[scale=0.3]{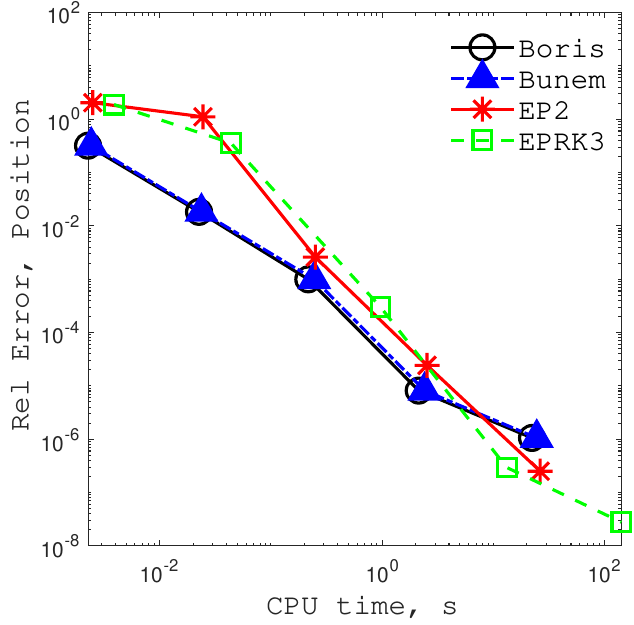} \\[1em]
\multicolumn{3}{c}{Potential Hills $\frac{|V''|}{B} = 1$} \\
Quadratic & Cubic & Quartic \\
\includegraphics[scale=0.3]{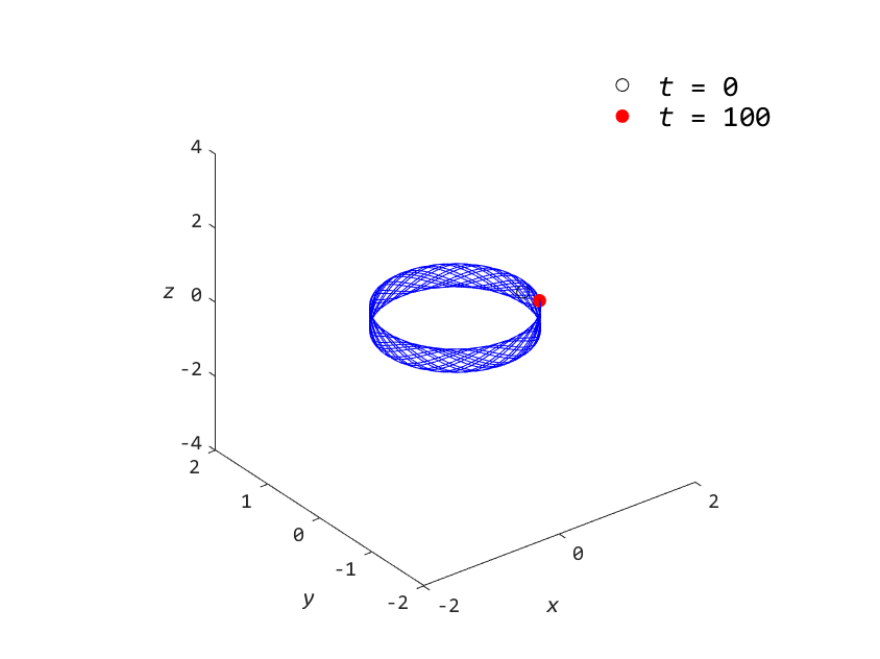} &
\includegraphics[scale=0.3]{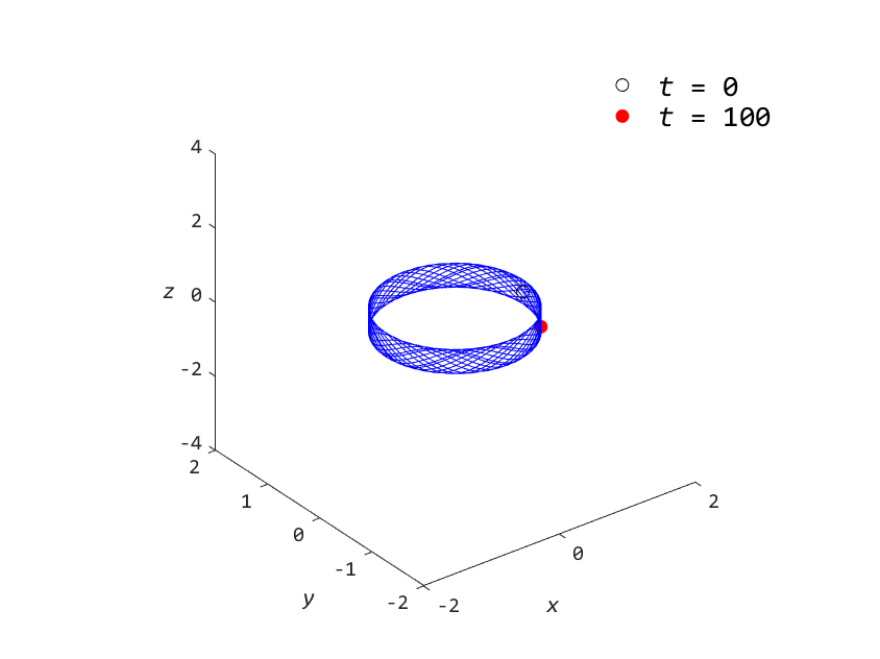} &
\includegraphics[scale=0.3]{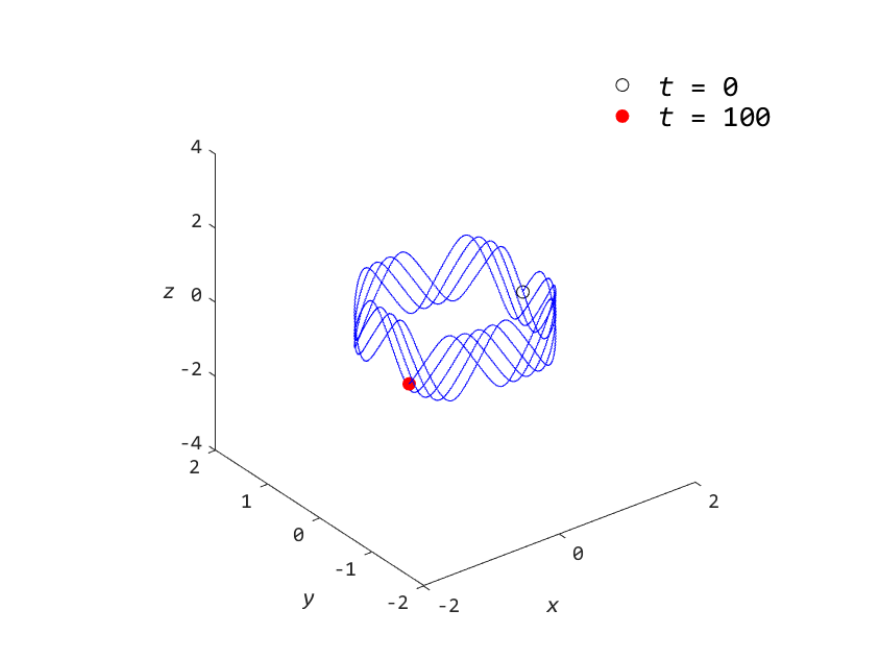} \\
\includegraphics[scale=0.3]{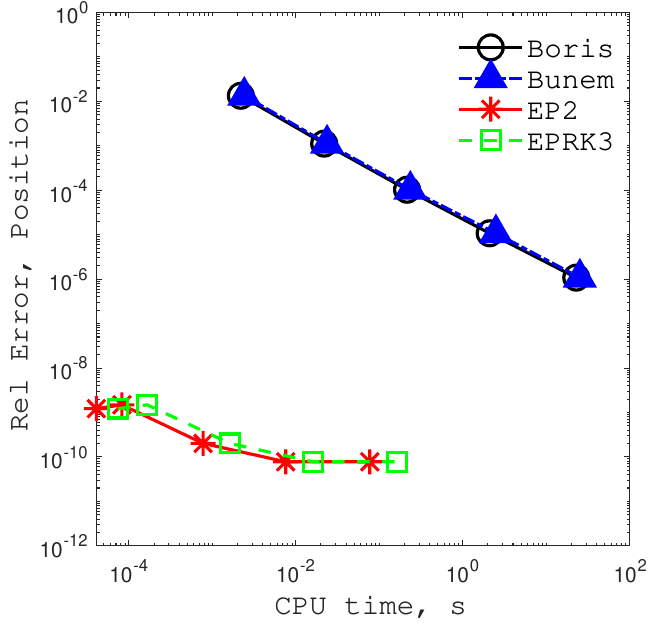} &
\includegraphics[scale=0.3]{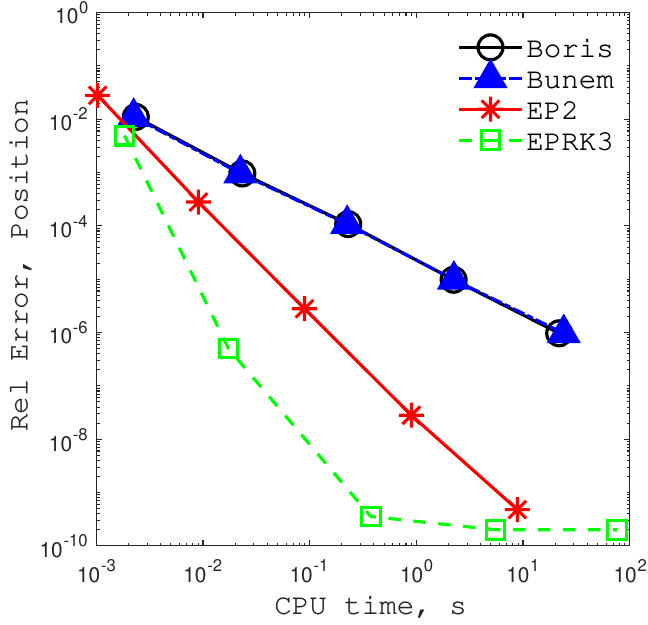} &
\includegraphics[scale=0.3]{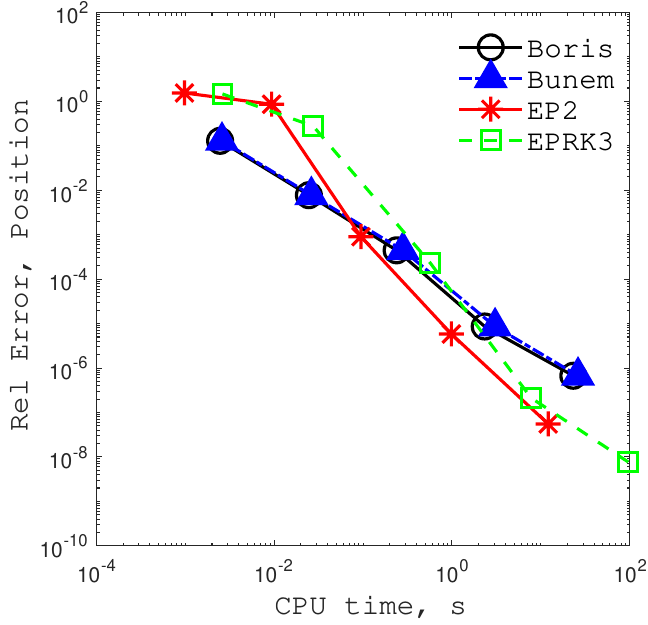}
\end{tabular}
\caption{Results for 3D test problems with $|V''|/B = 1$: potential well reference solution orbits (first row), potential well precision diagrams (second row), potential hill reference solution orbits (third row), and potential hill precision diagrams (fourth row). Boris/Buneman step sizes are $h = 10^{-2}, 10^{-3}, 10^{-4}, 10^{-5}, 10^{-6}$ for quadratic potential problems and $h = 10^{-3}, 10^{-4}, 10^{-5}, 10^{-6}, 10^{-7}$ for cubic/quartic potential problems. EP2/EPRK3 step sizes are $h = 100, 10, 1, 10^{-1}, 10^{-2}$ for quadratic potential problems and $h = 10^{-1}, 10^{-2}, 10^{-3}, 10^{-4}, 10^{-5}$ for cubic/quartic potential problems.}\label{3Dpot3}
\end{figure}

Figures \ref{3Dpot1}, \ref{3Dpot2}, and \ref{3Dpot3} show plots of the reference solution orbits and precision diagrams for test problems with $|V''|/B = 1/100$, $|V''|/B = 1/10$, and $|V''|/B = 1$, respectively.  Note that overall the comparative performance of the exponential methods with traditional particle pushers is similar for three-dimensional problems compared to two-dimensional cases.  As in the two-dimensional experiments the exponential methods perform well in the linear case and remain competitive for cubic and quartic potentials. A minor difference for the linear case performance between two-dimensional and three-dimensional is in the slight increase of the error for the largest steps sizes. This is the result of finite precision computations of large analytic formulas involved in evaluation of the eigenvalues of the Jacobian matrix and the polynomials in the Lagrange-Sylvester formula. This error can be reduced or eliminated, if needed, if the calculations are performed using software packages that double the precision of the calculations.  

\begin{figure}[h!]
\centering
\begin{tabular}{ccc}
\multicolumn{3}{c}{Potential Wells, $\frac{|V''|}{B} = \frac{1}{100}$} \\[0.5em]
Quadratic & Cubic & Quartic \\
\includegraphics[scale=0.3]{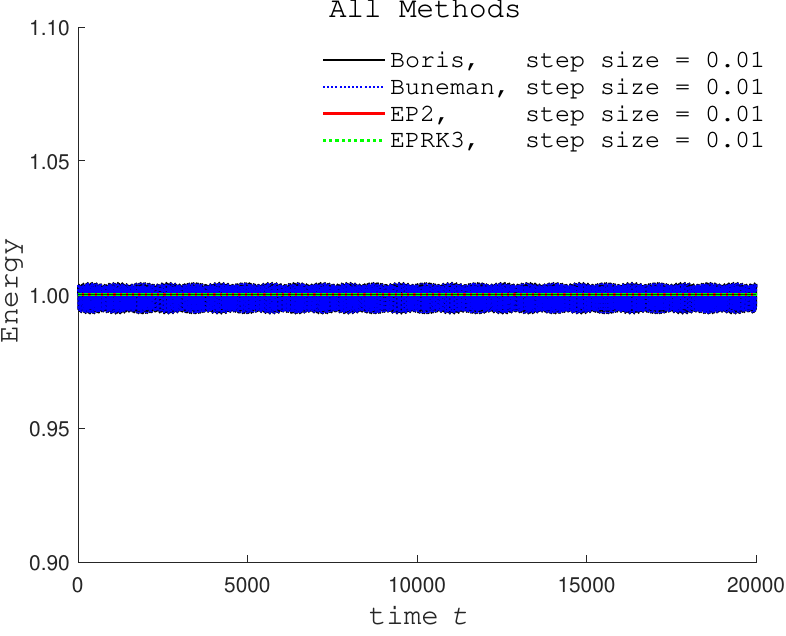} & \includegraphics[scale=0.3]{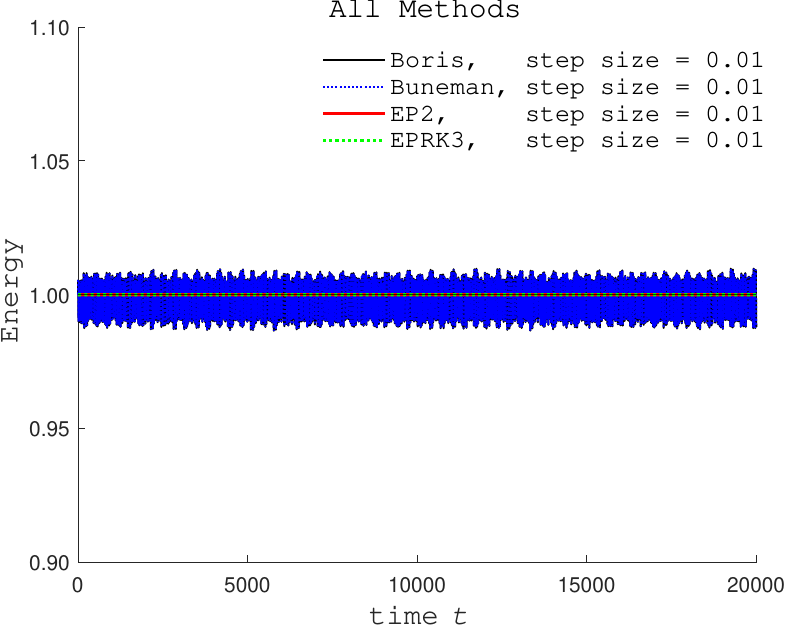} & \includegraphics[scale=0.3]{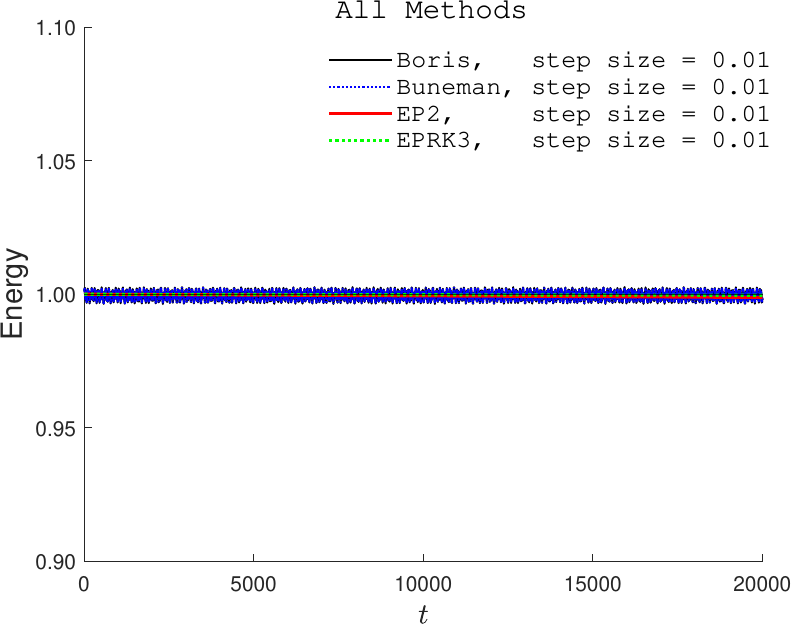} \\
\includegraphics[scale=0.3]{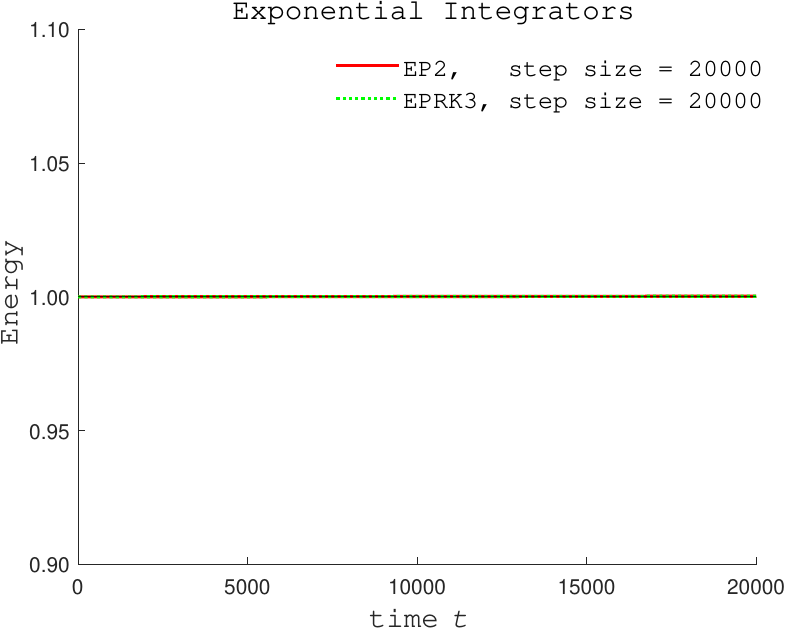} & \includegraphics[scale=0.3]{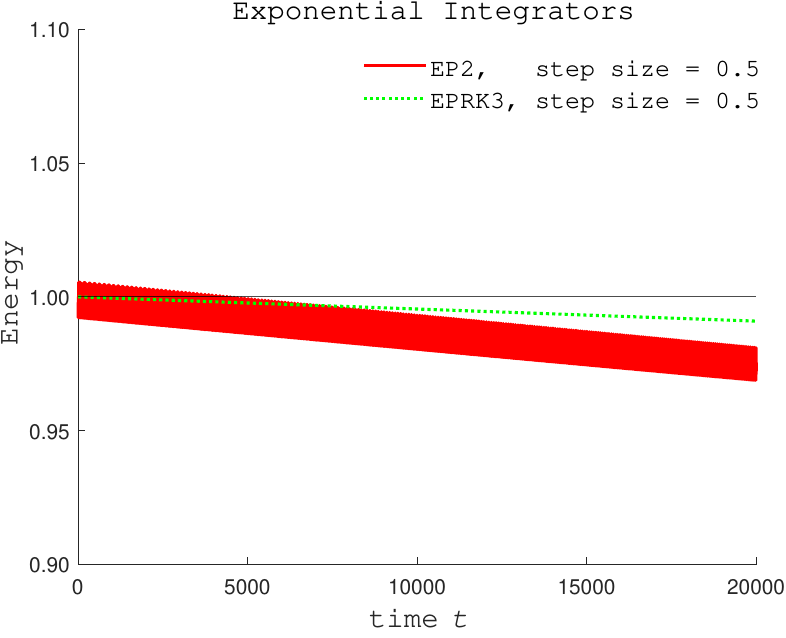} & \includegraphics[scale=0.3]{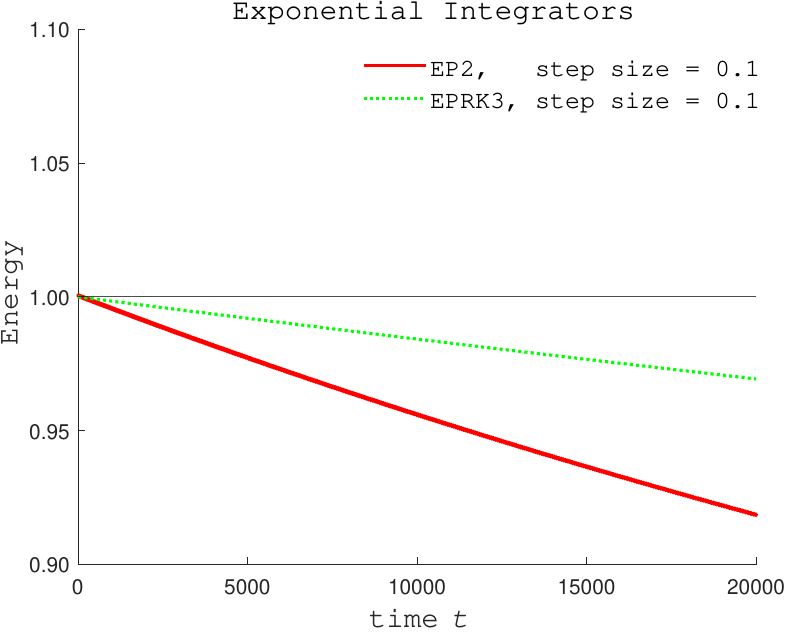} \\[1.5em]
\multicolumn{3}{c}{Potential Hills, $\frac{|V''|}{B} = \frac{1}{100}$} \\[0.5em]
Quadratic & Cubic & Quartic \\
\includegraphics[scale=0.3]{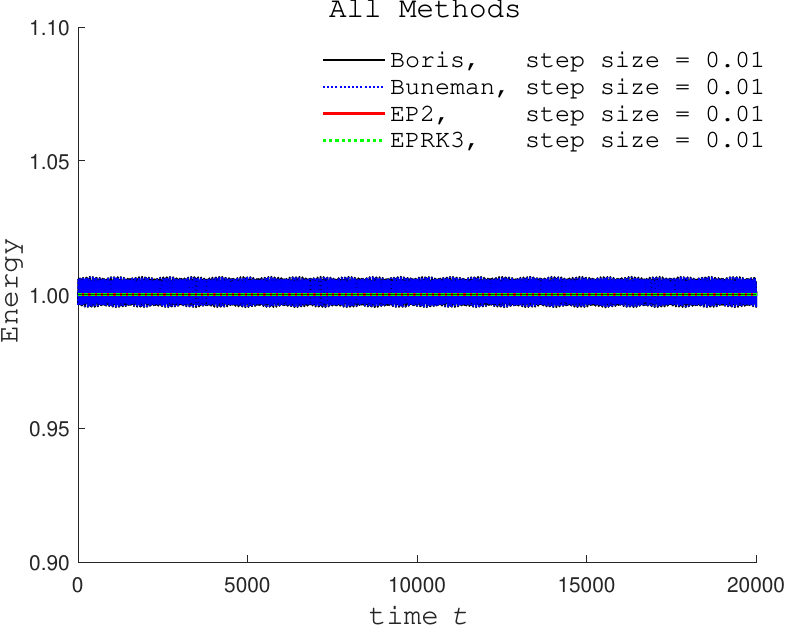} & \includegraphics[scale=0.3]{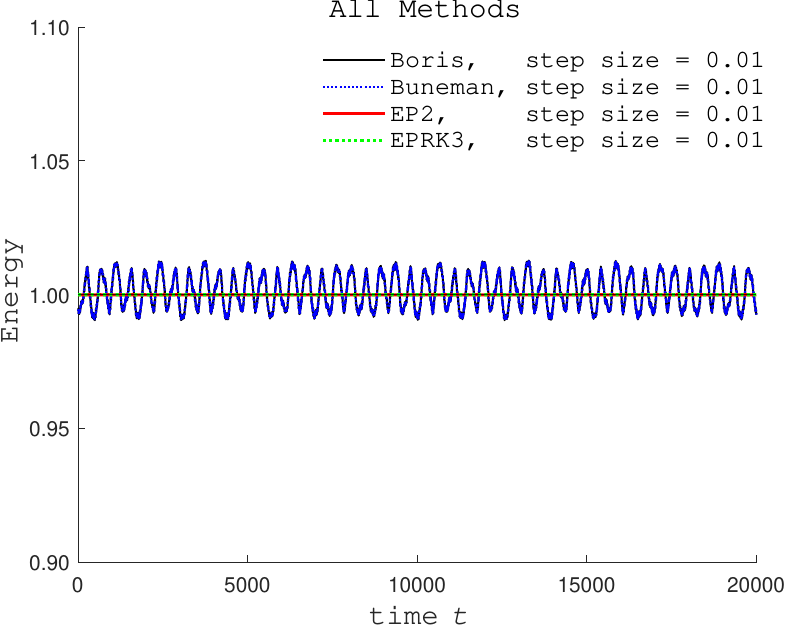} & \includegraphics[scale=0.3]{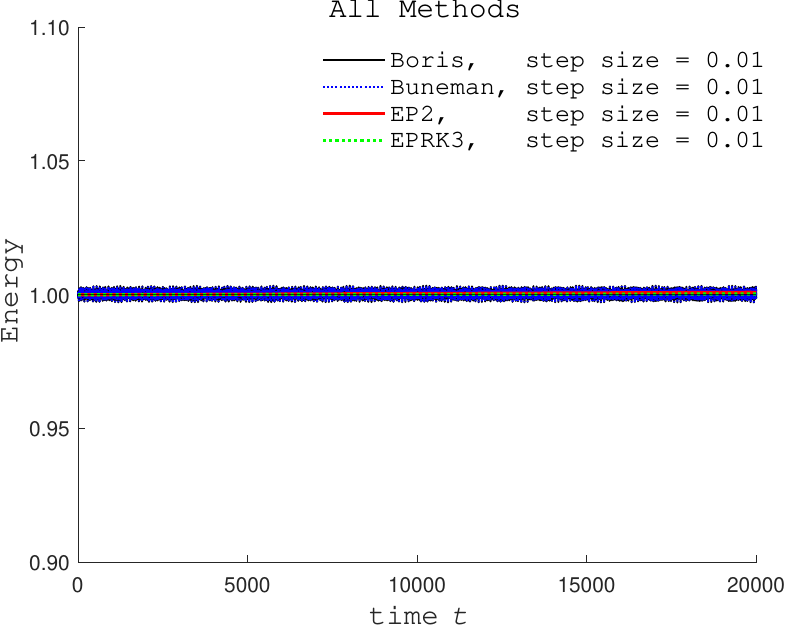} \\
\includegraphics[scale=0.3]{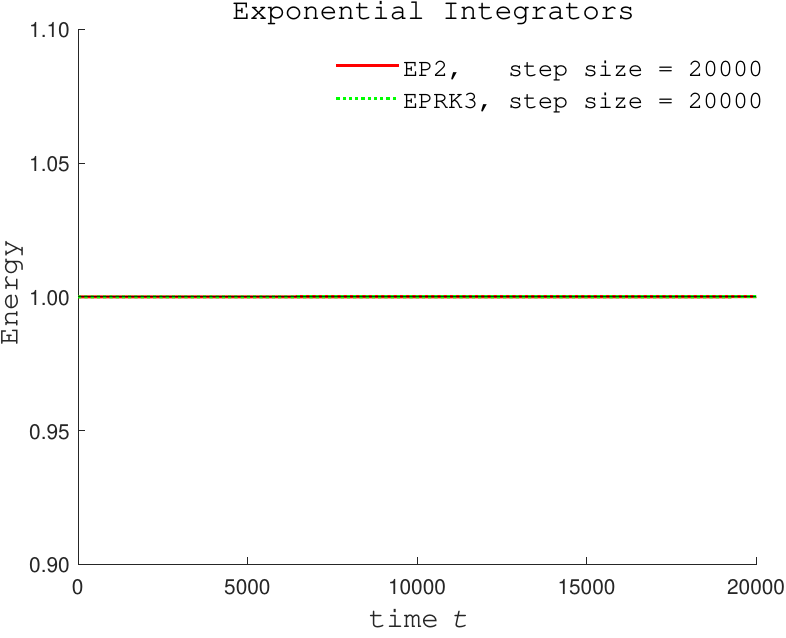} & \includegraphics[scale=0.3]{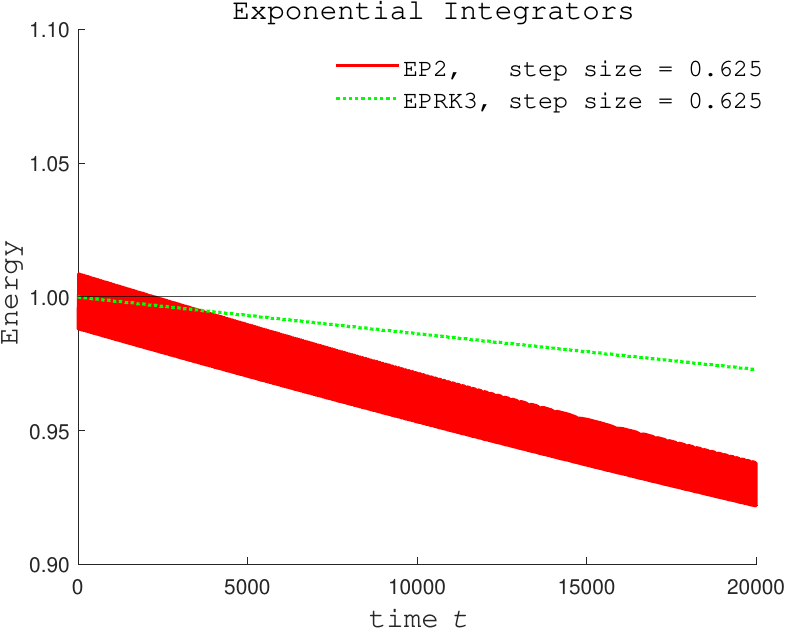} & \includegraphics[scale=0.3]{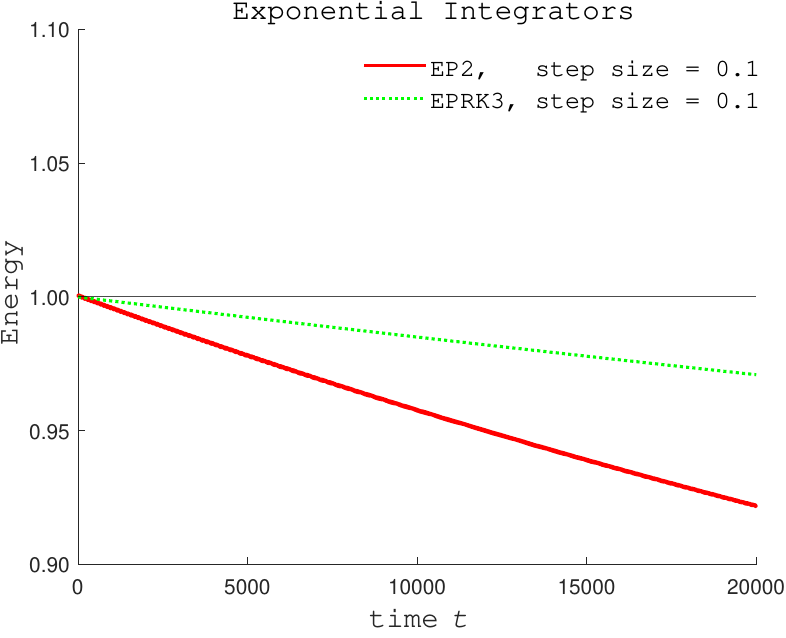}
\end{tabular}
\caption{Energy of 3D test problems with $|V''|/B = 1/100$}\label{3DEnergy1}
\end{figure}

\begin{figure}[h!]
\centering
\begin{tabular}{ccc}
\multicolumn{3}{c}{Potential Wells, $\frac{|V''|}{B} = \frac{1}{10}$} \\[0.5em]
Quadratic & Cubic & Quartic \\
\includegraphics[scale=0.3]{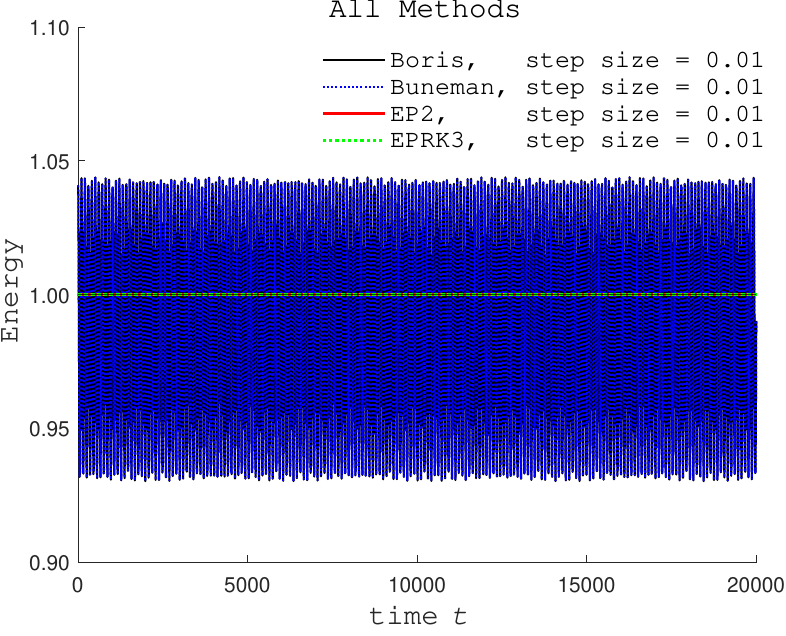} & \includegraphics[scale=0.3]{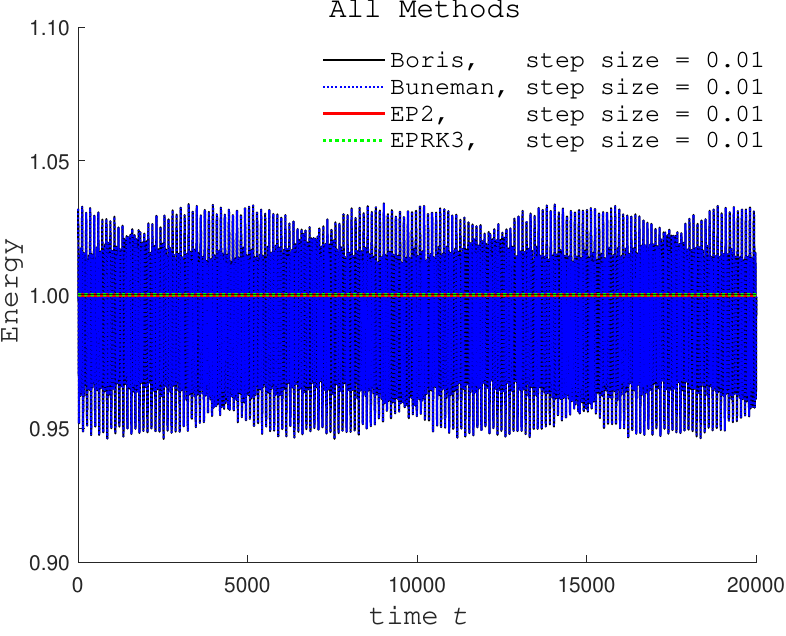} & \includegraphics[scale=0.3]{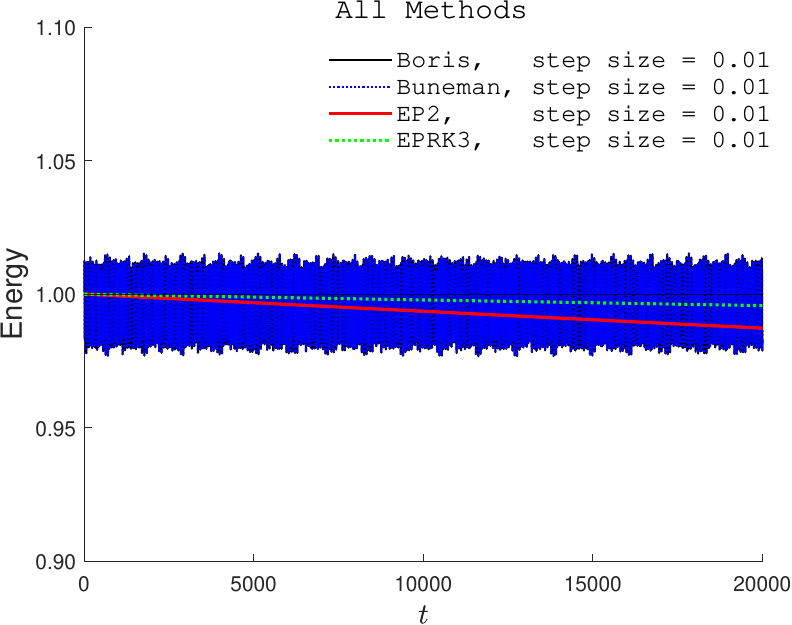} \\
\includegraphics[scale=0.3]{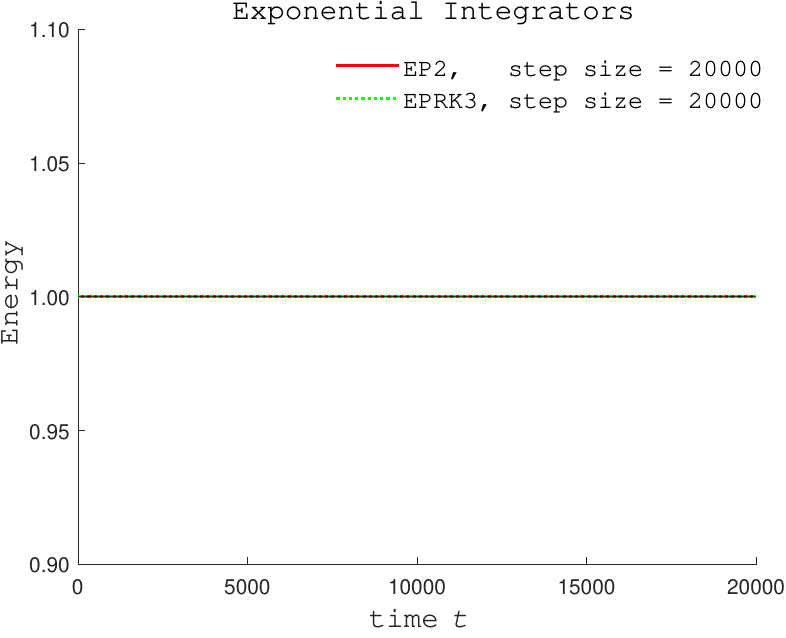} & \includegraphics[scale=0.3]{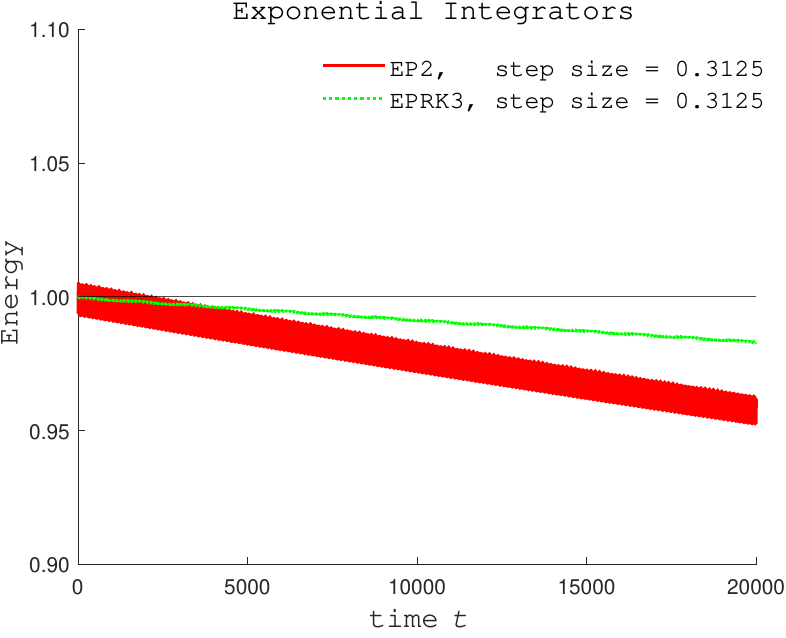} & \includegraphics[scale=0.3]{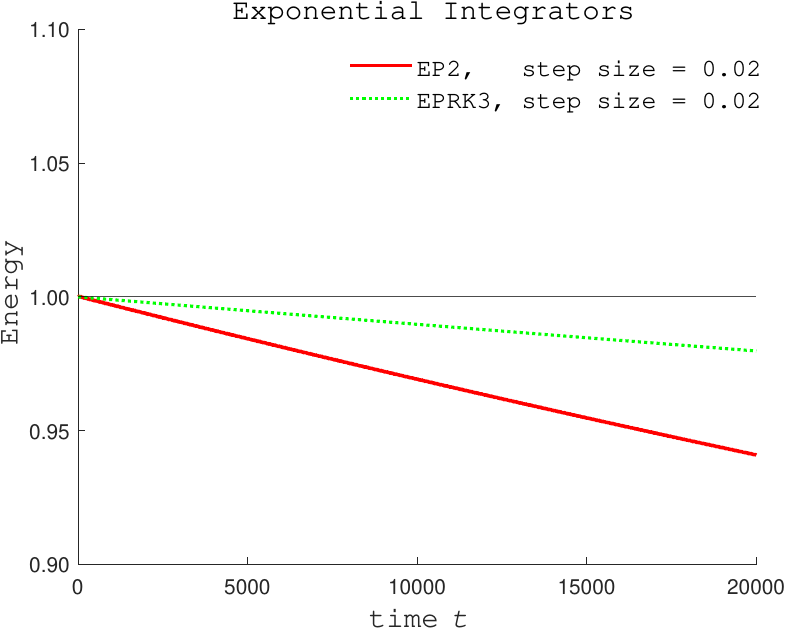} \\[1.5em]
\multicolumn{3}{c}{Potential Hills, $\frac{|V''|}{B} = \frac{1}{10}$} \\[0.5em]
Quadratic & Cubic & Quartic \\
\includegraphics[scale=0.3]{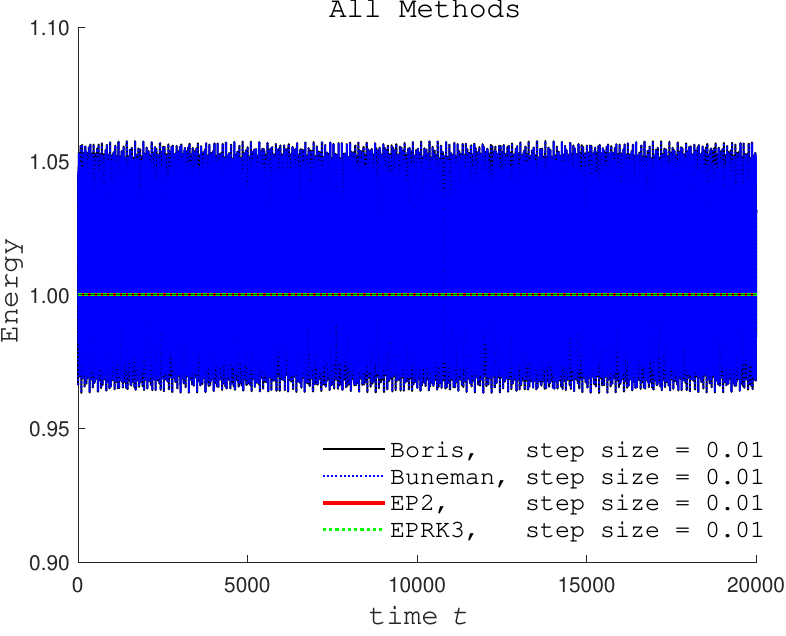} & \includegraphics[scale=0.3]{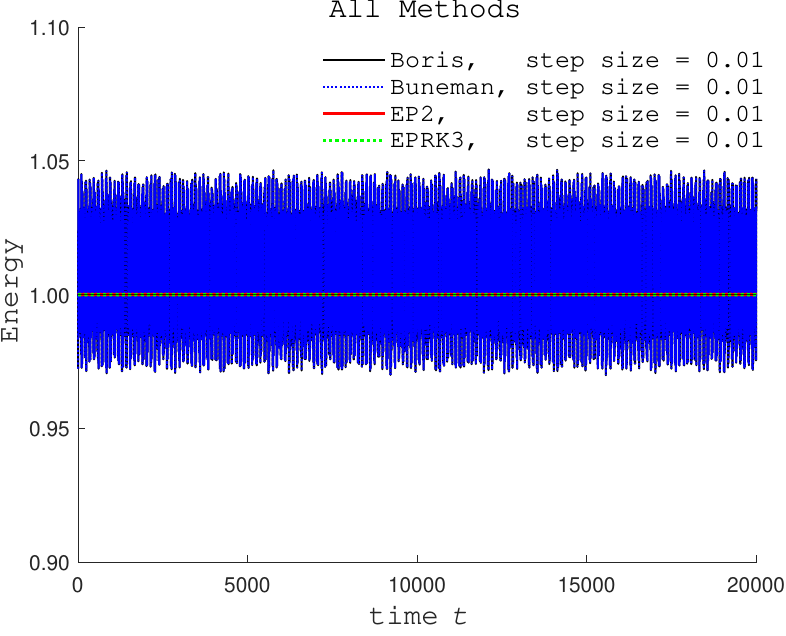} & \includegraphics[scale=0.3]{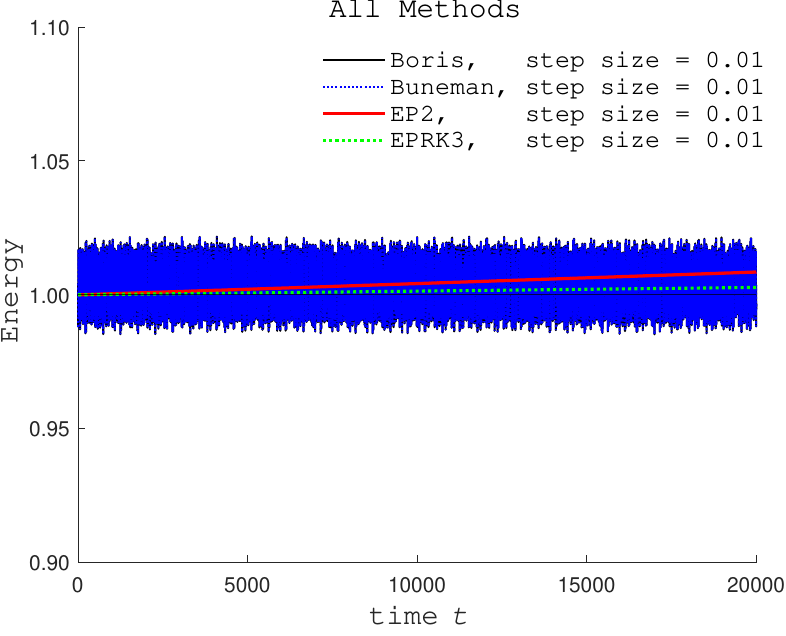} \\
\includegraphics[scale=0.3]{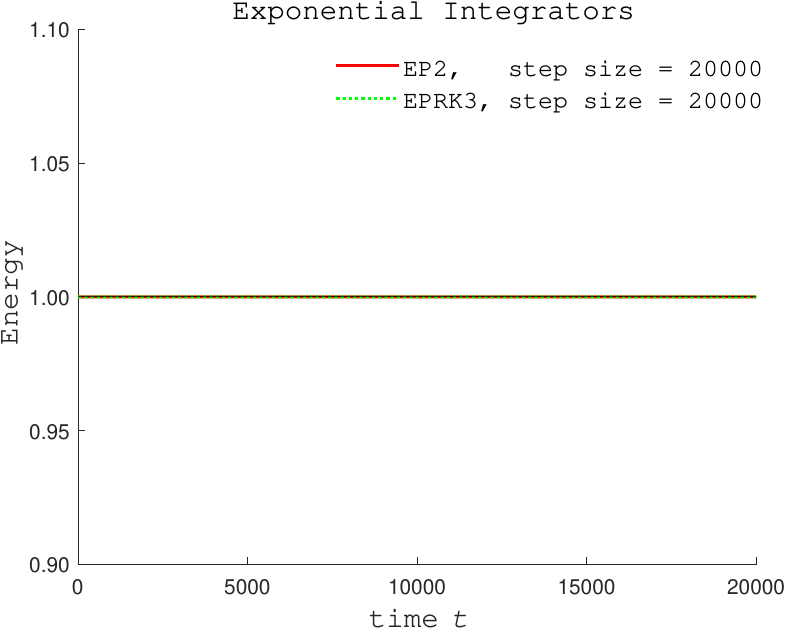} & \includegraphics[scale=0.3]{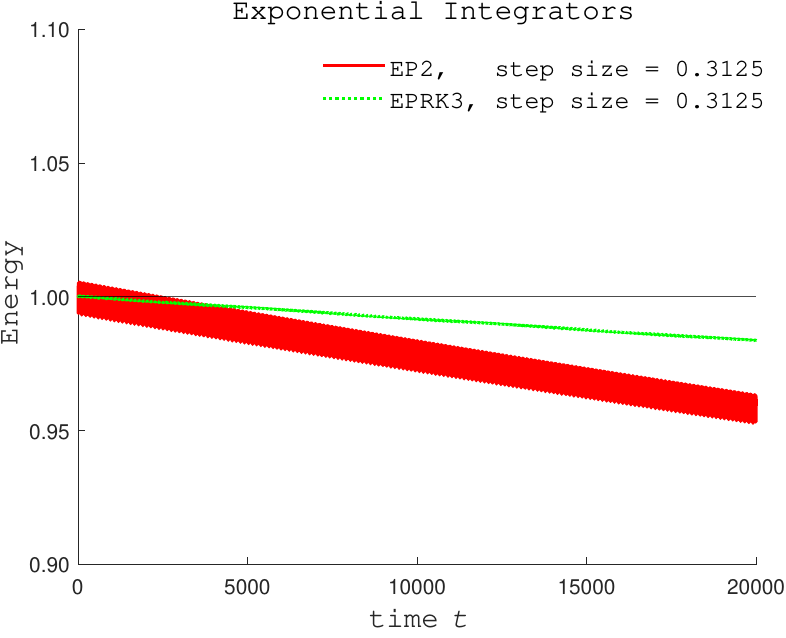} & \includegraphics[scale=0.3]{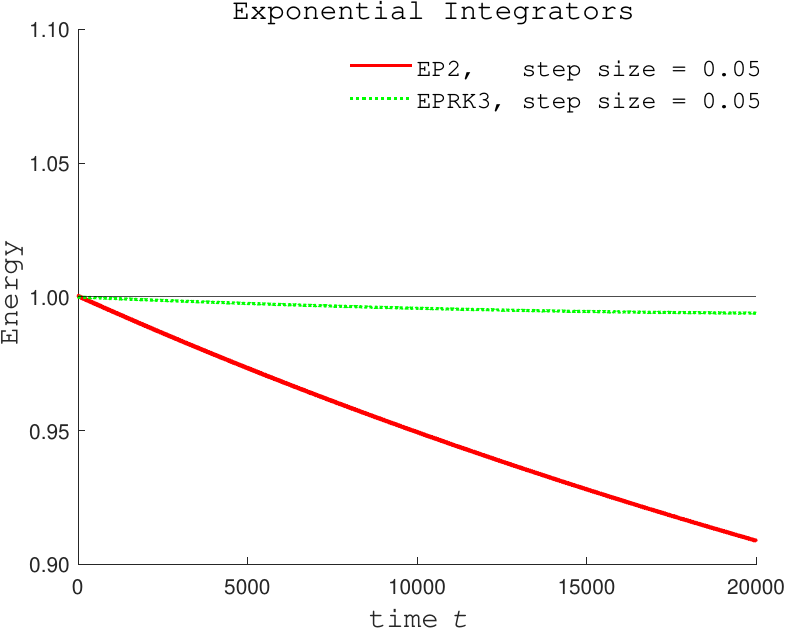}
\end{tabular}
\caption{Energy of 3D test problems with $|V''|/B = 1/10$}\label{3DEnergy2}
\end{figure}

\begin{figure}[h!]
\centering
\begin{tabular}{ccc}
\multicolumn{3}{c}{Potential Wells, $\frac{|V''|}{B} = 1$} \\[0.5em]
Quadratic & Cubic & Quartic \\
\includegraphics[scale=0.3]{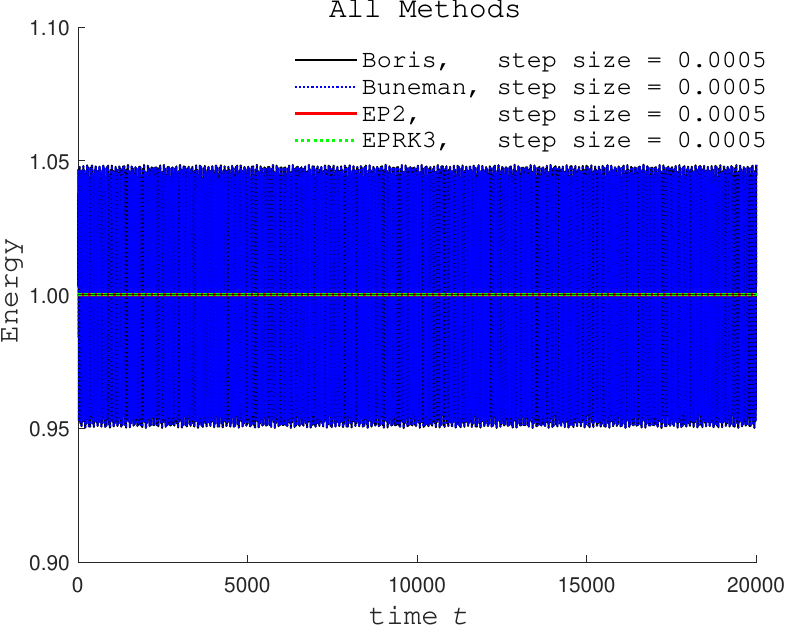} & \includegraphics[scale=0.3]{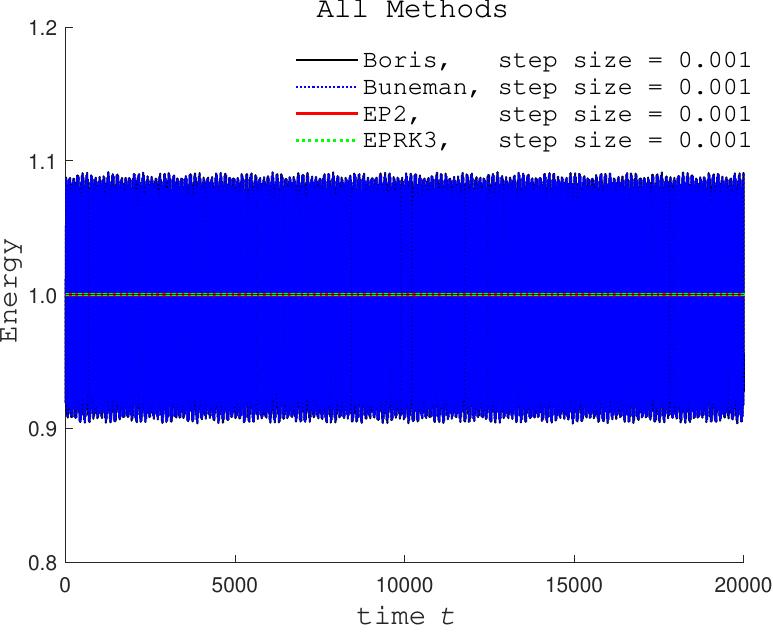} & \includegraphics[scale=0.3]{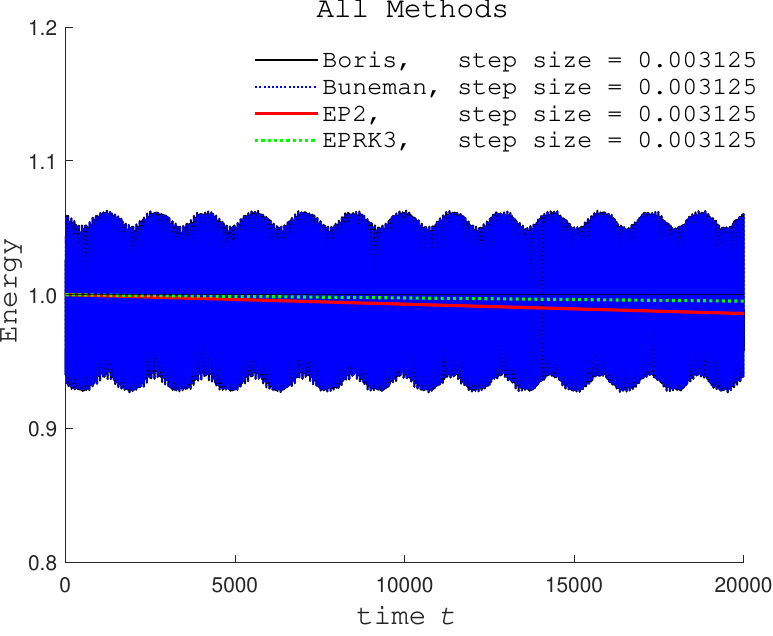} \\
\includegraphics[scale=0.3]{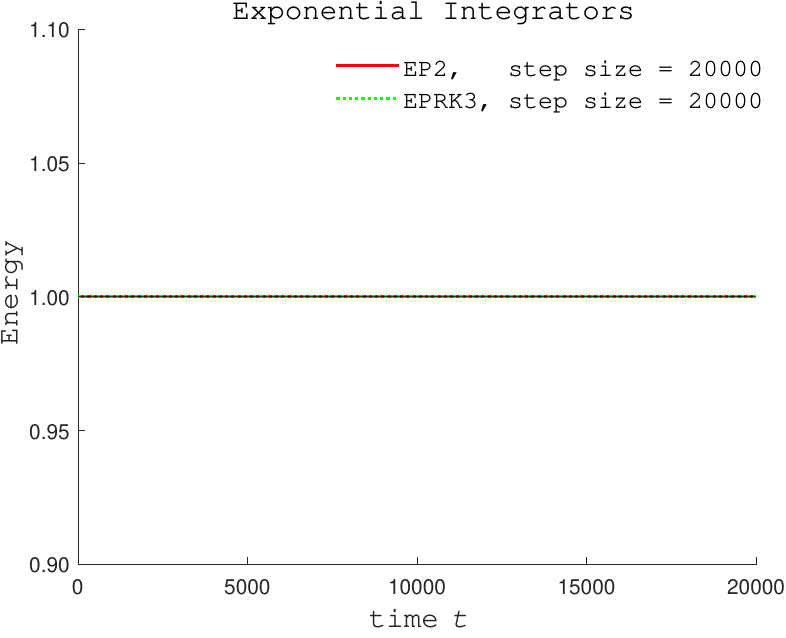} & \includegraphics[scale=0.3]{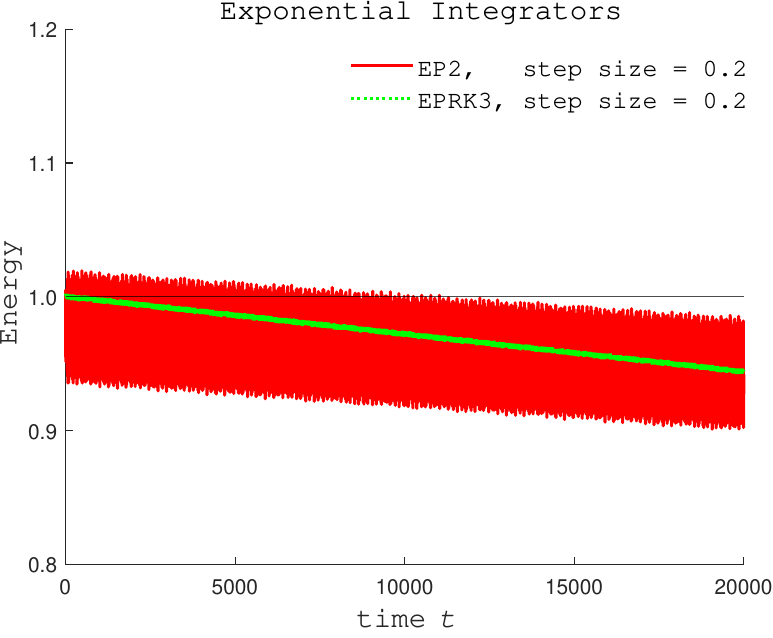} & \includegraphics[scale=0.3]{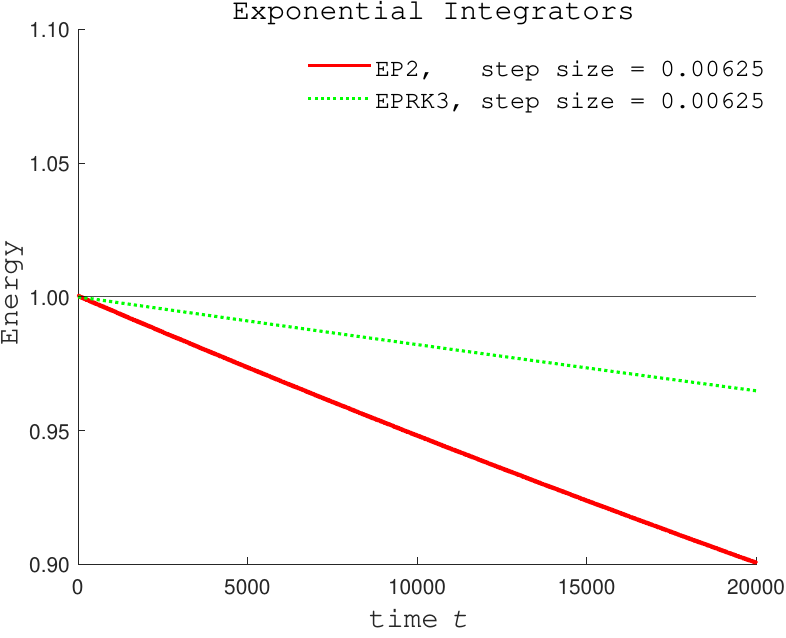} \\[1.5em]
\multicolumn{3}{c}{Potential Hills, $\frac{|V''|}{B} = 1$} \\[0.5em]
Quadratic & Cubic & Quartic \\
\includegraphics[scale=0.3]{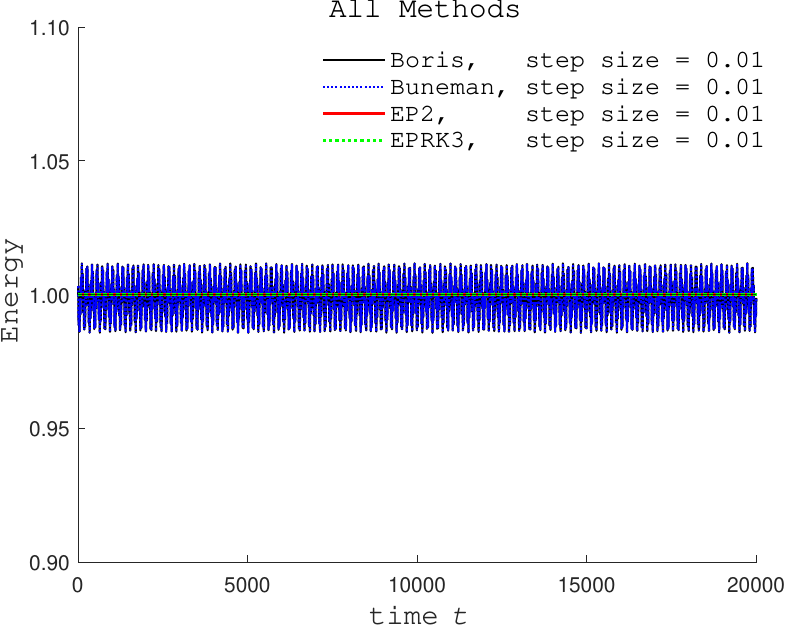} & \includegraphics[scale=0.3]{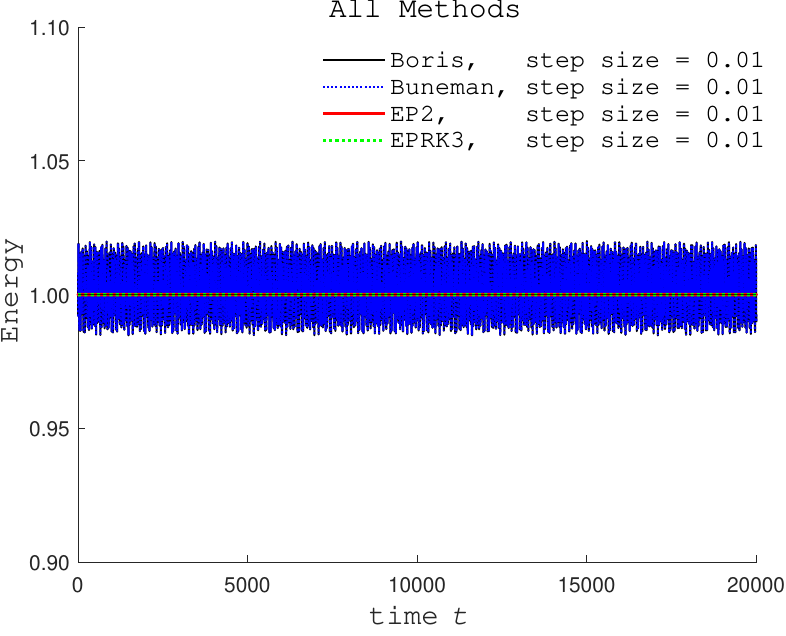} & \includegraphics[scale=0.3]{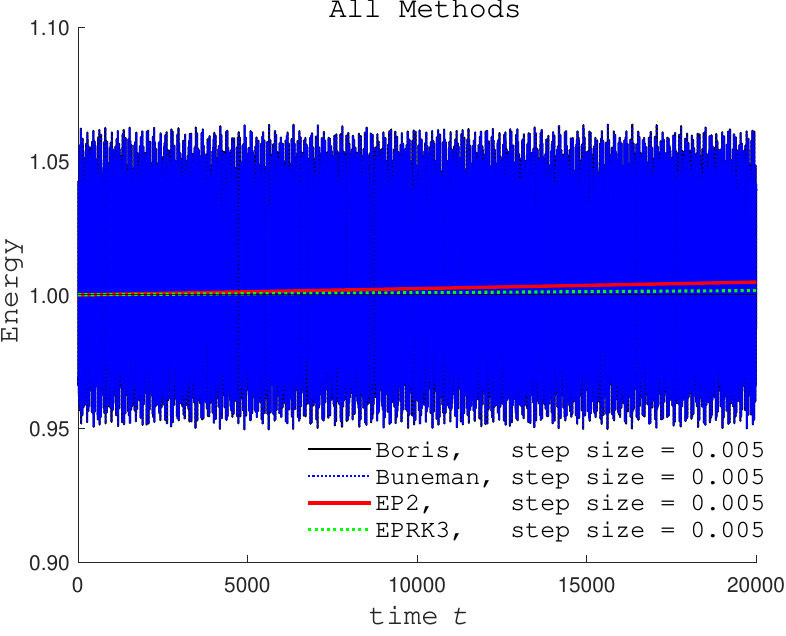} \\
\includegraphics[scale=0.3]{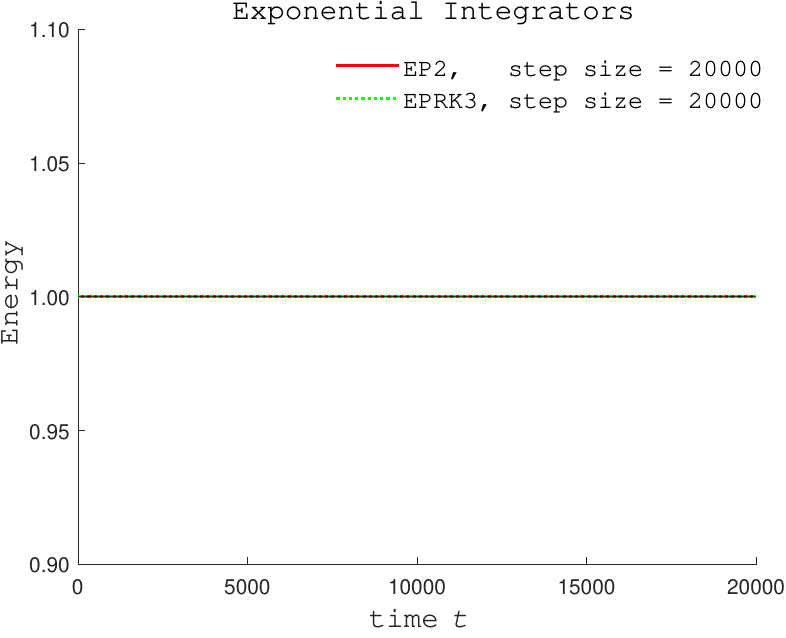} & \includegraphics[scale=0.3]{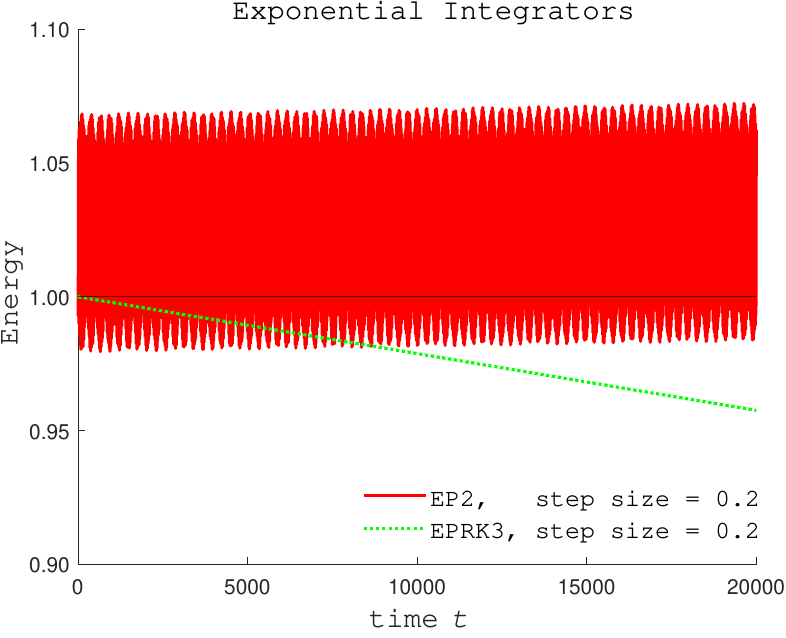} & \includegraphics[scale=0.3]{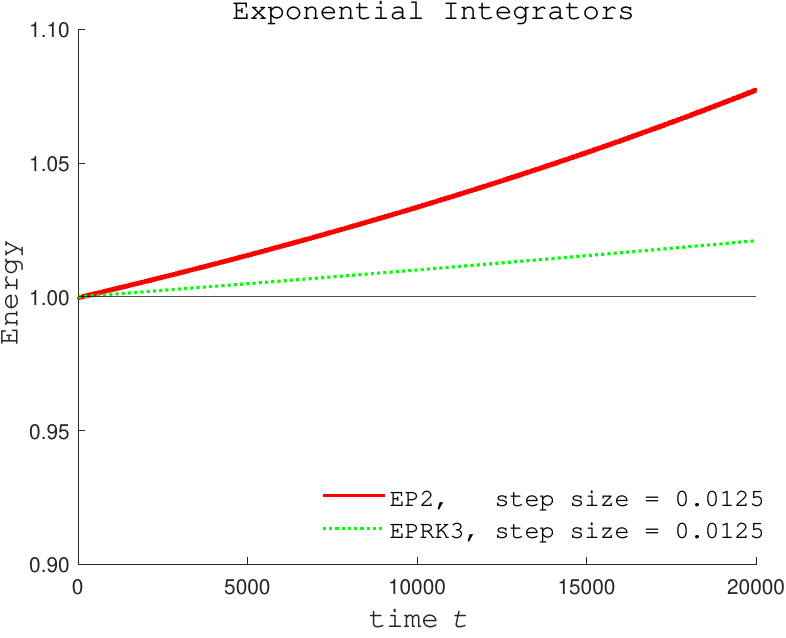}
\end{tabular}
\caption{Energy of 3D test problems with $|V''|/B = 1$}\label{3DEnergy3}
\end{figure}

The energy plots for the test problems with $|V''|/B = 1/100$, $|V''|/B = 1/10$, and $|V''|/B = 1$ are shown in figures \ref{3DEnergy1}, \ref{3DEnergy2}, and \ref{3DEnergy3} are also similar to the two-dimensional case. As in the two-dimensional experiments, the accuracy of the energies of the system computed with the exponential integrators drifts over long time intervals and the magnitude of the drift depends on the time step size. Again, the EPRK3 integrator performs better than the EP2 integrator by exhibiting both less drift and less variation in the computed energies indicating that higher order methods indeed yield more accurate solutions.  Thus, comparative performance of all methods is consistent across two- and three-dimensional problems and the numerical results are aligned with theoretically expected performance.  

\newpage
\vfill
\begin{table}[h!]
\centering
\begin{tabular}{ll|c|c|c|}
\\[0.5em]
\cline{3-5}
& & & & \\[-0.75em]
& & $|V''|=1$ & $|V''|=10$ & $|V''|=100$ \\[0.5em]
\hline
\multicolumn{1}{|c|}{} & \multicolumn{1}{|c|}{} & & & \multicolumn{1}{c|}{} \\[-0.5em]
\multicolumn{1}{|c|}{\multirow{6}{4.5em}{Quadratic Well}} & \multicolumn{1}{|c|}{\multirow{2}{*}{$V$}} & $\frac{1}{2}(x^2 + y^2)$ & $5(x^2 + y^2)$ & \multicolumn{1}{c|}{$50(x^2 + y^2)$} \\[0.25em]
 \multicolumn{1}{|c|}{} & \multicolumn{1}{|c|}{} & $+ \frac{1}{20}z^2 - \frac{1}{2}$ & $+ \frac{1}{2}z^2 - 5$ & $+ 5z^2 - 50$ \\[0.75em]
\cline{2-5}
\multicolumn{1}{|c|}{} & \multicolumn{1}{|c|}{} & & & \multicolumn{1}{|c|}{} \\[-0.5em]
\multicolumn{1}{|c|}{} & \multicolumn{1}{|c|}{$\bm{E}$} & $-\begin{bmatrix}
\phantom{\frac{1}{10}}x \\ \phantom{\frac{1}{10}}y \\ \frac{1}{10}z
\end{bmatrix}$ & $-\begin{bmatrix}
10x \\ 10y \\ \phantom{10}z
\end{bmatrix}$ & \multicolumn{1}{c|}{$-\begin{bmatrix}
100x \\ 100y \\ \phantom{0}10z
\end{bmatrix}$} \\[2em]
\hline
\multicolumn{1}{|c|}{} & \multicolumn{1}{|c|}{} & & & \multicolumn{1}{c|}{} \\[-0.5em]
\multicolumn{1}{|c|}{\multirow{8}{4.25em}{Cubic Well}} & \multicolumn{1}{|c|}{\multirow{3.5}{*}{$V$}} & $-\frac{5}{6} + x^2 + y^2$ & $-\frac{11}{3} + 3(x^2 + y^2)$ & \multicolumn{1}{c|}{$47(x^2 + y^2)$} \\[0.25em]
\multicolumn{1}{|c|}{} & \multicolumn{1}{|c|}{} & $- \frac{1}{6}(x^3 + y^3)$ & $+ \frac{2}{3}(x^3 + y^3)$ & $-48 + x^3 + y^3$ \\[0.25em]
\multicolumn{1}{|c|}{} & \multicolumn{1}{|c|}{} & $+ \frac{1}{10}z^2 - \frac{1}{60}z^3$ & $+ \frac{3}{10}z^2 + \frac{1}{15}z^3$ & $+ \frac{1}{10}(47z^2 + z^3)$ \\[0.75em]
\cline{2-5}
\multicolumn{1}{|c|}{} & \multicolumn{1}{|c|}{} & & & \multicolumn{1}{c|}{} \\[-0.5em]
\multicolumn{1}{|c|}{} & \multicolumn{1}{|c|}{$\bm{E}$} & $-\begin{bmatrix}
2x - \frac{1}{2}x^2 \\[0.25em] 2y - \frac{1}{2}y^2 \\[0.25em] \frac{1}{5}z - \frac{1}{20}z^2
\end{bmatrix}$ & $-\begin{bmatrix}
6x - 2x^2 \\[0.25em] 6y - 2y^2 \\[0.25em] \frac{3}{5}z + \frac{1}{5}z^2
\end{bmatrix}$ & \multicolumn{1}{c|}{$-\begin{bmatrix}
94x + 3x^2 \\[0.25em] 94y + 3y^2 \\[0.25em] \frac{47}{5}z + \frac{3}{10}z^2
\end{bmatrix}$} \\[2.25em]
\hline
\multicolumn{1}{|c|}{} & \multicolumn{1}{|c|}{} & & & \multicolumn{1}{c|}{} \\[-0.5em]
\multicolumn{1}{|c|}{\multirow{6}{4em}{Quartic Well}} & \multicolumn{1}{|c|}{\multirow{2}{*}{$V$}} & $\frac{1}{12}(x^4 + y^4)$ & $\frac{5}{6}(x^4 + y^4)$ & \multicolumn{1}{c|}{$\frac{25}{3}(x^4 + y^4)$} \\[0.25em]
\multicolumn{1}{|c|}{} & \multicolumn{1}{|c|}{} & $+ \frac{1}{120}z^4 - \frac{1}{12}$ & $+ \frac{1}{12}z^4 - \frac{5}{6}$ & $+ \frac{5}{6}z^4 - \frac{2}{3}$ \\[0.75em]
\cline{2-5}
\multicolumn{1}{|c|}{} & \multicolumn{1}{|c|}{} & & & \multicolumn{1}{|c|}{} \\[-0.5em]
\multicolumn{1}{|c|}{} & \multicolumn{1}{|c|}{$\bm{E}$} & $-\frac{1}{3}\begin{bmatrix}
\phantom{\frac{1}{10}}x^3 \\[0.25em] \phantom{\frac{1}{10}}y^3 \\[0.25em] \frac{1}{10}z^3
\end{bmatrix}$ & $-\frac{1}{3}\begin{bmatrix}
10x^3 \\[0.25em] 10y^3 \\[0.25em] \phantom{10}z^3
\end{bmatrix}$ & \multicolumn{1}{c|}{$-\frac{1}{3}\begin{bmatrix}
100x^3 \\[0.25em] 100y^3 \\[0.25em] \phantom{0}10z^3
\end{bmatrix}$} \\[2.25em]
\hline
\end{tabular}
\caption{Electric scalar potential wells and corresponding electric fields for 3D model test problems}\label{3DWellConfig}
\end{table}
\vfill

\newpage
\vfill
\begin{table}[h!]
\centering
\begin{tabular}{ll|c|c|c|}
\\[0.5em]
\cline{3-5}
& & & & \\[-0.75em]
& & $|V''|=1$ & $|V''|=10$ & $|V''|=100$ \\[0.5em]
\hline
\multicolumn{1}{|c|}{} & \multicolumn{1}{|c|}{} & & & \multicolumn{1}{c|}{} \\[-0.5em]
\multicolumn{1}{|c|}{\multirow{6}{4.5em}{Quadratic Hill}} & \multicolumn{1}{|c|}{\multirow{2}{*}{$V$}} & $-\frac{1}{2}(x^2 + y^2)$ & $-5(x^2 + y^2)$ & \multicolumn{1}{c|}{$-50(x^2 + y^2)$} \\[0.25em]
 \multicolumn{1}{|c|}{} & \multicolumn{1}{|c|}{} & $+ \frac{1}{20}z^2 + \frac{1}{2}$ & $+ \frac{1}{2}z^2 + 5$ & $+ 5z^2 + 50$ \\[0.75em]
\cline{2-5}
\multicolumn{1}{|c|}{} & \multicolumn{1}{|c|}{} & & & \multicolumn{1}{|c|}{} \\[-0.5em]
\multicolumn{1}{|c|}{} & \multicolumn{1}{|c|}{$\bm{E}$} & $\begin{bmatrix}
\phantom{\frac{1}{10}}x \\ \phantom{\frac{1}{10}}y \\ \frac{1}{10}z
\end{bmatrix}$ & $\begin{bmatrix}
10x \\ 10y \\ \phantom{10}z
\end{bmatrix}$ & \multicolumn{1}{c|}{$\begin{bmatrix}
100x \\ 100y \\ \phantom{0}10z
\end{bmatrix}$} \\[2em]
\hline
\multicolumn{1}{|c|}{} & \multicolumn{1}{|c|}{} & & & \multicolumn{1}{c|}{} \\[-0.5em]
\multicolumn{1}{|c|}{\multirow{8}{4.25em}{Cubic Hill}} & \multicolumn{1}{|c|}{\multirow{3.5}{*}{$V$}} & $\frac{5}{6} + x^2 + y^2$ & $\frac{11}{3} - 3(x^2 + y^2)$ & \multicolumn{1}{c|}{$-47(x^2 + y^2)$} \\[0.25em]
\multicolumn{1}{|c|}{} & \multicolumn{1}{|c|}{} & $+ \frac{1}{6}(x^3 + y^3)$ & $- \frac{2}{3}(x^3 + y^3)$ & $+ 48 - x^3 - y^3$ \\[0.25em]
\multicolumn{1}{|c|}{} & \multicolumn{1}{|c|}{} & $+ \frac{1}{10}z^2 - \frac{1}{60}z^3$ & $+ \frac{3}{10}z^2 + \frac{1}{15}z^3$ & $+ \frac{1}{10}(47z^2 + z^3)$ \\[0.75em]
\cline{2-5}
\multicolumn{1}{|c|}{} & \multicolumn{1}{|c|}{} & & & \multicolumn{1}{c|}{} \\[-0.5em]
\multicolumn{1}{|c|}{} & \multicolumn{1}{|c|}{$\bm{E}$} & $\begin{bmatrix}
\phantom{-}2x - \frac{1}{2}x^2 \\[0.25em] \phantom{-}2y - \frac{1}{2}y^2 \\[0.25em] -\frac{1}{5}z + \frac{1}{20}z^2
\end{bmatrix}$ & $\begin{bmatrix}
\phantom{-}6x - 2x^2 \\[0.25em] \phantom{-}6y - 2y^2 \\[0.25em] -\frac{3}{5}z - \frac{1}{5}z^2
\end{bmatrix}$ & \multicolumn{1}{c|}{$\begin{bmatrix}
\phantom{-}94x + 3x^2 \\[0.25em] \phantom{-}94y + 3y^2 \\[0.25em] -\frac{47}{5}z - \frac{3}{10}z^2
\end{bmatrix}$} \\[2.25em]
\hline
\multicolumn{1}{|c|}{} & \multicolumn{1}{|c|}{} & & & \multicolumn{1}{c|}{} \\[-0.5em]
\multicolumn{1}{|c|}{\multirow{6}{4em}{Quartic Hill}} & \multicolumn{1}{|c|}{\multirow{2}{*}{$V$}} & $-\frac{1}{12}(x^4 + y^4)$ & $-\frac{5}{6}(x^4 + y^4)$ & \multicolumn{1}{c|}{$-\frac{25}{3}(x^4 + y^4)$} \\[0.25em]
\multicolumn{1}{|c|}{} & \multicolumn{1}{|c|}{} & $+ \frac{1}{120}z^4 + \frac{1}{12}$ & $+ \frac{1}{12}(z^4 + 1)$ & $+ \frac{5}{6}z^4 + \frac{25}{3}$ \\[0.75em]
\cline{2-5}
\multicolumn{1}{|c|}{} & \multicolumn{1}{|c|}{} & & & \multicolumn{1}{|c|}{} \\[-0.5em]
\multicolumn{1}{|c|}{} & \multicolumn{1}{|c|}{$\bm{E}$} & $\frac{1}{3}\begin{bmatrix}
\phantom{-\frac{1}{10}}x^3 \\[0.25em] \phantom{-\frac{1}{10}}y^3 \\[0.25em] -\frac{1}{10}z^3
\end{bmatrix}$ & $\frac{1}{3}\begin{bmatrix}
10x^3 \\[0.25em] 10y^3 \\[0.25em] -z^3
\end{bmatrix}$ & \multicolumn{1}{c|}{$\frac{1}{3}\begin{bmatrix}
\phantom{1}100x^3 \\[0.25em] \phantom{1}100y^3 \\[0.25em] -10z^3
\end{bmatrix}$} \\[2.25em]
\hline
\end{tabular}
\caption{Electric scalar potential hills and corresponding electric fields for 3D model test problems}\label{3DHillConfig}
\end{table}
\vfill

\newpage

\newpage

\newpage

\newpage

\newpage

\newpage

\newpage
\section{Conclusion and Future Work}
In this paper we proposed an alternative approach to the numerical simulation of charged particle dynamics using exponential integrators. An integral part of this algorithm is taking advantage of the low dimensionality of the particle pushing problem and using an analytic method to compute matrix $\varphi$ functions needed at each step of an exponential scheme. We showed that exponential integrators can be competitive compared to traditional particle pushers when the problem is strongly magnetized. As expected, exponential integrators offer dramatic computational advantages for cases where electric fields are generated by quadratic electric potentials. Since the problem is linear in this case, an exponential integrator with accurate evaluation of matrix $\varphi$ functions computes a very accurate solution with the error coming primarily from the finite-precision computation of $\varphi$ that involves the eigenvalue solver and the interpolation polynomial of the function. Compared to the traditional Boris and Buneman algorithms, for these linear problems, we showed that exponential integrators could bring approximately six orders of magnitude gains in computational speed and three orders of magnitude improvements in accuracy simultaneously. For nonlinear problems with the cubic electric potentials we still saw significant computational savings, though not as dramatic as for quadratic problems. To obtain the solution at the same accuracy level, exponential integrators exhibited savings in computation time of about two orders of magnitude for two-dimensional problems and at least an order of magnitude for three-dimensional problems compared to traditional methods. The quartic potentials yielded comparable performance between exponential integrators and Boris and Buneman schemes. We also see that higher order exponential methods can improve the computational performance. These points indicate that for highly nonlinear problems like those with a quartic electric potential, it is important to pay particular attention to approximation of  the nonlinear integral in the exact solution \eqref{intform2} when constructing an exponential integrator. This is further evidenced by the grad $B$ drift experiments. Exploring different approximations of the nonlinear integral to develop better performing exponential methods for highly nonlinear problems will be one of the research directions we plan to pursue in the future.  

Of course, the exponential methods we used have not been designed to be energy preserving and, indeed, a drift in energy is observed in the numerical results. However, we showed improvements in the accuracy of the computed energy as the order of an integrator is increased.  This result warrants further research into development of exponential methods of higher order that would potentially exhibit better energy preservation.

Additionally, we showed that for a linear $\bm{E}\times\bm{B}$ drift problem the exponential integrators accurately compute the gyroradius regardless of the step size value as expected. By contrast, both the Boris and Buneman pushers artificially enlarge the gyroradius for a large step size relative to the gyroradius.

Investigating the performance of the exponential particle pushers as they are embedded within an overall PIC integrator is another research direction we plan to pursue. For example, low order spatial discretizations of the electric field can result in the potentials dominated by the quadratic terms. It would be interesting to study whether the computational advantages of exponential integrators for such quadratically dominated potentials would persist even if we account for the redefining of potentials as the particles cross the cell boundaries.  

To summarize, we have shown preliminary results that offer some evidence of the numerical advantages of the new numerical approach we propose.  This work also highlighted possible directions for improving the exponential integration-based methods making them more suitable for highly nonlinear particle pushing problems and we plan to pursue these directions in our future research. 



\newpage
\appendix
\section{Conventional Particle Pushing}
The conventional framework to numerical particle pushing approximates the Newtonian equations of motion \eqref{Newtonform} with the finite-difference model
\begin{subequations}
\begin{align}
&\begin{array}{ccl}
\dfrac{\bm{x}_{n+1} - \bm{x}_n}{h} & = & \bm{v}_{n+1/2}, \label{xdot}\\[-2em]
\phantom{\dfrac{\bm{v}_{n+1/2} - \bm{v}_{n-1/2}}{h}}
\end{array} \\[0.75em]
&\begin{array}{ccl}
\dfrac{\bm{v}_{n+1/2} - \bm{v}_{n-1/2}}{h} & = & \dfrac{q}{m}\left(\bm{E}_n + \dfrac{\bm{v}_{n+1/2} + \bm{v}_{n-1/2}}{2} \times \bm{B}_n\right), \label{Lorentz}
\end{array}
\end{align}
\end{subequations}
where $h$ is a fixed time step size and the subscripts $n$, $n \rm{\pm} 1/2$, $n \rm{+} 1$ denote times $t_n$, $t_n \rm{\pm} h/2$, $t_n \rm{+} h$, respectively. Position and the electromagnetic fields are computed at integer time nodes while velocity is computed at half-integer time nodes. This staggering of position and velocity by one-half time step gives a leapfrog-like, centered-difference, time reversible scheme with second-order accuracy. Observe that the second equation \eqref{Lorentz} is implicit in $\bm{v}_{n+1/2}$ and, hence, numerically stable. However, the step size $h$ must be sufficiently small such that the electric field $\bm{E}$ and magnetic field $\bm{B}$ are approximately constant over the time interval $[t_n, t_n + h]$ to yield accurate solutions.

It follows from equation \eqref{xdot} that the finite-difference model approximates the second derivative of position with the centered difference formula:
\[
\frac{d^2\bm{x}}{dt^2} \approx \frac{\bm{x}_{n+1} - 2\bm{x}_n + \bm{x}_{n-1}}{h^2}.
\]
In order for this model to properly capture harmonic motion (such as uniform gyromotion), Birdsall and Langdon \cite{Birdsall} point out this numerical framework must obey the step size restriction
\[
h < \frac{2}{|\omega|}, \qquad\omega = \frac{qB}{m}.
\]

Dynamic propagation of the particle state in the finite-difference model is as follows. The update formula for the particle position is given by a simple rearrangement of \eqref{xdot}:
\[
\bm{x}_{n+1} = \bm{x}_n + h\cdot\bm{v}_{n+1/2}.
\]
For the Lorentz force equation \eqref{Lorentz}, note that the right-hand side is composed of an electric push term and a magnetic rotation term due to the electric field $\bm{E}_n$ and magnetic field $\bm{B}_n$, respectively. Also observe that the updated velocity $\bm{v}_{n+1/2}$ is given implicitly, which requires inversion of the equation to get an explicit expression for $\bm{v}_{n+1/2}$. Two common algorithms to resolve these tasks and update the particle velocity are the Buneman \cite{Buneman} and the Boris \cite{Boris} particle pushers.

\subsection{Buneman Particle Pusher}
The Buneman particle pushing algorithm decomposes the action of the electric field $\bm{E}$ on particle velocity into components parallel and perpendicular to the magnetic field $\bm{B}$. In the presence of an electric field $\bm{E}$ and a magnetic field $\bm{B}$, the particle experiences a so-called $\bm{E}\times\bm{B}$ drift velocity (perpendicular to both fields)
\[
\bm{v}_\text{drift} = \frac{\bm{E} \times \bm{B}}{B^2}, \quad B = \|\bm{B}\|.
\]
The Buneman algorithm subtracts this drift from the particle velocities at time nodes $t_{n-1/2}$ and $t_{n+1/2}$ thereby defining two intermediate velocities:
\begin{align*}
\bm{v}^- & = \bm{v}_{n-1/2} - \bm{v}_{\text{drift}}, \\
\bm{v}^+ & = \bm{v}_{n+1/2} - \bm{v}_{\text{drift}}.
\end{align*}
Substituting $\bm{v}_{n-1/2}$ and $\bm{v}_{n+1/2}$ into equation \eqref{Lorentz} then yields
\[
\frac{\bm{v}^+ - \bm{v}^-}{h} = \frac{q}{m}\left(\bm{E}_\parallel + \frac{\bm{v}^+ + \bm{v}^-}{2} \times \bm{B}\right).
\]
The above formula is composed of acceleration parallel to the magnetic field (the $\bm{E}_\parallel$ term) and a rotation of the velocity perpendicular to the magnetic field (the cross product term). For a uniform magnetic field with magnitude $B$, the angle of magnetic rotation over time step $h$ is
\[
\theta = h\omega, \qquad \omega = \frac{qB}{m}.
\]
The Buneman algorithm updates the velocity from $\bm{v}^-$ to $\bm{v}^+$ by the formula
\[
\bm{v}^+ = \cos\theta\,\bm{v}^- - \sin\theta\left(\frac{\bm{B}}{B}\times\bm{v}^-\right).
\]
As a historical note, the Bunemam algorithm was introduced in 1967 during which time the evaluation of transcendental functions was computationally expensive. To reduce computational cost, the algorithm makes use of the small angle approximation
\[
w = \frac{h}{2}\cdot\frac{qB}{m} \approx \tan\left(\frac{\theta}{2}\right).
\]
Then, using half-angle trigonometric identities, $\sin\theta$ and $\cos\theta$ are computed as follows:
\begin{align*}
s & = \frac{2\cdot w}{1 + w^2} = \sin\theta, \\[0.5em]
c & = \frac{1 - w^2}{1 + w^2} = \cos\theta.
\end{align*}

\begin{algorithm}
\caption{Buneman Velocity Push}\label{BunemPush}
\KwInput{$h$, $q$, $m$, $\bm{B}_n$, $B = \|\bm{B}_n\|$, $\bm{E}_n$, $\bm{v}_{n-1/2}$} \vspace*{0.5em}

\KwOutput{$\bm{v}_{n+1/2}$}
\vspace*{1em}
\begin{algorithmic}[1]
\State $\bm{v}_{\rm{drift}} \gets \dfrac{\bm{E}_n\times\bm{B}_n}{B^2}$ \vspace*{1em}
\State $\bm{v}^- \gets \bm{v}_{n-1/2} - \bm{v}_{\rm{drift}}$ \vspace*{1em}
\State $w \gets \dfrac{h}{2}\cdot\dfrac{qB}{m}$ \vspace*{1em}
\State $s \gets \dfrac{2\cdot w}{1 + w^2}$ \vspace*{1em}
\State $c \gets \dfrac{1 - w^2}{1 + w^2}$ \vspace*{1em}
\State $\bm{v}^+ \gets c\cdot\bm{v}^- -  s\cdot\left(\dfrac{\bm{B}_n}{B}\times\bm{v}^-\right)$ \vspace*{1em}
\State $\bm{v}_{n+1/2} \gets \bm{v}^+ + \bm{v}_{\rm{drift}}$ \end{algorithmic}
\end{algorithm}

\newpage
\subsection{Boris Particle Pusher}
\begin{figure}[h!]
\begin{center}
\begin{tikzpicture}
\draw[->] (0,0)--(0.25,2) node[above left] {$\bm{v}_{n-1/2}$};
\draw[->] (0,0)--(2,2) node[above right] {$\bm{v}^-$};
\draw[blue,->] (0.35,2)--(1.9,2);
\draw[blue,very thin,-] (1,2.1)--(1,3) node[above] {i. First-half electric push};
\draw[blue,thin,<-] (0.75,-0.75) arc (-45:45:1.05) node[left] at (1,0) {$\theta$};
\draw[blue,very thin,-] (1.2,0)--(3,0) node[right] {ii. Magnetic rotation};
\draw[->] (0,0)--(2,-2) node[below left] {$\bm{v}^+$};
\draw[blue,->] (2.05,-2)--(3.6,-2);
\draw[blue,very thin,-] (2.825,-2.1)--(2.825,-3) node[below] {iii. Second-half electric push};
\draw[->] (0,0)--(3.75,-2) node[below right] {$\bm{v}_{n+1/2}$};
\draw[gray,very thin,dashed,->] (0.35,1.9)--(3.6,-1.8);
\draw[->] (-2.1,1.5)--(-1.5,1.5) node[below left]{$\bm{E}$};
\draw (-2,-1.5) circle (0.1);
\fill (-2,-1.5) circle (1pt) node[right] {$\bm{B}$};
\end{tikzpicture}
\end{center}
\caption{Boris velocity update}\label{Borisvupdate}
\end{figure}
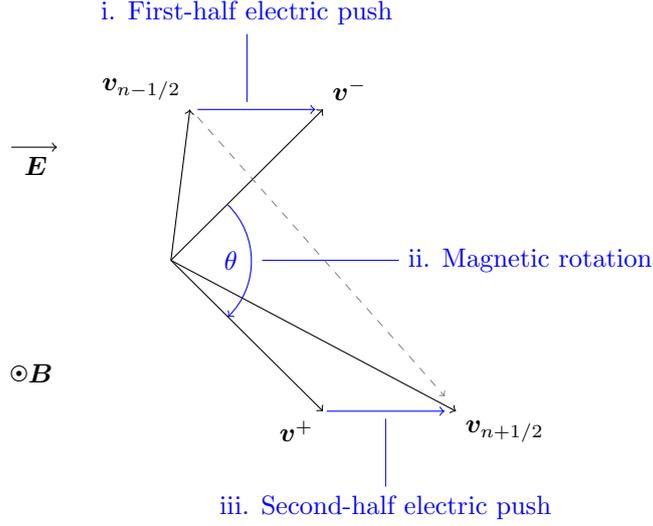

The Boris algorithm takes an alternative approach to the velocity update by decoupling the electric push and magnetic rotation in \eqref{Lorentz}. The discussion presented here is taken from \cite{Birdsall}. The algorithm defines two intermediate velocities $\bm{v}^-$ and $\bm{v}^+$ by the relations
\begin{align*}
\bm{v}_{n-1/2} & = \bm{v}^- - \frac{h}{2}\cdot\frac{q}{m}\cdot\bm{E}_n, \\[0.25em]
\bm{v}_{n+1/2} & = \bm{v}^+ + \frac{h}{2}\cdot\frac{q}{m}\cdot\bm{E}_n.
\end{align*}
Substituting the above expressions into the Lorentz force equation \eqref{Lorentz} cancels the $\bm{E}_n$ term resulting in the magnetic rotation equation
\begin{equation}\label{Brotation}
\frac{\bm{v}^+ - \bm{v}^-}{h} = \frac{q}{m}\left(\frac{\bm{v}^+ + \bm{v}^-}{2}\times\bm{B}_n\right).
\end{equation}
Thus, the actions due to the electric field $\bm{E}_n$ and the magnetic field $\bm{B}_n$ are decoupled and velocity is updated in a Strang-like splitting scheme as follows:
\begin{table}[h!]
\centering
\begin{tabular}{lll}
i. & First-half electric push & $\bm{v}^- = \bm{v}_{n-1/2} + \dfrac{h}{2}\cdot\dfrac{q}{m}\cdot\bm{E}_n$; \\[1em]
ii. & Magnetic rotation & $\dfrac{\bm{v}^+ - \bm{v}^-}{h} = \dfrac{q}{m}\left(\dfrac{\bm{v}^+ + \bm{v}^-}{2}\times\bm{B}_n\right)$; \\[1em]
iii. & Second-half electric push & $\bm{v}_{n+1/2} = \bm{v}^+ + \dfrac{h}{2}\cdot\dfrac{q}{m}\cdot\bm{E}_n$.
\end{tabular}
\end{table}
Figure \ref{Borisvupdate} illustrates an example of the velocity update in a configuration where the magnetic field is pointing out of the plane of the page, the electric field points from left to right, and the initial particle velocity $\bm{v}_{n-1/2}$ is perpendicular to the magnetic field.

\vfill

\begin{figure}[h!]
\begin{center}
\begin{tikzpicture}
\draw[thick,->] (0,0)--(2.12,-1.5) node at (1.3,-1.3) {$\bm{v}^+$};
\draw[thick,->] (0,0)--(2.12,1.5) node at (1.3,1.3) {$\bm{v}^-$};
\draw[thin,<-] (1.414/2,-0.5) arc (-30:0:1) node[right] at (0.8,-0.35) {$\alpha = \frac{\theta}{2}$};
\draw[thin] (1.414/2,0.5) arc (30:0:1) node[right] at (0.8,0.35) {$\alpha = \frac{\theta}{2}$};
\draw[thick,->] (2.12,1.5)--(3,0) node[right] at (2.6,0.75) {$\bm{v}^-\times w\,\hat{\bm{B}}$};
\draw[ultra thin] (1.885,1.33)--(2.033,1.083)--(2.284,1.245);
\draw[thick,->] (0,0)--(3,0) node[below] at (2.5,0) {$\bm{v'}$};
\draw (5,0) circle (0.1);
\fill (5,0) circle (1pt) node[right] {$\bm{B}$};
\end{tikzpicture}
\end{center}
\caption{Vector $\bm{v}'$ bisects the magnetic rotation angle $\theta$. A right triangle is formed by the vectors $\bm{v}^-$, $\bm{v}'$, and $\bm{v}^-\times w\,\hat{\bm{B}}$ for some scalar $w$.}\label{vprime}
\end{figure}
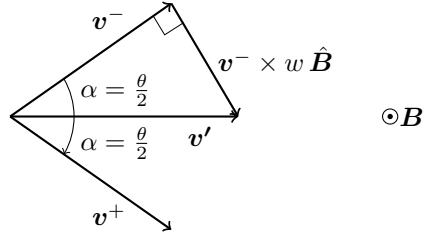

Observe that the magnetic rotation equation \eqref{Brotation} is an implicit expression in $\bm{v}^+$ and, therefore, requires inversion to get an explicit expression for $\bm{v}+$. The Boris algorithm achieves this inversion as follows. First, an intermediate velocity $\bm{v}'$ is defined to be the vector that bisects the magnetic rotation angle $\theta$ in the plane perpendicular to the magnetic field $\bm{B}$. Furthermore, $\bm{v}'$ is specified such that a right triangle is formed with $\bm{v}'$ as the hypotenuse and $\bm{v}^-$ as one of the legs. This implies that there exist a scalar $w$ such that the other leg of the triangle is given by
\[
\bm{v}^- \times w\,\hat{\bm{B}},
\]
where $\hat{\bm{B}}$ is the unit vector in the direction of $\bm{B}$; see figure \ref{vprime}. Letting $\alpha = \theta/2$, we see that
\[
\tan \alpha = \frac{|\bm{v}^- \times w \hat{\bm{B}}|}{|\bm{v}^-|} = \frac{|\bm{v}^-|w}{|\bm{v}^-|} = w.
\]
Hence, by straightforward vector addition
\[
\bm{v}' = \bm{v}^- + \bm{v}^- \times \tan\alpha \, \hat{\bm{B}}, 
\quad \alpha = \frac{\theta}{2} = \frac{h}{2}\cdot\frac{qB}{m}.
\]

\begin{figure}[h!]
\begin{center}
\begin{tikzpicture}
\draw[thick,->] (0,0)--(2.121,-1.5) node[below] {$\bm{v}^+$};
\draw[thick,->] (0,0)--(2.121,1.5) node[above] {$\bm{v}^-$};
\draw[thin,<-] (1.414/2,-0.5) arc (-30:0:1) node[right] at (0.8,-0.35) {$\alpha$};
\draw[thick,->] (0,0)--(3,0) node[right] {$\bm{v'}$};
\draw[thin] (1.414/2,0.5) arc (30:0:1) node[right] at (0.8,0.35) {$\alpha$};
\draw[ultra thin] (1.846,0)--(1.846,0.275)--(2.121,0.275);
\draw[thick,<-] (2.121,-1.5)--(2.121,1.5) node[right] at (2.1212,-0.75) {$\bm{v}'\times u\,\hat{\bm{B}} = \bm{v}^+ - \bm{v}^-$};
\draw (5,1) circle (0.1);
\fill (5,1) circle (1pt) node[right] {$\bm{B}$};
\end{tikzpicture}
\end{center}
\caption{Vector $\bm{v}'\times u\,\hat{\bm{B}}$ is equal to $\bm{v}^+ - \bm{v}^-$ and perpendicular to both $\bm{v}'$ and $\bm{B}$.}\label{sinalpha}
\end{figure}
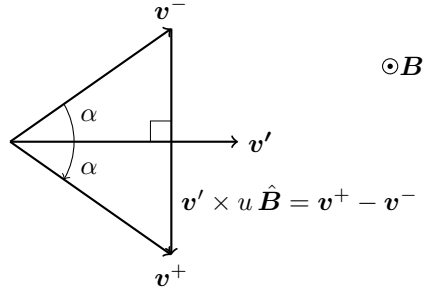

The algorithm next solves for the vector $\bm{v}^+ - \bm{v}^-$ by making use of the fact that it is perpendicular to both $\bm{v}'$ and $\bm{B}$. Hence, there exist some scalar $u$ such that
\[
\bm{v}^+ - \bm{v}^- = \bm{v}' \times u\,\hat{\bm{B}}.
\]
To find the value of $u$, refer to figure \ref{sinalpha} and observe that
\[
\sin\alpha = \frac{\frac{1}{2}|\bm{v}'\times u\,\hat{\bm{B}}|}{|\bm{v}^+|} = \frac{|\bm{v}'|u}{2|\bm{v}^+|}.
\]
Solving for $u$ gives
\[
u = \frac{2|\bm{v}^+|\sin\alpha}{|\bm{v}'|}.
\]
Substituting $\bm{v}' = \bm{v}^- + \bm{v}^- \times \tan\alpha \hat{\bm{B}}$ and making use of the fact $|\bm{v}^+| = |\bm{v}^-|$ ($\bm{v}^+$ is $\bm{v}^-$ rotated by angle $\theta$)
\[
u = \frac{2\sin\alpha}{\sqrt{1 + \tan^2\alpha}} = \frac{2\tan\alpha}{1 + \tan^2\alpha}.
\]
Thus, the update from $\bm{v}^-$ to $\bm{v}^+$ is given by the formula
\[
\bm{v}^+ = \bm{v}^- + \bm{v}' \times \frac{2\tan\alpha}{1 + \tan^2\alpha}\hat{\bm{B}}.
\]

Similar to the Buneman algorithm, the Boris algorithm was introduced at a time (1970) when the evaluation of transcendental functions was computationally expensive. Therefore, implementations of the Boris algorithm typically use the small angle approximation\
\[
w = \alpha \approx \tan\alpha
\]
in step 2 above.

\begin{algorithm}
\caption{Boris Velocity Push}\label{BorisPush}
\KwInput{$h$, $q$, $m$, $\bm{B}_n$, $B = \|\bm{B}_n\|$, $\bm{E}_n$, $\bm{v}_{n-1/2}$} \vspace*{0.5em}
\KwOutput{$\bm{v}_{n+1/2}$} \vspace*{1em}
\begin{algorithmic}[1]
\State$\bm{v}^- \gets \bm{v}_{n-1/2} + \dfrac{h}{2}\cdot\dfrac{q}{m}\cdot\bm{E}_n$ \vspace*{1em}
\State $w \gets \dfrac{h}{2}\cdot\dfrac{qB}{m}$ \vspace*{1em}
\State $\bm{v}' \gets \bm{v}^- + \bm{v}^- \times \dfrac{1}{B}\cdot w\cdot\bm{B}_n$ \vspace*{1em}
\State $\bm{v}^+ \gets \bm{v}^- + \bm{v}' \times \dfrac{1}{B}\cdot\dfrac{2\cdot w}{1 + w^2}\cdot\bm{B}_n$ \vspace*{1em}
\State $\bm{v}_{n+1/2} \gets \bm{v}^+ + \dfrac{h}{2}\cdot\dfrac{q}{m}\cdot\bm{E}_n$
\end{algorithmic}
\end{algorithm}

\newpage
\section{Proof of Theorem \ref{matfuncthm} (Lagrange-Sylvester Interpolation Formula)}
\subsection*{Case: $A$ has $N$ distinct eigenvalues.}
If $A$ has $N$ distinct eigenvalues $\lambda_1,\lambda_2,\ldots,\lambda_N$, then its characteristic polynomial satisfies
\begin{align*}
\det(\lambda I - A) &= (\lambda - \lambda_1)(\lambda - \lambda_2)\cdots(\lambda - \lambda_N) \\
&= \lambda^N \, + \, \alpha_{N-1}\lambda^{N-1} \, + \, \ldots \, + \, \alpha_1\lambda \, + \, \alpha_0 \\
&= 0.
\end{align*}
Solving for $\lambda^N$ gives
\begin{equation}\label{N}
\lambda^N = -\alpha_{N-2}\,\lambda^{N-1} \, - \, \ldots \, - \, \alpha_1\,\lambda^2 \, - \, \alpha_0\,\lambda.
\end{equation}
In other words, $\lambda^N$ can be expressed in terms of $\lambda, \lambda^2, \ldots, \lambda^{N-1}$, i.e. a polynomial of (at most) degree $N\rm{-}1^\text{th}$.
\par\noindent
Multiplying equation \eqref{N} by $\lambda$ gives
\[
\lambda^{N+1} = -\alpha_{N-2}\,\lambda^{N} \, - \, \ldots \, - \, \alpha_1\,\lambda^3 \, - \, \alpha_0\,\lambda^2.
\]
Substituting equation \eqref{N} into the right-hand side and grouping powers of $\lambda$ yields
\[
\lambda^{N+1} = \alpha_{N-1}^{(1)}\,\lambda^{N-1} \, + \, \alpha_{N-2}^{(1)}\,\lambda^{N-2} \, + \, \ldots \, + \,  \alpha_2^{(1)}\,\lambda^2 \, + \, \alpha_1^{(1)}\,\lambda,
\]
for some coefficients $\alpha_1^{(1)},\alpha_2^{(1)},\ldots,\alpha_{N-1}^{(1)}$. It follows from induction that for any $k = 0,1,2,\ldots$,
\begin{equation}\label{lambda_N}
\lambda^{N+k} = \alpha_1^{(k)}\,\lambda \, + \, \alpha_2^{(k)}\,\lambda^2 \, + \, \ldots \, + \, \alpha_{N-1}^{(k)}\,\lambda^{N-1}.
\end{equation}
That is, $\lambda^{N+k}$ can always be expressed in terms of $\lambda, \lambda^2, \ldots, \lambda^{N-1}$, i.e. a polynomial of (at most) degree $N\rm{-}1^\text{th}$.
\par
Since $f(\lambda)$ is an analytic function, it has a convergent series expansion:
\begin{align*}
f(\lambda) &= c_0 \, + \, c_1\,\lambda \, + \, c_2\,\lambda^2 \, + \, \ldots \\
&= c_0 \, + \, c_1\,\lambda \, + \, \ldots \, + \, c_{N-1}\,\lambda^{N-1} \, + \, c_N\,\lambda^N \, + \, \ldots \, + \, c_{N+k}\,\lambda^{N+k} \, + \, \ldots \\
&= c_0 \, + \, c_1\,\lambda \, + \, \ldots \, + \, c_{N-1}\,\lambda^{N-1} \, + \, \sum_{k=0}^\infty c_{N+k}\,\lambda^{N+k}.
\end{align*}
From equation \eqref{lambda_N}, each term inside the summation can be expressed by a polynomial of (at most) $N\rm{-}1$\textsuperscript{th} degree. Making the substitutions and grouping powers of $\lambda$ gives the polynomial
\[
p(\lambda) = a_0 \, + \, a_1\,\lambda \, + \, \ldots \, + \, a_{N-1}\,\lambda^{N-1} = f(\lambda),
\]
for some coefficients $a_0,a_1,\ldots,a_{N-1}$. In other words, $f(\lambda)$ can be expressed by some polynomial $p(\lambda)$ of (at most) degree $N \rm{-} 1$.

To find an explicit expression for this polynomial, observe that
\begin{equation}\label{interp1}
p(\lambda_j) = f(\lambda_j)
\end{equation}
must hold true for each eigenvalue $\lambda_j$. This yields a system of $N$ linearly independent equations in $N$ coefficients $a_0,a_1,\ldots,a_{N-1}$. Hence, $p(\lambda)$ is the unique polynomial that interpolates $f(\lambda)$ on the spectrum of $A$.

Extending the series representation of $f(\lambda)$ to the matrix argument $A$, we have
\[
f(A) = c_0\,I + c_1\,A + c_2\,A^2 + \ldots.
\]
This implies that
\[
p(A) = a_0\,I \, + \, a_1\,A \, + \, \ldots \, + \, a_{N-1}\,A^{N-1} = f(A).
\]
 
\subsection*{Case: $A$ has repeated eigenvalues.}
For the case when $A$ has repeated eigenvalues, suppose $A$ has eigenvalues
\[
\lambda_1,\lambda_2,\ldots,\lambda_m
\]
with respective multiplicities
\[
r_1,r_2,\ldots,r_m,\]
where
\[
m \leq N \quad\text{and}\quad r_1 \, + \, r_2 \, + \, \ldots \, + \, r_m = N.
\]
Then the characteristic polynomial of $A$ satisfies
\begin{align*}
\det(\lambda I - A) &= (\lambda - \lambda_1)^{r_1}\,(\lambda - \lambda_2)^{r_2}\,\cdots\,(\lambda - \lambda_m)^{r_m} \\
&= \lambda^{N-1} \, + \, \alpha_{N-2}\,\lambda^{N-2} \, + \, \ldots \, + \, \alpha_1\,\lambda \, + \, \alpha_0 \\
&= 0.
\end{align*}
Following the same argument as in the previous case, for any $k = 0,1,\ldots$ $\lambda^{N+k}$ can be expressed by a polynomial of (at most) $N-1$ degree polynomial in $\lambda$. Hence, $f(\lambda)$ can be expressed by a polynomial of (at most) degree $N \rm{-} 1$:
\[
p(\lambda) = a_0 \, + \, a_1\,\lambda \, + \, \ldots \, + \, a_{N-1}\,\lambda^{N-1} = f(\lambda)
\]
for some coefficients $a_0,a_1,\ldots,a_{N-1}$.

For each eigenvalue $\lambda_j$ with multiplicity $r_j$, we make the observation that
\begin{equation}\label{interp2}
\begin{array}{cccl}
p(\lambda_j) & = & f(\lambda_j) & \text{interpolation condition,} \\[0.25em]
p'(\lambda_j) & = & f'(\lambda_j) & \text{1st osculating condition,} \\[0.25em]
p''(\lambda_j) & = & f''(\lambda_j) & \text{2nd osculating condition,} \\[0.25em]
\vdots & \vdots & \vdots & \qquad\vdots \\[0.5em]
p^{(r_j-1)}(\lambda_j) & = & f^{(r_j-1)}(\lambda_j) & r_j\rm{-}1\textsuperscript{th}\text{ osculating condition,}
\end{array}
\end{equation}
where the superscript denotes the order of the derivative with respect to $\lambda$. This establishes a system of $N$ linearly independent equations in $N$ coefficients $a_0,a_1,\ldots,a_{N-1}$. Hence, $p(\lambda)$ is the unique interpolation polynomial that satisfies $r_k \rm{-} 1$ osculating conditions for each eigenvalue $\lambda_j$ of multiplicity $r_j$. Consequently,
\[
p(A) = a_0\,I \, + \, a_1\,A \, + \, \ldots \, + \, a_{N-1}\,A^{N-1} = f(A).
\]
\qedsymbol

\vfill
\section*{Acknowledgements}
This work was supported in part by the Department of Energy, Contract DE-AC52 07NA27344 and the National Science Foundation, Award No. 1840265 and Award No. 2012875. The authors also thank Hubertus von Bremen, Valentin Dallerit and Jeffrey Parker for their helpful discussions.



\end{document}